\documentclass[10pt,epsfig]{article} 

\usepackage{lscape}
\usepackage{enumerate}
\usepackage{float}
\usepackage{subfig}
\usepackage{color}
\usepackage[table]{xcolor}
\usepackage{slashed}
\usepackage{multirow}
\usepackage{cite}

\let\counterwithin\relax
\usepackage{chngcntr}
\usepackage{amssymb, amsmath,mathrsfs}

\usepackage{graphics}
\usepackage{graphicx}
\usepackage{epsf}
\usepackage{epsfig}
\usepackage{float}
\usepackage{makecell}
\usepackage{multirow}
\usepackage{color}
\usepackage{xcolor}
\usepackage{simplewick}
\usepackage{longtable}

\usepackage[utf8]{inputenc}
\usepackage{amsmath}
\usepackage{amssymb}
\usepackage{subfig}
\usepackage[normalem]{ulem}
\usepackage[force]{feynmp-auto}
\usepackage{tikz}
\usepackage{tikz-feynhand}
\usepackage{simpler-wick}
\usepackage{mathtools}
\usepackage{amsmath}
\usepackage{diagbox}
\usepackage{mmacells}

\newcommand\undermat[2]{
	\makebox[0.5pt][l]{$\smash{\underbrace{\phantom{%
					\begin{matrix}#2\end{matrix}}}_{ \let\scriptstyle\textstyle\text{\large $#1$}}}$}#2}
\newcommand\overmat[2]{
	\makebox[-1pt][l]{$\smash{\overbrace{\phantom{%
					\begin{matrix}#2\end{matrix}}}^{ \let\scriptstyle\textstyle\text{\large $#1$}}}$}#2}    
\usepackage{tikz}
\usepackage[vcentermath]{youngtab}
\usepackage{slashed} 
\usepackage[font=small]{caption} 
\usepackage{times} 

\usepackage{bm}       
\usepackage{bbm} 

\usepackage{relsize}  

\usepackage{autobreak}  

\usepackage{makeidx} 
\usepackage{bbding} 
\usepackage{listings} 
\usepackage{ytableau}

\long\def\rpl#1!!#2!!{\textcolor{red}{#1} \textcolor{blue}{#2}}

\usepackage[top=1in, left=0.95in, bottom=1.1in, right=0.95in]{geometry}
\def\baselinestretch{1.27}
\usepackage[toc]{appendix}

\newcommand{\beq}{\begin {equation}}  
\newcommand{\eeq}{\end   {equation}} 
\newcommand{\bea}{\begin {eqnarray}} 
\newcommand{\eea}{\end   {eqnarray}}  
\newcommand{\baa}{\begin {array}   } 
\newcommand{\eaa}{\end   {array}   }     
\newcommand{\bit}{\begin {itemize} }
\newcommand{\eit}{\end   {itemize} }
\newcommand{\be }{\begin {equation}} 
\newcommand{\ee }{\end   {equation}}

\newcommand{\mc}[1]{\mathcal{#1}}

\newcommand{\ket}[1]{| #1 \rangle}
\newcommand{\bra}[1]{\langle #1 |}
\newcommand{\vev}[1]{ \left\langle {#1}  \right\rangle }

\newcommand{\ie}{{\text{i.e.}}~}

\newcommand{\eq}[1]{\begin{equation}\begin{split} #1 \end{split}\end{equation}}
\newcommand{\eqs}[1]{\begin{align} #1 \end{align}}

\newcommand{\4}{\!\!\!\!/\,}
\newcommand{\comment}[1]{}

\newcommand{\mbf}[1]{\mathbf{#1}}
\newcolumntype{M}[1]{>{\centering\arraybackslash}m{#1}}
\newcolumntype{N}{@{}m{0pt}@{}}

\allowdisplaybreaks
\begin{document}

\begin{center}

{\Large \textbf  {Complete UV Resonances of the Dimension-8 SMEFT Operators}}\\[10mm]

Hao-Lin Li$^{c}$\footnote{haolin.li@uclouvain.be}, Yu-Han Ni$^{a, b}$\footnote{niyuhan@itp.ac.cn}, Ming-Lei Xiao$^{d,e,f}$\footnote{xiaomlei@mail.sysu.edu.cn}, Jiang-Hao Yu$^{a, b, g, h, i}$\footnote{jhyu@itp.ac.cn}\\[10mm]

\noindent 
$^a${\em \small CAS Key Laboratory of Theoretical Physics, Institute of Theoretical Physics, Chinese Academy of Sciences,    \\ Beijing 100190, P. R. China}  \\
$^b${\em \small School of Physical Sciences, University of Chinese Academy of Sciences,   Beijing 100049, P.R. China}   \\
$^c${\em \small Centre for Cosmology, Particle Physics and Phenomenology (CP3), Universite Catholique de Louvain,\\
Chem. du Cyclotron 2, 1348, Louvain-la-neuve, Belgium}\\
$^d${\em \small High Energy Physics Division, Argonne National Laboratory, Argonne, IL 60439, USA}\\
$^e${\em \small Department of Physics and Astronomy, Northwestern University, Evanston, IL 60208, USA}\\
$^f${\em \small School of Science, Sun Yat-Sen University, Shenzhen 518100, P. R. China} \\
$^g${\em \small Center for High Energy Physics, Peking University, Beijing 100871, China} \\
$^h${\em \small School of Fundamental Physics and Mathematical Sciences, Hangzhou Institute for Advanced Study, UCAS, Hangzhou 310024, China} \\
$^i${\em \small International Centre for Theoretical Physics Asia-Pacific, Beijing/Hangzhou, China}\\
[10mm]

\date{\today}   
          
\end{center}

\begin{abstract}

The effective field theory approach parameterizes the low energy behaviors of all possible ultraviolet (UV) theories in a systematic way. One of the most important tasks is thus to find the connection between the effective operators and their UV origins. The redundancy relations among operators make the connection very subtle, hence we proposed the J-basis prescription to illuminate the correspondence between operators and their UV resonances in the bottom-up way. 
In this work, we work out the dimension-8 J-basis operators in the standard model effective field theory (SMEFT), and find all the 146 (82) tree-level UV resonances along with their couplings up to mass dimension 5 (4). Furthermore, we point out a few subtleties on operator generation via field redefinition and on the UV Lagrangian for generic spin resonances. We also provide a data base storing our results and a \texttt{Mathematica} notebook for extracting those results for the reader's conveinence.

\end{abstract}

\newpage

\setcounter{tocdepth}{4}
\setcounter{secnumdepth}{4}

\tableofcontents

\setcounter{footnote}{0}

\def\baselinestretch{1.5}
\counterwithin{equation}{section}

\newpage

\section{Introduction}

%
%
%
%
%
%

The standard model effective field theory (SMEFT) provides the most general parametrization on all possible Lorentz-invariant new physics, encoded in the Wilson coefficients of the effective operators, see Ref.~\cite{Isidori:2023pyp, Falkowski:2023hsg} for recent review. These effective operators are written based on the standard model field building blocks, following the Lorentz and gauge symmetry. They are organized order by order via the canonical mass dimension and they form the complete and independent basis at each order, and the operator basis has been enumerated up to
dimension 9 and higher~\cite{Weinberg:1979sa, Buchmuller:1985jz, Grzadkowski:2010es, Lehman:2014jma, Liao:2016hru, Li:2020gnx, Murphy:2020rsh, Li:2020xlh, Liao:2020jmn, Harlander:2023psl}, for generic procedure to all order, see Ref.~\cite{Li:2022tec}.
After obtaining the complete basis of effective operators, the next question would be what kinds of new physics should be identified for each type of effective operators. Current experimental data show null results on new physics searches, which indicates experimental constraints on
the Wilson coefficients of the SMEFT operators, e.g. Ref.~\cite{Ellis:2020unq} for a global analysis. Once future data exhibit significant deviation from the SM prediction, the corresponding effective operator should be specified and their Wilson coefficients can thus be determined.

There are usually two ways to find the connection between the effective operators and their possible UV origins: the top-down approach and the bottom-up approach. 
In the top-down approach, a well-motivated UV model is specified with field contents and symmetry, and there are scale separation between new particles and SM particles. Then one performs the matching procedure by integrating out the heavy degrees of freedom in this model, and obtains the non-vanishing Wilson coefficients for some effective operators, which is a subset of the complete and independent operators. In this approach, it is easy to systematically identify all possible effective operators corresponding to single UV model. However, it is quite difficult to find all the possible UV origins of effective operator without performing an exhaustive search of various possible UV models, which is very time-consuming and error-prone due to the large variety of UV models. 
The work along this direction includes the classification of the tree-level and loop-level generated operators ~\cite{Arzt:1994gp, Einhorn:2013kja, Gargalionis:2020xvt, DasBakshi:2021xbl, Naskar:2022rpg, Banerjee:2022thk, Cepedello:2022pyx, Banerjee:2023iiv}, the complete tree-level and one-loop dictionary for the dimension-6 operators~\cite{deBlas:2017xtg,Guedes:2023azv}, and the dimension-7 operators~\cite{Li:2023cwy}, etc.

In the bottom-up approach, the aim is to find all the possible UV origins of an effective operator without writing down explicit UV Lagrangian, and it can be realized by examining the Lorentz and gauge quantum numbers for the contact on-shell amplitudes generated by the J-basis operators, proposed in Ref.~\cite{Li:2022abx}. This approach is made possible by introducing the J-basis of the effective operators \cite{Jiang:2020rwz,Li:2022abx,Shu:2021qlr}, together with the whole mechanism of Young tensor method \cite{Li:2020gnx,Li:2020xlh,Li:2020tsi,Li:2020zfq,Li:2021tsq,Li:2022tec} including the operator reduction procedure. It is well-known that a lot of equivalence relations exist among the effective operators, and thus we have a large variety of different operator bases to choose from. One way to choose the basis, which results in the J-basis, is to find the eigenstates of all the symmetry groups, including the Poincar\'e group and the internal symmetry groups, when their algebra act on a ``partition'' of the external particles. The $J$ in the name J-basis comes from the frame-independent angular momenta of an operator with respect to a partition defined by the Casimir $W^2$ of the Poincar\'e group. From the perspective of looking for UV origins of effective operators, the J-basis has the privilege that the quantum numbers of the possible UV resonances in a particular tree partition are fixed by the corresponding quantum numbers of the amplitude and the associated operator. Therefore, it becomes straightforward to deduce the possible resonances in the UV origin of an operator. 


In this paper, we for the first time clarify the conventions for the couplings involving massive particles of spin $J\geq 1$.
First, for these higher-spin massive fields in the Lagrangian, there are different choices of gauge for their couplings to the other fields that lead to different conclusions regarding whether they are responsible for certain low-energy effective operators. In particular, the source they couple to can also be decomposed into different irreducible tensors with certain spins, while the components with mismatched spins can always be converted to operators without these fields via a field redefinition. For example, the current coupling to a vector field can be decomposed into a conserved current and a total derivative, the latter being a spin-0 component and can be removed. This prescription is related to the on-shell amplitude/operator correspondence \cite{Ma:2019gtx,Shadmi:2018xan,Li:2020gnx}, as the couplings with non-conserved current does not produce on-shell spin-1 particle. It also leads to the justification of our J-basis algorithm, since the spin-0 coupling generates $J=0$ amplitude and the corresponding $J=0$ effective operator at low energy, not a $J=1$ operator indicated by the resonance as it cannot be generated on-shell. 
Second, we enumerate the tree-level couplings involving massive particles of spin $J\geq 1$ in the UV theory in terms of the external particle species. We adopt the canonically normalized fields for all the particles, so that the bosons are all dimension-$1$ and the fermions are all dimension-$3/2$. Based on these tree-level couplings, we could classify the effective operators into tree-level generated and loop-induced, which is an important piece of information in phenomenology.
Thirdly, we resolve the subtleties related to the generation of the operator by field redefinitions. We build a diagrammatic correspondence between the operator generation via field redefinition and the substitution of the heavy resonance with SM particles of the same quantum numbers in the tree diagram. In this way, from a seed UV theory obtained by our ordinary J-basis analysis, one can recursively remove relevant heavy resonances in the diagram to obtain a series of UV theories that can generate the operator under analysis via field redefinition after the top-down matching.

For the convenience of our readers, we have created a database in the form of a \texttt{Mathematica} package that stores all possible tree-level UV theories for each dimension-8 operator in the arXiv webpage. Each UV theory is defined by a list of new fields and new interactions, with interactions colored in red if their possible minimal dimension is greater than 4 according to our convention. Using this database, we have identified a total of 146 (82) UV resonances that can participate in the generation of dimension-8 operators at the tree level when the maximum allowed interaction dimension is set to 5 (4). Among these resonances, 88 (35) are new and not previously involved in the tree-level generation of dimension-6 and dimension-7 operators~\cite{deBlas:2017xtg,Li:2022abx}.
However, for those restricted to a maximum dimension-4 UV interactions, all 35 of the new resonances alone are incapable of generating dimension-8 operators. They must accompany at least one other resonance that also contributes to dimension-6 or dimension-7 operators. 

This paper is organized as follows. In section~\ref{sec:2}, we review the basic knowledge of our Young Tensor method, with emphasis on the method of construction of j-basis operator. In section.~\ref{sec:3}, we explain the meaning of renormalizable operators and tree-level interactions in an on-shell convention, and list all the amplitude form for renormalizable 3-pt and 4-pt interactions involving different number of massive particles. In section.~\ref{sec:3} we also explain how to include those UV theories that generate a given operator via field redefinition.
In section.~\ref{sec:4}, we summarize the procedure for obtaining a complete set of UV theories for a given operator type using the updated J-basis analysis 
and introduce a database with auxiliary functions to extract all the UV theories related to a given operator in the form of a \texttt{Mathematica} package. In the meantime, we list the operator types that must be generate with at least one new UV resonance, the list of UV theoreies that does not generate dimension 6 operators, and finally the detailed information for the J-basis for each operator type.
We conclude in section.~\ref{sec:con}.

\section{Brief Review on J-basis and UV Correspondence}\label{sec:2}


In our previous work \cite{Li:2022abx} we introduced a systematic method to obtain UV resonances by construction of j-basis of SMEFT. In this section, we briefly review the whole construction process. 

\subsection{Operator-Amplitude Correspondence and Operator Bases} 

\subsubsection{Operator-Amplitude Correspondence}

There is a one-to-one map between contact amplitudes and irrelevant EFT operators, also shown in \cite{Li:2022abx,Li:2020gnx,Li:2020xlh,Ma:2019gtx}. A following problem comes from the covariant derivative $D$, causing a condition that one operator probably contributes to several vertices. Therefore we only pay attention to the leading vertex with minimal fields, treating $D$ as $\partial$, and $F_{\mu\nu}$ as $\partial_\mu A_\nu -\partial_\nu A_\mu$. The two other redundancies of operators, integration by part (IBP) and equation of motion (EoM), can be erased automatically by on-shell amplitude basis construction. To be more specific, the on-shell operator-amplitude correspondence is an isomorphism between the linear spaces of irrelevant operators and independent contact amplitudes. Thus the key idea is (i) to find the map between the building blocks of operators and amplitudes, (ii) to construct an independent amplitude basis, so that we obtain an independent operator basis. 

We shall start with the building blocks of operators and amplitudes, which are the representations of Lorentz group $SO(3,1) \sim SL(2,\mathbb{C}) = SU(2)_l \times SU(2)_r$ and gauge group $SU(3)_C \otimes SU(2)_W \otimes U(1)_Y$. The operator blocks of the Lorentz symmetry are
\begin{equation}
\begin{array}{c|c|c|c|c|c|c}
\mbox{Irrep} & (0,0) & (\frac{1}{2},0) & (0,\frac{1}{2}) & (1,0) & (0,1) & (\frac{1}{2},\frac{1}{2}) \\
\hline
\mbox{Operator block} & \phi & \psi & \psi^\dagger & {F_L}= \frac{1}{2} (F+i\Tilde{F}) & {F_R}= \frac{1}{2} (F-i\Tilde{F}) & D
\end{array}. 
\end{equation}
Meanwhile the amplitudes are expressed in spinor-helicity formalism, where a massless particle $i$ of helicity $h_i$ contributes $\lambda_i^{r_i-h_i} \Tilde{\lambda}_i^{r_i+h_i}$. With a proper set of $\{h_i,r_i\}$, a Lorentz singlet can be determined. Also, since $p_i=\lambda_i \Tilde{\lambda}_i \sim (\frac{1}{2},\frac{1}{2}) \sim D$, the following translation can be easily inferred
\begin{equation}
\begin{array}{c|c|c|c|c|c}
\mbox{Amplitude Blocks} & \lambda_i^{r_i} \Tilde{\lambda}_i^{r_i} & \lambda_i^{r_i+1/2} \Tilde{\lambda}_i^{r_i-1/2} & \lambda_i^{r_i-1/2} \Tilde{\lambda}_i^{r_i+1/2} & \lambda_i^{r_i+1} \Tilde{\lambda}_i^{r_i-1} & \lambda_i^{r_i-1} \Tilde{\lambda}_i^{r_i+1} \\
\hline
\mbox{Operator Blocks} & D^{r_i} \phi_i & D^{r_i-1/2} \psi_i & D^{r_i-1/2} \psi^\dagger_i & D^{r_i-1} F_L & D^{r_i-1} F_R 
\end{array}.
\end{equation}
Such a translation implies that the covariant derivative commutator (CDC) is reduced, which satisfies the choice of leading vertex, because the helicity indices of the derivatives on a field are symmetric without contraction, or there will be $[D,D]=F$ or EoM with the corresponding amplitude as $0$.

\subsubsection{Operator Bases}

In \cite{Li:2020gnx,Li:2020xlh}, a complete and independent operator basis is given by the semi-standard Young tableau (SSYT) and generalized Littlewood-Richardson (LR) rule for Lorentz and gauge group respectively. 

The Young diagram of $SL(2,\mathbb{C}) \times U(N)$ is
\newline\newline
\begin{equation}
[\mathcal{M}]_{N,\tilde{n},n}= \arraycolsep=0pt\def\arraystretch{1}
 \rotatebox[]{90}{\text{$N-2$}} 
 \left\{
 \begin{array}{cccccc}
  \yng(1,1) &\ \ldots{}&\ \yng(1,1)& \overmat{n}{\yng(1,1)&\ \ldots{}\  &\yng(1,1)} \\
  \vdotswithin{}& & \vdotswithin{}&&&\\
  \undermat{\tilde{n}}{\yng(1,1)\ &\ldots{}&\ \yng(1,1)} &&&
 \end{array}
 \right. ,\label{eq:yt}
\end{equation}
\newline\newline
where $N$ is the number of particles. The number of columns for $SL(2,\mathbb{C}) \times U(N)$ irrep $n$ and conjugate irrep $\Tilde{n}$ are determined by
\begin{eqnarray}\label{eq:Lorentz-y}
n = \frac{k}{2} + \sum_{h_i<0}h_i \ ,\quad \tilde{n} = \frac{k}{2} + \sum_{h_i>0}h_i \ , \\
\# i = \tilde{n} - 2 h_i \ ,
\end{eqnarray}
where $k$ is the number of derivative, $h_i$ is the helicity of particle $i$, and $\# i$ is the number of label $i$ appearing in the SSYTs. The SSYTs are translated into amplitudes as
\begin{eqnarray}
\langle ij\rangle \sim \young(i,j),\quad 
[ij] \sim \mathcal{E}^{k_1...k_{N-2}ij} \tilde{\lambda}_i\tilde{\lambda}_j \sim \rotatebox[]{90}{\text{$N-2$}} \left\{ \begin{array}{c}
\ytableausetup{centertableaux, boxsize=2em} \begin{ytableau} k_1 \\ k_2 \end{ytableau}
 \\
\vdotswithin{}\\
\begin{ytableau}
\scriptstyle k_{N-2}
\end{ytableau}
\end{array}\right. ,
\end{eqnarray}
and the full amplitude is the product of all columns. Once the parameters are determined, the SSYTs give a complete and independent amplitude basis, free from IBP, EoM and Schouten identity redundancies. 

For example, the helicity category of type $D {e_{\mathbb{C}}}^\dagger H Q^3 $ is $\{-\frac{1}{2},-\frac{1}{2}, -\frac{1}{2},0,+\frac{1}{2}\}$, and there is 1 momentum. The shape of SSYTs and the number of each particle label are determined by Eq.~\eqref{eq:Lorentz-y}, leading to the Lorentz y-basis 
\begin{eqnarray}\label{eq:ex.y}
\mathcal{B}^y = \begin{pmatrix}
\langle12\rangle \langle34\rangle [45] \\ -\langle13\rangle \langle23\rangle [35] \\ \langle13\rangle \langle24\rangle [45]
\end{pmatrix}. 
\end{eqnarray}

The gauge y-basis construction is based on the tensor representation of $SU(N)$ gauge group, where the fields are written in fundamental indices only. For fields $\mathcal{F}$ in anti-fundamental or ajoint representations, we convert them into those with $N-1$ of $N$ fundamental indices respectively
\begin{eqnarray}\label{eq:tofundamental}
\mathcal{F}^A&\rightarrow& \mathcal{F}_{a_1...a_{N-1}i}\equiv \mathcal{F}^A (T^A)^{a_N}_i \epsilon_{a_1...a_N} \sim \ytableausetup{boxsize=2.5em} \begin{ytableau}
a_1 & i \\ \none[\vdots] \\ a_{M-1}
\end{ytableau};\\
\mathcal{F}^a&\rightarrow& \mathcal{F}_{a_1...a_{N-1}}\equiv \mathcal{F}^a \epsilon_{a_1...a_{N-1}a} \sim \begin{ytableau}
a_1 \\ \none[\vdots] \\ a_{N-1}
\end{ytableau}.
\end{eqnarray}
Then we apply the generalized L-R rule to Young tableaux to construct a series of $SU(N)$ singlets, where each column represents an $\epsilon$ tensor, and each expression within a y-basis is the product of the $\epsilon$ tensors derived from its Young tableau. 

Let us study the $SU(2)_W$ structure of type $D {e_{\mathbb{C}}}^\dagger H Q^3 $. There are 4 doublets 
\begin{eqnarray}
{Q}_{i_1}, {Q}_{i_2}, {Q}_{i_3}, H_{i_4}. 
\end{eqnarray}
According to the generalized L-R rule, the $SU(2)_W$ y-basis is constructed as
\begin{eqnarray}\label{eq:m-basis}
\ytableausetup{boxsize=1.2em}
\begin{ytableau}
i_1
\end{ytableau} \mathop{\rightarrow}\limits^{\begin{ytableau}
i_2
\end{ytableau}} \begin{ytableau}
i_1 & i_2
\end{ytableau} \mathop{\rightarrow}\limits^{\begin{ytableau}
i_3
\end{ytableau}} \begin{ytableau}
i_1 & i_2 \\ i_3
\end{ytableau} \mathop{\rightarrow}\limits^{\begin{ytableau}
i_4
\end{ytableau}} \begin{ytableau}
i_1 & i_2 \\ i_3 & i_4 
\end{ytableau} \sim \mathcal{T}^y_2 = \epsilon^{i_1 i_3} \epsilon^{i_2 i_4}, \\
\mathcal{T}^y_1 = \epsilon^{i_1 i_2} \epsilon^{i_3 i_4}. 
\end{eqnarray}

The y-basis is not necessarily composed of monomials if there were ajoint representation, because of Eq.~\eqref{eq:tofundamental}, but we can always choose a group of independent and complete monomials from the y-basis, namely the monomial basis (m-basis) \cite{Li:2020gnx,Li:2020xlh,Li:2022tec}. In the former example $D {e_{\mathbb{C}}}^\dagger H Q^3 $, we are lucky to find out that its $SU(2)_W$ y-basis is composed of monomials. Along with the $SU(3)_C$ structure, the m-basis of type $D {e_{\mathbb{C}}}^\dagger H Q^3 $ is
\begin{eqnarray}
\mathcal{O}^m = \begin{pmatrix}
\mathcal{O}^m_1 \\ \mathcal{O}^m_2 \\ \mathcal{O}^m_3 \\ \mathcal{O}^m_4 \\ \mathcal{O}^m_5 \\ \mathcal{O}^m_6
\end{pmatrix} = \begin{pmatrix}
i \epsilon^{abc} \epsilon^{ik} \epsilon^{jl} (D_\mu H_l) ({Q_s}_{ck} \sigma^\mu {e_{\mathbb{C}}}^\dagger_t) ({Q_p}_{ai} {Q_r}_{bj}) \\ i \epsilon^{abc} \epsilon^{ik} \epsilon^{jl} H_l ({Q_p}_{ai} \sigma^\mu {e_{\mathbb{C}}}^\dagger_t) ({Q_r}_{bj} (D_\mu {Q_s}_{ck})) \\ i \epsilon^{abc} \epsilon^{ik} \epsilon^{jl} (D_\mu H_l) ({Q_r}_{bj} \sigma^\mu {e_{\mathbb{C}}}^\dagger_t) ({Q_p}_{ai} {Q_s}_{ck}) \\ i \epsilon^{abc} \epsilon^{ij} \epsilon^{kl} (D_\mu H_l) ({Q_s}_{ck} \sigma^\mu {e_{\mathbb{C}}}^\dagger_t) ({Q_p}_{ai} {Q_r}_{bj}) \\ i \epsilon^{abc} \epsilon^{ij} \epsilon^{kl} H_l ({Q_p}_{ai} \sigma^\mu {e_{\mathbb{C}}}^\dagger_t) ({Q_r}_{bj} (D_\mu {Q_s}_{ck})) \\ i \epsilon^{abc} \epsilon^{ij} \epsilon^{kl} (D_\mu H_l) ({Q_r}_{bj} \sigma^\mu {e_{\mathbb{C}}}^\dagger_t) ({Q_p}_{ai} {Q_s}_{ck})
\end{pmatrix}, 
\end{eqnarray}
where $\{a,b,c\}$ are $SU(3)_C$ fundamental indices, $\{i,j,k,l\}$ $SU(2)$ fundamental, and $\{p,r,s,t\}$ flavor. 

In some cases it is inevitable to deal with repeated fields. If we want to have a flavor specified independent basis (discussed in \cite{Li:2022tec}), we need to rearrange the operators into irreps of $SU(n_f)$ flavor group. The process from m-basis to flavor-specified basis (f-basis) requires 2 steps. First we construct irreps of permutation group $S_m$ and obtain the permutation basis (p-basis), where $m$ is the number of repeated fields. According to Schur-Weyl duality, any irrep of $S_m$ is also an irrep of $SU(n_f)$. Then we select the operators that span different $SU(n_f)$ tensor spaces of each irrep, and obtain the f-basis. 

For example, type $D {e_{\mathbb{C}}}^\dagger H Q^3 $ has 3 repeated $Q$s. The irreps of $SU(n_f)$ ($S_3$) by $\{Q_r,Q_s,Q_t\}$ are $\{\young(rst),\young(rs,t),\young(r,s,t)\}$, with the tensors
\begin{eqnarray}
\mathcal{O}^p_{\tiny\yng(3)} &=& \mathcal{Y}[\young(rst)] \mathcal{O}^m_1, \\
\mathcal{O}^p_{\tiny\yng(2,1)} &=& \begin{pmatrix}
\mathcal{Y}[\young(rs,t)] \mathcal{O}^m_1 \\ (s \ t) \mathcal{Y}[\young(rs,t)] \mathcal{O}^m_1 \\ \mathcal{Y}[\young(rs,t)] \mathcal{O}^m_2 \\ (s \ t) \mathcal{Y}[\young(rs,t)] \mathcal{O}^m_2
\end{pmatrix}, \\ 
\mathcal{O}^p_{\tiny\yng(1,1,1)} &=& \mathcal{Y}[\young(r,s,t)] \mathcal{O}^m_1. 
\end{eqnarray}
Notice that $\mathcal{Y}[\young(rs,t)] $ and $(s\ t) \mathcal{Y}[\young(rs,t)]$ span the same linear space of irrep of $SU(n_f)$, so the f-basis is
\begin{eqnarray}
\mathcal{O}^f_{\tiny\yng(3)} &=& \mathcal{Y}[\young(rst)] \mathcal{O}^m_1, \\
\mathcal{O}^f_{\tiny\yng(2,1)} &=& \begin{pmatrix}
\mathcal{Y}[\young(rs,t)] \mathcal{O}^m_1 \\ \mathcal{Y}[\young(rs,t)] \mathcal{O}^m_2
\end{pmatrix}, \\ 
\mathcal{O}^f_{\tiny\yng(1,1,1)} &=& \mathcal{Y}[\young(r,s,t)] \mathcal{O}^m_1. 
\end{eqnarray}

So far the bases we have introduced do not directly predict the possible UV origins. From the group theory point of view, the UV particles generating the effective operators are the possible representations of the Poincar\'e group and the gauge group. This recognition leads to the partial wave expansion of amplitudes of a certain scattering process that gives the momentum $J$ and gauge quantum number $\mathbf{R}$ of the multi-particle state at a specific scattering channel. The traditional partial wave expansion mainly focuses on $2\rightarrow N$ scattering, but we can always generalize the concept to $M\rightarrow N$ scattering and multi-partite scattering, as is shown in \cite{Li:2022abx}, in order to obtain the complete possible tree-level UV origins. Along with the operator-amplitude correspondence, we can apply the partial wave expansion on a complete and independent operator basis under any possible topology, and gain the information of the resonances. The result is named j-basis. J-basis is an eigenbasis of the partial Casimir operators under the given topology. We shall discuss the details of j-basis construction after introducing the Casimir operators of Poincar\'e group and $SU(N)$ and the definition of partial Casimir operators in the next subsection. The key idea is to transfer the partial wave problem into a linear algebra problem about solving the eigenfunctions of partial Casimir operators. Fig.~\ref{fig:relation}, proposed in \cite{Li:2022abx}, gives the relations among all the bases mentioned above. 

\begin{figure}
    \centering
\tikzset{every picture/.style={line width=0.3pt}} 

\begin{tikzpicture}[x=0.75pt,y=0.75pt,yscale=-1,xscale=1]

\draw   (205.9,150.02) .. controls (205.9,131.13) and (225.87,115.82) .. (250.49,115.82) .. controls (275.12,115.82) and (295.09,131.13) .. (295.09,150.02) .. controls (295.09,168.9) and (275.12,184.21) .. (250.49,184.21) .. controls (225.87,184.21) and (205.9,168.9) .. (205.9,150.02) -- cycle ;
\draw [color={rgb, 255:red, 74; green, 144; blue, 226 }  ,draw opacity=1 ]   (296.89,166.11) -- (333.17,195.48) ;
\draw [shift={(334.73,196.73)}, rotate = 218.98] [color={rgb, 255:red, 74; green, 144; blue, 226 }  ,draw opacity=1 ][line width=0.75]    (10.93,-3.29) .. controls (6.95,-1.4) and (3.31,-0.3) .. (0,0) .. controls (3.31,0.3) and (6.95,1.4) .. (10.93,3.29)   ;
\draw [color={rgb, 255:red, 74; green, 144; blue, 226 }  ,draw opacity=1 ]   (205.9,162.93) -- (168.74,192.03) ;
\draw [shift={(167.16,193.27)}, rotate = 321.94] [color={rgb, 255:red, 74; green, 144; blue, 226 }  ,draw opacity=1 ][line width=0.75]    (10.93,-3.29) .. controls (6.95,-1.4) and (3.31,-0.3) .. (0,0) .. controls (3.31,0.3) and (6.95,1.4) .. (10.93,3.29)   ;
\draw [color={rgb, 255:red, 74; green, 144; blue, 226 }  ,draw opacity=1 ]   (253.65,101.11) -- (252.78,53.13) ;
\draw [shift={(252.75,51.13)}, rotate = 88.97] [color={rgb, 255:red, 74; green, 144; blue, 226 }  ,draw opacity=1 ][line width=0.75]    (10.93,-3.29) .. controls (6.95,-1.4) and (3.31,-0.3) .. (0,0) .. controls (3.31,0.3) and (6.95,1.4) .. (10.93,3.29)   ;
\draw [color={rgb, 255:red, 208; green, 2; blue, 27 }  ,draw opacity=1 ]   (340.13,192.4) -- (300.23,159.1) ;
\draw [shift={(298.69,157.82)}, rotate = 39.85] [color={rgb, 255:red, 208; green, 2; blue, 27 }  ,draw opacity=1 ][line width=0.75]    (10.93,-3.29) .. controls (6.95,-1.4) and (3.31,-0.3) .. (0,0) .. controls (3.31,0.3) and (6.95,1.4) .. (10.93,3.29)   ;
\draw [color={rgb, 255:red, 208; green, 2; blue, 27 }  ,draw opacity=1 ]   (161.76,185.47) -- (198.94,155.52) ;
\draw [shift={(200.5,154.27)}, rotate = 141.15] [color={rgb, 255:red, 208; green, 2; blue, 27 }  ,draw opacity=1 ][line width=0.75]    (10.93,-3.29) .. controls (6.95,-1.4) and (3.31,-0.3) .. (0,0) .. controls (3.31,0.3) and (6.95,1.4) .. (10.93,3.29)   ;
\draw [color={rgb, 255:red, 208; green, 2; blue, 27 }  ,draw opacity=1 ]   (243.74,52) -- (244.6,99.98) ;
\draw [shift={(244.64,101.98)}, rotate = 268.97] [color={rgb, 255:red, 208; green, 2; blue, 27 }  ,draw opacity=1 ][line width=0.75]    (10.93,-3.29) .. controls (6.95,-1.4) and (3.31,-0.3) .. (0,0) .. controls (3.31,0.3) and (6.95,1.4) .. (10.93,3.29)   ;
\draw   (214.01,17.34) -- (284.28,17.34) -- (284.28,46.8) -- (214.01,46.8) -- cycle ;
\draw   (334.73,191.53) -- (404.99,191.53) -- (404.99,221) -- (334.73,221) -- cycle ;
\draw   (88.79,193.27) -- (159.06,193.27) -- (159.06,222.73) -- (88.79,222.73) -- cycle ;
\draw  [draw opacity=0][line width=3]  (121.13,169.5) .. controls (121.08,168.92) and (121.03,168.35) .. (120.98,167.77) .. controls (116.82,114.08) and (149.78,65.96) .. (199,46.67) -- (246.57,158.77) -- cycle ; \draw  [color={rgb, 255:red, 248; green, 231; blue, 28 }  ,draw opacity=1 ][line width=3]  (121.13,169.5) .. controls (121.08,168.92) and (121.03,168.35) .. (120.98,167.77) .. controls (116.82,114.08) and (149.78,65.96) .. (199,46.67) ;  
\draw  [draw opacity=0][line width=3]  (333.02,246.86) .. controls (332.77,247.08) and (332.53,247.3) .. (332.28,247.52) .. controls (281.37,293.02) and (201.73,290.17) .. (154.39,241.15) .. controls (153.44,240.16) and (152.5,239.16) .. (151.58,238.14) -- (246.57,158.77) -- cycle ; \draw  [color={rgb, 255:red, 248; green, 231; blue, 28 }  ,draw opacity=1 ][line width=3]  (333.02,246.86) .. controls (332.77,247.08) and (332.53,247.3) .. (332.28,247.52) .. controls (281.37,293.02) and (201.73,290.17) .. (154.39,241.15) .. controls (153.44,240.16) and (152.5,239.16) .. (151.58,238.14) ;  
\draw  [draw opacity=0][line width=3]  (300.82,49.44) .. controls (301.47,49.73) and (302.12,50.04) .. (302.77,50.35) .. controls (350.12,73.06) and (376.06,120.65) .. (372.02,168.49) -- (246.57,158.77) -- cycle ; \draw  [color={rgb, 255:red, 248; green, 231; blue, 28 }  ,draw opacity=1 ][line width=3]  (300.82,49.44) .. controls (301.47,49.73) and (302.12,50.04) .. (302.77,50.35) .. controls (350.12,73.06) and (376.06,120.65) .. (372.02,168.49) ;  
\draw    (418.71,206.27) -- (467.86,206.27) ;
\draw [shift={(469.86,206.27)}, rotate = 180] [color={rgb, 255:red, 0; green, 0; blue, 0 }  ][line width=0.75]    (10.93,-3.29) .. controls (6.95,-1.4) and (3.31,-0.3) .. (0,0) .. controls (3.31,0.3) and (6.95,1.4) .. (10.93,3.29)   ;
\draw [shift={(416.71,206.27)}, rotate = 0] [color={rgb, 255:red, 0; green, 0; blue, 0 }  ][line width=0.75]    (10.93,-3.29) .. controls (6.95,-1.4) and (3.31,-0.3) .. (0,0) .. controls (3.31,0.3) and (6.95,1.4) .. (10.93,3.29)   ;
\draw   (477.06,190.67) -- (547.33,190.67) -- (547.33,220.13) -- (477.06,220.13) -- cycle ;

\draw (214.95,137.88) node [anchor=north west][inner sep=0.75pt]   [align=left] {{\Large y-basis}};
\draw (217.99,22) node [anchor=north west][inner sep=0.75pt]   [align=left] {{\large m-basis}};
\draw (93.97,197.93) node [anchor=north west][inner sep=0.75pt]   [align=left] {{\large j-basis}};
\draw (339.76,197.93) node [anchor=north west][inner sep=0.75pt]   [align=left] {{\large p-basis}};
\draw (481.25,195.33) node [anchor=north west][inner sep=0.75pt]   [align=left] {{\large f-basis}};
\draw (251.23,65.68) node [anchor=north west][inner sep=0.75pt]   [align=left] {\textcolor[rgb]{0.29,0.56,0.89}{monomial}};
\draw (188.85,184.23) node [anchor=north west][inner sep=0.75pt]    {$\textcolor[rgb]{0.29,0.56,0.89}{W^{2},\mathbb{C}_n}$};
\draw (267.04,179.21) node [anchor=north west][inner sep=0.75pt]   [align=left] {\textcolor[rgb]{0.29,0.56,0.89}{permute}};
\draw (420.78,187.35) node [anchor=north west][inner sep=0.75pt]   [align=left] {\begin{minipage}[lt]{32.21pt}\setlength\topsep{0pt}
\begin{center}
S-W
\end{center}
duality
\end{minipage}};

\end{tikzpicture}
    
    \caption{Relations among the different bases~\cite{Li:2022tec}. The red arrow means that the basis can always be expressed as linear combination of the y-basis through a reduce procedure. The yellow circle indicates that the m-basis, the p-basis and the j-basis are related by linear transformations. The operators in the p-basis are tensors of the symmetric group $S_m$, and those in the f-basis are tensors of $SU(n_f)$ group, where $m$ is the number of repeated fields and $n_f$ is the flavor number of each repeated field. These two basis are connected by the Schur-Weyl duality.}
    \label{fig:relation}
\end{figure}

\subsection{Pauli-Lubanski and Gauge Casimir Projection}\label{sec:PLGCP}

To do the partial wave expansion, first we need to determine all the scattering channels, which requires the concept of partition. Let us start with the traditional $2\rightarrow N$ scattering. This process involves $N+2$ particles totally, with the incoming channel $\mathcal{I}_1 =\{12\}$ and the outgoing one $\mathcal{I}_2= \{3...(N+2)\}$, so its partition is $\{12|3...(N+2)\}$. Notice that the partition satisfies 2 conditions that (i) $\mathcal{I}_1 \cup \mathcal{I}_2 = \{1,2,3,...,(N+2)\} $; and (ii) the 2 channels do not overlap $\mathcal{I}_1 \cap \mathcal{I}_2 = \varnothing $. This specific example also satisfies $\mathcal{I}_1 = \bar{\mathcal{I}}_2 $, where $\bar{\mathcal{I}}_i \equiv \{1,...,N,N+1,N+2\} - \mathcal{I}_i $, which is the feature of a 2-partite partition. 

Then we generalize 2-partite partitions to multi-partite partitions. For simplicity, we treat all particles as incoming ones, and suppose there are totally $N$ external particles. Intuitively we can obtain a multi-partite partition by dividing the $N$ external particles into several channels, satisfying the generalized conditions of (i) and (ii) mentioned in the last paragraph. To be more formal, an $m$-partite partition of $N$ external particles is denoted as $N|m \equiv \mathcal{I}_{1,...,m} \subset \{1,...,N\} $, with requirement (i) and the generalized one of (ii), (ii$'$)
\begin{eqnarray}
\mathcal{I}_i \cap \mathcal{I}_j = \varnothing \mbox{ or } {\mathcal{I}}_i \cap \bar{\mathcal{I}}_j = \varnothing, 
\end{eqnarray}
that any 2 channels are either non-overlapping or contain one another.
All the possible partitions of particles $\{1,...,N\}$ form a space $\mathbbm{P}_N$ with $S_N$ symmetry. If we apply the quotient set $\mathbbm{P}_N/S_N$, the result is a set of all possible topologies, usually labeled by diagrams. 
For example when $N=6$, one of all the possible partitions is $\{12|123|456|56\}$. Channel $\{12\}\subset \{123\}$ and $\{12\}\cap\{456\} = \varnothing $ satisfies the requirement (ii$'$). In this case, $\{12|123|456|56\}$ and $\{12|124|356|56\}$ belong to topology \begin{fmffile}{6-2}
\begin{fmfgraph*}(40,20)
\fmfpen{thin} \fmfsurroundn{e}{10} \fmfvn{}{g}{4} 
\fmf{plain}{e7,g1} \fmf{plain}{g1,e5} \fmf{plain}{g1,g2} \fmf{plain}{e4,g2} \fmf{phantom}{g2,e8} \fmf{phantom}{g3,e9} \fmf{plain}{g2,g3} \fmf{plain}{e3,g3} \fmf{plain}{g3,g4} \fmf{plain}{e2,g4} \fmf{plain}{g4,e10}
\end{fmfgraph*}
\end{fmffile}. 

The introducing of multi-partite partitions provides favorable conditions of defining partial Casimir operators. Casimir operators label the irreps of their group/algebra, which can be very useful to partial wave expansion. A partial Casimir operator thus labels the irreps of a specific channel, so that the resonance information of all scattering channels can be obtained if we consider all the partial Casimir operators. 

\subsubsection{Lorentz j-basis}

There are 2 Casimir operators within the Poincar\'e algebra, momentum $\mathbf{P}^2$ and Pauli-Lubanski operator $\mathbf{W}^2$, which label irreps of the Poincar\'e group
\begin{eqnarray}
\mathbf{W}^2|P,J,\sigma\rangle &=& -P^2 J (J+1) |P,J,\sigma\rangle, \\
\mathbf{P}^2|P,J,\sigma\rangle &=& P^2 |P,J,\sigma\rangle. 
\end{eqnarray}
The partial wave expansion of an $N$-pt scattering amplitude $\mathcal{M}(\{\Psi\}_{N|m})$ in partition $N|m$ is in the form
\begin{eqnarray}
\label{eq:pw_general}
    \mathcal{M}(\{\Psi\}_{N|m}) = \sum_{J_1,\dots,J_m} \sum_a \mathcal{M}^{J_1,\dots,J_m}_a(\{s\}_{N|m}) \overline{\mathcal{B}}^{J_1,\dots,J_m}_a(\{\Psi\}_{N|m}), 
\end{eqnarray}
where $\{\Psi\}_{N|m}$ denotes the states of $N$ particles divided into partition $N|m$, $J_i$ denotes the total angular momentum of channel $\mathcal{I}_i$, $\mathcal{M}^{J_1,\dots,J_m}_a(\{s\}_{N|m})$ contains the physical information of the scattering process and is irrelevant to the partial wave, $a$ is the possible degeneracy, and $\overline{\mathcal{B}}^{J_1,\dots,J_m}_a(\{\Psi\}_{N|m})$ is the partial wave basis
\begin{eqnarray}
\overline{\mathcal{B}}^{J_1,\dots,J_m}_a(\{\Psi\}_{N|m}) = (\prod\limits_{i=1}^{m} \langle P_i,J_i,\sigma_i|\{\Psi\}_{N_i}\rangle) b_a^{J_1,\dots,J_m}. 
\end{eqnarray}
If we only pay attention to channel $\mathcal{I}_i$, we can always act $\mathbf{W}^2$ on state $|P_i,J_i,\sigma_i\rangle$ 
\begin{eqnarray}
\langle P_i,J_i,\sigma_i| \mathbf{W}^2 |\{\Psi\}_{N_i}\rangle \equiv W_{\mathcal{I}_i}^2 \langle P_i,J_i,\sigma_i|\{\Psi\}_{N_i}\rangle = -P_i^2 J_i (J_i+1) \langle P_i,J_i,\sigma_i|\{\Psi\}_{N_i}\rangle, 
\end{eqnarray}
where $W_{\mathcal{I}_i}^2$ is defined as the partial Casimir operator of $\mathbf{W}^2 $ only concerning channel $\mathcal{I}_i$. The definition is essential, because an amplitude is always a Lorentz singlet, but part of the external particles in a specific channel may form a non-trivial irrep that can be detected by the partial Casimir operator of this channel. Since $W^2_{\mathcal{I}_i}$ is defined in the subspace $|\{\Psi\}_{N_i}\rangle$ in $|\{\Psi\}_{N}\rangle$, it must satisfy 
\begin{eqnarray}
W_{\mathcal{I}_i}^2 \langle P_j,J_j,\sigma_j|\{\Psi\}_{N_j}\rangle =0 \ ( \forall i\neq j) .
\end{eqnarray}
Therefore
\begin{eqnarray}\label{eq:W2B}
W^2_{\mathcal{I}_i} \overline{\mathcal{B}}^{J_1,\dots,J_m}_a(\{\Psi\}_{N|m}) = -s_{\mathcal{I}_i} J_i (J_i+1) \overline{\mathcal{B}}^{J_1,\dots,J_m}_a(\{\Psi\}_{N|m}),  
\end{eqnarray}
where $s_{\mathcal{I}_i} = P^2_{i} $ is the Mandelstam variable of channel $\mathcal{I}_i$. We apply the spinor-helicity formalism, and the form of $W^2$ given in \cite{Jiang:2020rwz} is thus
\eq{\label{eq:W2casimir}
 W^2 &=\frac{1}{8} P^2(\mathrm{Tr}[M^2] +\mathrm{Tr}[\widetilde{M}^2]) -\frac14 \mathrm{Tr}[P^\intercal M P\widetilde{M}] ,     
}
where $P=P_\mu \sigma^\mu_{\alpha\dot\alpha}$, $P^\intercal=P_\mu\bar\sigma^{\mu\dot\alpha\alpha}$, and $M$, $\widetilde{M}$ are the chiral components of the Lorentz generator $ M_{\mu\nu} \sigma^\mu_{\alpha\dot{\alpha}} \sigma^\nu_{\beta\dot{\beta}} =\epsilon_{\alpha\beta}\widetilde{M}_{\dot{\alpha}\dot{\beta}} +\tilde{\epsilon}_{\dot{\alpha}\dot{\beta}} M_{\alpha\beta} $. 

Eq.~\eqref{eq:W2B} is an eigenfunction of $W^2_{\mathcal{I}_i}$, so a natural question emerges whether there is a basis to solve the eigenfunction, so that the partial wave expansion becomes a linear algebra problem. Luckily we have already introduced how to construct a complete and independent basis classified by dimension and external particles. To sum up, the process of partial wave expansion consists of 5 steps: 
\begin{description}
\item[1.] find out the linear space spanned by a complete and independent basis (usually the y-basis) at dimension $d$, $V^d \equiv [\mathcal{B}^{y,d}]$; 

\item[2.] determine the partition $N|m$ and its corresponding partial Casimir operators $\{W^2_{\mathcal{I}_i}\}$; 

\item[3.] act each partial Casimir operator on the basis $\{W^2_{\mathcal{I}_i} \mathcal{B}^{y,d} \}$ and obtain a series of equations; 

\item[4.] solve the eigenfunction of each $W^2_{\mathcal{I}_i}$ and obtain the corresponding irreps; 

\item[5.] find the linear intersections of the irreps' spaces of each $W^2_{\mathcal{I}_i}$ and obtain a series of eigenbasis of $\{W^2_{\mathcal{I}_i}\}$, which is named as the Lorentz j-basis. 
\end{description}

We still study type $D {e_{\mathbb{C}}}^\dagger H Q^3 $ as an example and construct its Lorentz j-basis of partition $\{12|125|34\}$ \begin{fmffile}{5x}
\begin{fmfgraph*}(40,20)
\fmfpen{thin} \fmfleft{e1,e2} \fmflabel{1}{e1} \fmflabel{2}{e2} \fmflabel{5}{e3} \fmflabel{4}{e4} \fmflabel{3}{e5} \fmftop{e3} \fmfbottom{n} \fmfv{}{g1} \fmfv{}{g2} \fmfv{}{g3} \fmfright{e5,e4}
\fmf{plain}{e1,g1} \fmf{plain}{g1,e2} \fmf{plain}{g1,g2} \fmf{plain}{g2,e3} \fmf{phantom}{g2,n} \fmf{plain}{g2,g3} \fmf{plain}{e4,g3} \fmf{plain}{g3,e5}
\end{fmfgraph*}
\end{fmffile}.\newline \newline Eq.~\eqref{eq:ex.y} gives its Lorentz y-basis. Since there are 2 resonances, we only need to focus on channel $\{12\}$ and $\{34\}$. Act $W^2_{\{12\}}$ on $\mathcal{B}^y$ and 
\begin{eqnarray}
W^2_{\{12\}} \mathcal{B}^y = -s_{12} \mathcal{W}_{\{12\}} \mathcal{B}^y, \ \mathcal{W}_{\{12\}} = \begin{pmatrix}
0 & 0 & 0 \\ 0 & 2 & 0 \\ -1 & 0 & 2
\end{pmatrix}. 
\end{eqnarray}
Diagonalize the matrix $\mathcal{W}_{\{12\}}$
\eq{
\mathcal{K}^{jy}_{\{12\}} = \begin{pmatrix}
-1 & 0 & 2 \\ 0 & 1 & 0 \\ 1 & 0 & 0
\end{pmatrix}, \quad \mathcal{K}^{jy}_{\{12\}} \mathcal{W}_{\{12\}} (\mathcal{K}^{jy}_{\{12\}})^{-1} & = \mathrm{diag}(2,2,0) \\ & \Rightarrow J_{\{12\}} = (1,1,0), 
}
where $\mathcal{B}^j_{\mathcal{I}} = \mathcal{K}^{jy}_{\mathcal{I}} \mathcal{B}^y $. The amplitude space $[\mathcal{B}^y]$ is decomposed into 2 subspaces of irreps $J_{\{12\}}=1$ and $J_{\{12\}}=0$ respectively.
Similarly for channel $\{34\}$, the transformation matrix $\mathcal{K}^{jy}_{\{34\}}$ and the corresponding spins $J_{\{34\}}$ are
\eq{
\mathcal{K}^{jy}_{\{34\}} = \begin{pmatrix}
-1&1&2 \\ 0&-1&1 \\ 1&0&0
\end{pmatrix},\quad J_{\{34\}} = (\frac{3}{2}, \frac{1}{2}, \frac{1}{2}). 
}
So far there are 2 kinds of decomposition of space $[\mathcal{B}^y]$, $[J_{\{12\}}=1] \oplus [J_{\{12\}}=0] $ and $[J_{\{34\}}=\frac{3}{2}] \oplus [J_{\{34\}}=\frac{1}{2}] $, derived from ${W}^2_{\{12\}}\mathcal{B}^y$ and $W^2_{\{34\}}\mathcal{B}^y$ respectively. Then we find out the linear intersections of the subspaces
\begin{align}
[J_{\{12\}}=1] \bigcap [J_{\{34\}}=\frac{3}{2}] =& \mathrm{Span}[\begin{pmatrix}
1&-1&-2
\end{pmatrix} \mathcal{B}^y], \\ 
[J_{\{12\}}=1] \bigcap [J_{\{34\}}=\frac{1}{2}] =& \mathrm{Span}[\begin{pmatrix}
-1&-2&2
\end{pmatrix} \mathcal{B}^y], \\
[J_{\{12\}}=0] \bigcap [J_{\{34\}}=\frac{1}{2}] =& \mathrm{Span}[\begin{pmatrix}
-1&0&0
\end{pmatrix} \mathcal{B}^y].  
\end{align}
Therefore the Lorentz j-basis of partition $\{12|125|34\}$ is
\begin{eqnarray}
\mathcal{B}^j_{\{12|125|34\}} = \mathcal{K}^{jy}_{\{12|125|34\}} \mathcal{B}^y = \begin{pmatrix}
1&-1&-2 \\ -1&-2&2 \\ -1&0&0
\end{pmatrix} \mathcal{B}^y,\ (J_{\{12\}}, J_{\{34\}}) = \begin{pmatrix}
(1, \frac{3}{2}) \\ (1, \frac{1}{2}) \\ (0,\frac{1}{2})
\end{pmatrix}. 
\end{eqnarray}

\subsubsection{Gauge j-basis}

The partial wave expansion for gauge group shares a similar principle with the Poincar\'e group, that the Casimir operator labels the irreps of its group
\begin{eqnarray}
\mathop{\mathbbm{C}_n}_{\mathbbm{S}} \circ \mathcal{T}(\mathbf{R};\mathbbm{S})_{I_1...I_N} = \mathop{C_n}\limits_{\mathbbm{S}}(\mathbf{R}) \mathcal{T}(\mathbf{R};\mathbbm{S})_{I_1...I_N}
\end{eqnarray}
where $\mathbbm{C}_n$ is an Casimir operator, $\mathbbm{S}\subset \{1,...,N\} $ labels a channel of particles with direct product of irreps $\otimes \{\mathbf{r}_i|i\in\mathbbm{S}\}$, $\mathcal{T}(\mathbf{R};\mathbbm{S})_{I_1...I_N}$ is a tensor in a partial wave basis of the irrep $\mathbf{R}$ and $\otimes \{\mathbf{r}_i|i\in\mathbbm{S}\}$, and $C_n(\mathbf{R})$ is the eigenvalue of $\mathbf{R}$. 

Since we mainly focus on the Standard Model gauge group, we only give the Casimir operators of $SU(2)$ and $SU(3)$
\begin{eqnarray}\label{eq:C2C3def}
\mathbbm{C}_2&=& \mathbbm{T}^a\mathbbm{T}^a,\ \text{for both $SU(2)$ and $SU(3)$,}\label{eq:CasimirC2}\\
\mathbbm{C}_3 &=& d^{abc}\mathbbm{T}^a\mathbbm{T}^b\mathbbm{T}^c,\ \text{for $SU(3)$ only},\label{eq:CasimirC3}
\end{eqnarray}
where $\mathbbm{T}$'s are generators of their corresponding groups. The generator of direct product of irreps derived from the exponential map of $SU(N)$ is in the form
\begin{eqnarray}
\mathbbm{T}^{A}_{\otimes\{\mathbf{r}_i\}} = \sum_{i=1}^{N} T^A_{\mathbf{r}_i} \prod_{j\neq i} E_{\mathbf{r}_j}, 
\end{eqnarray}
where $E_{\mathbf{r}_j}$ is the identity matrix of irrep ${\mathbf{r}_j}$, and $T_{\mathbf{r}_i}^A $ is the generator of irrep $\mathbf{r}_i$. Such a generator acting on a tensor in the space of $\otimes\{\mathbf{r}_i\}$ gives 
\begin{eqnarray}
\mathbbm{T}^{A}_{\otimes \{\mathrm{r}_i\}} \circ \Theta_{I_1...I_i...I_N} = \sum_{i=1}^{N} (T^A_{\mathbf{r}_i})_{I_i}^{J} \Theta_{I_1...J...I_N} .
\end{eqnarray}
To realize the partial Casimir operator $\mathop{\mathbbm{C}_n}\limits_{\mathbbm{S}} $ we define the partial generator of $i \in \mathbbm{S}$
\begin{eqnarray}
{\mathop{\mathbbm{T}}\limits_{\mathbbm{S}}}^{A} \circ \Theta_{I_1...I_N} = \sum_{i \in \mathbbm{S}} (T^A_{\mathbf{r}_i})_{I_i}^{J} \Theta_{I_1...J...I_N}. 
\end{eqnarray}

We have shown the construction of the Lorentz j-basis of type $D {e_{\mathbb{C}}}^\dagger H Q^3 $, and now we construct its $SU(2)_W$ j-basis of partition $\{12|125|34\}$, following the same pattern as the construction of Lorentz j-basis. First we act the partial Casimir operator on the gauge y-basis (also the gauge m-basis) in Eq.~\eqref{eq:m-basis}
\begin{eqnarray}
\mathop{\mathbbm{C}_2}\limits_{\{12\}} \circ \mathcal{T}^y_1 &=& \underset{\{12\}}{\mathbbm{T}}^I \circ \underset{\{12\}}{\mathbbm{T}}^I \circ \mathcal{T}^y_1 = \underset{\{12\}}{\mathbbm{T}}^I \circ ({\tau^I}_{j}^{i_1} \epsilon^{j i_3} \epsilon^{i_2 i_4} +{\tau^I}_{j}^{i_2} \epsilon^{i_1 i_3} \epsilon^{j i_4} ) \nonumber \\ &=& \epsilon^{i_1 i_3} \epsilon^{i_2 i_4} +\epsilon^{i_1 i_4} \epsilon^{i_2 i_3} = 2\mathcal{T}^y_1 -\mathcal{T}^y_2, \\
\mathop{\mathbbm{C}_2}\limits_{\{12\}} \circ \mathcal{T}^y &=& (\mathop{C_2}\limits_{\{12\}})^{\mathtt{T}} \mathcal{T}^y = \begin{pmatrix}
2 & -1 \\ 0 & 0 
\end{pmatrix} \begin{pmatrix}
\mathcal{T}^y_1 \\ \mathcal{T}^y_2
\end{pmatrix} 
\end{eqnarray}
and obtain the matrix $(\mathop{C_2}\limits_{\{12\}})^{\mathtt{T}} $. Then we diagonalize $(\mathop{C_2}\limits_{\{12\}})^{\mathtt{T}} $ and gain the transformation matrix ${\cal K}^{jy}_{\{12\}}$ from y(m)-basis to j-basis of channel $\{12\}$
\begin{eqnarray}
{\cal K}^{jy}_{\{12\}} (\mathop{C_2}\limits_{\{12\}})^{\mathtt{T}} ({\cal K}^{jy}_{\{12\}})^{-1} = {\rm diag}\{2,0\},\ \text{with } {\cal K}^{jy}_{\{12\}}= \begin{pmatrix}
-2 & 1 \\ 0 & 1
\end{pmatrix}. 
\end{eqnarray}
The process for channel $\{34\}$ is the same. Finally we find out the linear intersections of irreps in each channel, and the result is 
\begin{eqnarray}
\mathcal{T}^j_{\{12|125|34\}} = \mathcal{K}^{jy}_{\{12|125|34\}} \mathcal{T}^y = \begin{pmatrix}
2 & -1 \\ 0 & -1
\end{pmatrix} \begin{pmatrix}
\mathcal{T}^y_1 \\ \mathcal{T}^y_2
\end{pmatrix} \mbox{ with } (\mathbf{R}_{\{12\}}, \mathbf{R}_{\{34\}}) = \begin{pmatrix}
(\mathbf{3},\mathbf{3}) \\ (\mathbf{1},\mathbf{1})
\end{pmatrix}. 
\end{eqnarray}

\section{UV Resonance and Effective Operators}\label{sec:3}


\subsection{Higher Spin Resonances the On-shell Way}\label{sec:HSR}


In the Young tensor method, the building blocks for particles with spin $J\ge1$ are the field strength, such as $F_{\mu\nu}$ for spin-1 particles and the Weyl tensor $C_{\mu\nu\rho\lambda}$ for spin-2 particles. However, when we construct the UV theory with such higher spin massive particles, the basic building blocks are usually taken to be $V_\mu$ and $h_{\mu\nu}$ for spin-1 and spin-2 particles, which involves longitudinal polarizations that are not allowed in the massless theories due to the gauge symmetries. 
These building blocks allow for larger freedom of constructing operators, but more redundancy relations are also introduced by the EOM. 
For example, the EOM of the vector field $V_\mu$ is
\eq{
    \partial^\nu F_{\mu\nu} + M^2 V_\mu + J_\mu = 0\ ,
}
where $J_\mu$ is the source of $V^\mu$. Applying $\partial^\mu$ we get another EOM
\eq{
    \partial^\mu V_\mu = -\frac{1}{M^2}\partial^\mu J_\mu \ .
}
The fact that there is no $V_\mu$ field on the right-hand side reflects the physical constraint on the polarization vector $k^\mu \epsilon_\mu = 0$, which indicates that the left-hand side does not create on-shell vector particle. It proves that we can always replace the $\partial^\mu V_\mu$ part in the operator by $V$-independent pieces. 
More specifically, suppose the source can be decomposed as
\eq{
    J_\mu = \partial_\mu\Phi + \mc{J}_\mu \ ,
}
where $\Phi$ is a scalar operator.
We can perform a field redefinition $V_\mu \to V_\mu - \frac{1}{M^2}\partial_\mu\Phi$, such that
\eq{\label{eq:f-redef}
    \mc{L}_{\rm UV} &= -\frac{1}{4}F_{\mu\nu}F^{\mu\nu} + \frac{1}{2}M^2V_\mu V^\mu + V_\mu J^\mu  \\
    \to \quad \tilde{\mc{L}}_{\rm UV} &= -\frac{1}{4}F_{\mu\nu}F^{\mu\nu} + \frac{1}{2}M^2V_\mu V^\mu - \frac{1}{2M^2}\partial_\mu\Phi \partial^\mu\Phi + V^\mu\mc{J}_\mu - \frac{1}{M^2}\partial^\mu\Phi\mc{J}_\mu \ .
}
In the new Lagrangian, there is no longer a coupling between $V$ and the scalar source $\Phi$. Letting $\mc{J}_\mu$ be a conserved current, the last term will also drop out after IBP.

From the on-shell perspective, the motivation of the above practice is twofold: first it makes sure that the amplitude-operator correspondence still holds, thus operators involving vector fields $V_\mu$ always correspond to scattering amplitudes involving the vector particle; second, the j-basis/UV resonance correspondence is kept valid. 
The latter can be understood by integrating out the heavy vector and examine the effective operators after matching. Without the replacement in eq.~\eqref{eq:f-redef}, the term $- \frac{1}{2M^2}\partial_\mu\Phi \partial^\mu\Phi$ is obtained through matching, and one may attribute it to the spin-1 resonance while it's a $J=0$ operator in the j-basis. On the contrary, if we start from $\tilde{\mc{L}}_{\rm UV}$, the term $- \frac{1}{2M^2}\partial_\mu\Phi \partial^\mu\Phi$ would be present in the UV which is irrelevant of the heavy vector, while the matching only produces the contact interaction of the conserved current $\mc{J}^\mu \mc{J}_\mu$, which is indeed $J=1$ operator in the j-basis.

The requirement can be extended to arbitrary spins. The key point is to decompose the tensor field into components with various spins, while only the component with correct spin could create the on-shell particle. For the spin-2 particle, the rank-2 tensor source can be decomposed as
\eq{
    T_{\mu\nu} = \Phi g_{\mu\nu} + (\partial_\mu J_\nu + \partial_\nu J_\mu) + \mc{T}_{\mu\nu}\ .
}
Like for the vector, the couplings to the scalar trace component $\Phi$ and the vectorial longitudinal component $J_\mu$ can be removed by field redefinition. In our on-shell prescription, we always choose the gauge that the spin-2 particle couples to the transverse-traceless source. This principle can be applied to arbitrary higher spins.

\comment{
Such ambiguity is sometimes referred to as a gauge choice, because it is very similar to the Lorentz gauge. 
It becomes confusing when we consider the coupling $V_\mu \partial^\mu \mc{O}$, because using EOM we would obtain
\eq{\label{eq:v0_eom}
    V_\mu \partial^\mu \mc{O} \ \simeq \ \frac{1}{M^2}\big[F_{\mu\nu} \partial^\mu\partial^\nu \mc{O} - J_\mu\partial^\mu \mc{O}\big]\ ,
}
where the first term vanishes due to the antisymmetry between $\mu,\nu$. Therefore on the right-hand side, there is no $V_\mu$ at all. It stems from the other part of the EOM $\partial^\mu V_\mu = -\frac{1}{M^2}\partial^\mu J_\mu$. Generically, it is only a choice of gauge to adopt the left or right forms of the operator.

From the on-shell point of view, such coupling does not contribute to amplitudes involving the on-shell vector particles, and should not be counted in the corresponding types. It turns out to be more problematic when we \emph{integrate out} the vector particle, one of the resulting effective operators is $-\frac{1}{M^2}\partial_\mu \mc{O} \partial^\mu \mc{O}$, which is a $J=0$ operator in the corresponding channel\footnote{A quick way to find the angular momentum $J$ is to extract the ``bridge'' across the channel \cite{} --- all the Lorentz contractions in the form of spinor contractions, the number of which being totally symmetric should be $2J$. Here by IBP, the operator can be written as $\mc{O}\Box\mc{O}$ which bears no Lorentz contractions across the channel, and must be a $J=0$ operator.}, inconsistent with the spin of $V_\mu$. Our story in the last section fails, because we were not starting from on-shell building blocks. Only when the resonance can be on-shell, do we obtain an effective operator with angular momentum $J$ consistent with the spin of the resonance. Thus in this case, what we had to do was to perform the conversion as in eq.~\eqref{eq:v0_eom} BEFORE the integrating out, so that the operator $-\frac{1}{M^2}\partial_\mu \mc{O} \partial^\mu \mc{O}$ was obtained as part of $-\frac{1}{M^2}J_\mu\partial^\mu \mc{O}$. The resulting EFT would be the same, but the interpretation of this operator is different --- it did not come from integrating out $V_\mu$, but already existed before that, and hence it is allowed to have $J=0$. 

To make it more specific, let's take a concrete example. Suppose we have a UV Lagrangian
\eq{
    \mc{L}_{\rm UV} \supset \frac{1}{2}V^\mu(g_{\mu\nu}\Box-\partial_\mu\partial_\nu + M^2)V^\nu + \frac{1}{2}(\partial_\mu\phi)^2 + g V_\mu\partial^\mu(\phi^2).
}
At energy $E\ll M$, integrating out the heavy vector $V_\mu$, one would get an IR Lagrangian as
\eq{
    \mc{L}_{\rm IR} \supset \frac{1}{2}(\partial_\mu\phi)^2 - \frac{g^2}{M^2} \partial^\mu(\phi^2)\partial_\mu(\phi^2)
}
where we got a dimension-six operator with four $\phi$ fields. 
Notice that the EOM of the UV Lagrangian includes
\eq{
    \partial^\mu V_\mu = -\frac{g}{M^2}\Box(\phi^2)
}
Using this, we can rewrite the UV Lagrangian as
\eq{
    \tilde{\mc{L}}_{\rm UV} \supset \frac{1}{2}V^\mu(g_{\mu\nu}\Box-\partial_\mu\partial_\nu + M^2)V^\nu + \frac{1}{2}(\partial_\mu\phi)^2 - \frac{g^2}{M^2} \partial^\mu(\phi^2)\partial_\mu(\phi^2)
}
such that there is no interaction term between $V_\mu$ and $\phi$ any more. It is possible because in the original $\mc{L}_{\rm UV}$ the interaction term does not contribute to on-shell amplitude for $V_\mu$ anyway. Therefore, starting from the altered UV Lagrangian $\tilde{\mc{L}}_{\rm UV}$, integrating out $V_\mu$ just amounts to remove its quadratic terms, and the same $\mc{L}_{\rm IR}$ is obtained. The key difference between the two interpretation is the source of the $\phi^4\partial^2$ interaction in the IR: in the latter ``on-shell'' interpretation, it does not come from integrating out $V_\mu$, hence it does not have to be a $J=1$ operator. One may be concerned about the higher-dimensional nature of this interaction term in $\tilde{\mc{L}}_{\rm UV}$, but remember that the theory with a massive vector $V_\mu$ is not UV complete by its own, thus it is no problem for it to include a higher dimensional term.
}

After choosing such a gauge, the correspondence between the UV resonance and the resulting low-energy operators is well defined without ambiguity. As explained previously, we can use the correspondence to find the UV origin of an effective operator. One remaining subtlety of the UV origin issue is when the 3-point UV couplings involved in the tree decomposition of the amplitude are considered ``loop-induced''. The typical example is the coupling between a vector resonance and two same-helicity fermions $\mc{B}(V_1,\psi_2,\psi_3) = \vev{{\bf1}2}\vev{{\bf1}3}$, which corresponds to the contribution of a dimension-5 operator $\mc{O}=\partial_\mu V_{1\nu} \bar{\psi}_2\sigma^{\mu\nu}\psi_3$. Therefore it is sometimes convenient to pick out those tree decompositions involving such couplings. 

The tree level UV couplings are mostly 3-point, with a few exceptions at 4-point. For clarity, we start with 3-point couplings with one spin-$S$ heavy particle, while the other two are treated as massless with helicities $(h_1,h_2)$. The on-shell amplitudes for such couplings have a general form \cite{Arkani-Hamed:2017jhn}
\eq{
    \mc{M}(h_1,h_2,S) \sim \vev{12}^{S-h_1-h_2}[1{\bf3}]^{S+h_1-h_2}[2{\bf3}]^{S-h_1+h_2}
}
where the little group indices on the spinors $|{\bf3}]$ are implicitly totally symmetric. The power $S-h_1-h_2$ can be negative, which means it should be replaced by $[12]^{-S+h_1+h_2}$, while the other two powers should be non-negative, implying the requirement $S\ge |h_1-h_2|$. This general form is written in the so-called \emph{chiral basis}, where only the square brackets of the massive particles are used, but the mass dimension of the corresponding operator $d_{\mc{O}}$ becomes illusive. 
For example, $\mc{M}(+1/2,-1/2,1)\sim \vev{12}[1{\bf3}]^2$ would directly correspond to the dimension-6 operator $(\partial_\mu\psi^\dagger_1)\bar\sigma_{\nu}\psi_2 F_3^{\mu\nu}$, which is actually equivalent to a dimension-4 operator as $\psi^\dagger_1\bar\sigma_{\mu}\psi_2 V_3^{\mu}$ by the EOM and IBP. 
It turns out that to obtain an equivalent form of the operator with lowest mass dimension, a more convenient basis would be the \emph{symmetric basis}, where the number of square brackets and angle brackets for the massive particles are equal for bosons, and differ by 1 for fermions. They correspond to the amplitudes in terms of polarization vectors $\epsilon_i^\mu \equiv \frac{\bra{{\bf i}}\sigma^\mu|{\bf i}]}{\sqrt{2}M}$ and $u = (\ket{{\bf i}} , |{\bf i}])^T$ as the solution of the Dirac equation, such as $\chi^\mu \simeq \epsilon^\mu u = $ for the spin-3/2 Rarita-Schwinger field and $h^{\mu\nu} \simeq \epsilon^\mu\epsilon^\nu$ for the spin-2 Fierz-Pauli field. Assuming that the fields are all canonically normalized, their mass dimensions would be precisely 1 more than their wave functions. In general we have \footnote{This formula only holds for canonically normalized kinetic terms for the massive fields, which means that the bosonic fields have mass dimension 1 while the fermionic fields have mass dimension 3/2. It is not the convention for gravitons, for instance, whose kinetic term has a factor of $M_{pl}^2$ and the field is dimensionless. However, for massive fields, it is natural to stick to this convention.}
\eq{
    d_{\mc{O}} = N + d_{\mc{M}}
}
Therefore, to have $d_{\mc{O}}\le 4$ for the tree-level couplings, the corresponding $N=3$-point amplitudes should satisfy $d_{\mc{M}}\le 1$ when the factors of $1/(\sqrt{2}M)$ are included. 
We use the convention for arbitrarily high spin fields such that the wave functions of the massive fields are
\eq{
&\text{spin-$S$ bosons:} \quad \epsilon_i^{(\mu_1}\epsilon_i^{\mu_2}\dots\epsilon_i^{\mu_S)} = \frac{1}{(\sqrt{2}M)^{2S}}\bra{{\bf i}}\sigma^{\mu_1}|{\bf i}]\bra{{\bf i}}\sigma^{\mu_2}|{\bf i}]\dots\bra{{\bf i}}\sigma^{\mu_S}|{\bf i}] \\
&\text{spin-$S$ fermions:} \quad \epsilon_i^{(\mu_1}\dots\epsilon_i^{\mu_{S-1/2})}u_i \sim \frac{1}{(\sqrt{2}M)^{2S-1}} \bra{{\bf i}}\sigma^{\mu_1}|{\bf i}]\dots\bra{{\bf i}}\sigma^{\mu_{S-1/2}}|{\bf i}]
\begin{pmatrix} \ket{{\bf i}} \\ |{\bf i}] \end{pmatrix}
}
all of which are in the symmetric basis.
For comparison, the amplitudes in the chiral basis correspond to fields with derivatives, for instance
\eq{
    i\sigma_{\mu\nu}\partial^\mu\chi^\nu \simeq \ket{{\bf i}}\bra{{\bf i}}\bra{{\bf i}}\ ,\quad (\partial_\mu h_{\nu\rho})\sigma^{\mu\nu} \simeq
    \frac{1}{\sqrt{2}M}\ket{{\bf i}}[{\bf i}|\sigma_\rho\ket{\bf i}\bra{\bf i}.
}
For the previous example, the lowest dimensional amplitude we can get is $\mc{M}(+1/2,-1/2,1) \sim \frac{1}{\sqrt{2}M}[1{\bf3}]\vev{2{\bf3}}$, which has mass dimension $d_{\mc{M}}=1$. 
We list below all the two-light-one-heavy couplings with $d_{\mc{M}}\le1$
\eq{
\begin{array}{c|c|r|r|l}
     \text{light particles} & \text{heavy particles} & \text{amplitude} & \text{operator} & \\
     \hline
     \varphi,\varphi & \phi & 1 & \varphi_1\varphi_2\phi & \\
     \psi,\psi & \phi & \vev{12} & (\psi_1\psi_2)\phi & \\
     \varphi,\psi & \chi & \vev{2{\bf3}} & \varphi(\psi\chi) & \\
     \varphi,\varphi & V & \frac{1}{\sqrt{2}M}\vev{2{\bf3}}[2{\bf3}] & \varphi_1\partial_{\mu}\varphi_2 V^\mu & \\
     \psi,\psi^\dagger & V & \frac{1}{\sqrt{2}M}[1{\bf3}]\vev{2{\bf3}} & \psi_1\sigma_\mu\psi_2^\dagger V^\mu & \\
\end{array}
}
Note that there are no heavy particles beyond $J=1$, as they are all considered loop-induced in this convention. For example, the coupling of a light fermion and scalar with a heavy spin-3/2 particle has the amplitude
\eq{
    \mc{M}(\varphi,\psi,\chi^{(3/2)}) = \frac{1}{\sqrt{2}M}\vev{1{\bf3}}\vev{2{\bf3}}[1{\bf 3}] \simeq (\partial_\mu\varphi)(\psi\chi^\mu) \ ,
}
which corresponds to a dimension-5 operator in terms of the Rarita-Schwinger field. We can also write down couplings with a heavy spin-2 particle such as
\eq{
    \mc{M}(\psi,\psi^\dagger,h^{(2)}) = \frac{1}{(\sqrt{2}M)^2}\vev{1{\bf3}}\vev{2{\bf3}}[2{\bf 3}]^2 \simeq (\psi\sigma_\mu \partial_\nu\psi^\dagger)h^{\mu\nu} \ ,
}
also corresponding to a dimension-5 operator.

To proceed, we look at the case with one light (massless) particle and two heavy particles. In the generic case, the massless particle can only be a scalar or a spin-1/2 fermion, while the spins of the two massive particles can differ by 0, 1 or 1/2. The amplitudes of the couplings are
\eqs{
    \mc{M}(\varphi,\Psi^{(S)},\Psi^{(S)}) &= \left\{
    \begin{array}{ll}
         \frac{1}{(\sqrt{2}M)^{2S}}\vev{{\bf23}}^S[{\bf23}]^S & S\in\mathbb{Z}\\
         \frac{1}{(\sqrt{2}M)^{2S-1}}\vev{{\bf23}}^{S+1/2}[{\bf23}]^{S-1/2} & S\in\mathbb{Z}+\frac{1}{2} 
    \end{array}\right. \label{eq:lhh_1}\\
    \mc{M}(\varphi,\Psi^{(S)},\Psi^{(S+1)}) &= \frac{1}{(\sqrt{2}M)^{2S+1}}\vev{1{\bf3}}[1{\bf3}]\vev{{\bf23}}^S[{\bf23}]^S \ ,\qquad S\in\mathbb{Z} \label{eq:lhh_2}\\
    \mc{M}(\psi,\Psi^{(S)},\Psi^{(S+1/2)}) &= \left\{
    \begin{array}{ll}
         \frac{1}{(\sqrt{2}M)^{2S}}\vev{1{\bf3}}\vev{{\bf23}}^S[{\bf23}]^S & S\in\mathbb{Z}\\
         \frac{1}{(\sqrt{2}M)^{2S}}\vev{1{\bf3}}\vev{{\bf23}}^{S-1/2}[{\bf23}]^{S+1/2} & S\in\mathbb{Z}+\frac{1}{2} 
    \end{array}\right. \label{eq:lhh_3}
}
A subtlety rises when the two heavy particles have the same mass, so that the $x$ factor can be introduced which is massless and carries $-1$ helicity of the massless particle. It is defined as
\eq{\label{eq:x_def}
    x|1] \equiv \frac{p_2\ket{1}}{M}
}
where the proportionality is guaranteed by the kinematics $[1|p_2\ket{1} = 2p_1\cdot p_2 = m_3^2-m_2^2=0$. It typically appears in the gauge interaction where the massless particle is the gauge boson. Consider a $|h|\ge1$ massless boson $A$ coupled to equal mass particles, the only couplings satisfying $d_{\mc{M}}\le1$ are\footnote{When the variable $x$ is chosen, only one of $\ket{1}$ and $|1]$ can be used as an independent variable due to the relation eq.~\eqref{eq:x_def}, thus the counterpart of eq.~\eqref{eq:lhh_2} is not present for the gauge interaction.}
\eq{
    \mc{M}(A^{(h)},\Psi^{(S)},\Psi^{(S)}) = x^h \times \left\{
    \begin{array}{ll}
         \frac{1}{(\sqrt{2}M)^{2S}}\vev{{\bf23}}^S[{\bf23}]^S & S\in\mathbb{Z}\\
         \frac{1}{(\sqrt{2}M)^{2S-1}}\vev{{\bf23}}^{S+1/2}[{\bf23}]^{S-1/2} & S\in\mathbb{Z}+\frac{1}{2} 
    \end{array}\right. 
}

The last case is the couplings among three heavy particles. 
Suppose the three spins are ordered $S_1\le S_2 \le S_3$, we can write down the general formula
\eq{
    & \mc{M}(\Psi^{(S_1)},\Psi^{(S_2)},\Psi^{(S_3)}) = \frac{1}{(\sqrt{2}M)^{S_1+S_2+S_3}} \times \\
    & \ \left\{
    \begin{array}{ll}
         \left(\vev{{\bf 12}}^{\frac{S_1+S_2-S_3}{2}}\vev{{\bf 23}}^{\frac{-S_1+S_2+S_3}{2}}\vev{{\bf 31}}^{\frac{S_1-S_2+S_3}{2}} \,\times\, c.c.\right)\,, & S_1+S_2+S_3 \in 2\mathbb{Z} \\
         \left(\vev{{\bf 12}}^{\frac{S_1+S_2-S_3+1}{2}}\vev{{\bf 23}}^{\frac{-S_1+S_2+S_3-1}{2}}\vev{{\bf 31}}^{\frac{S_1-S_2+S_3-1}{2}} \,\times\, c.c.\right) \times \bra{{\bf3}}p_1|{\bf3}]\,,
         & S_1+S_2+S_3 \in 2\mathbb{Z}+1
    \end{array}\right.
}
while we see that the positivity of the powers require $S_3\le S_1+S_2+1$.
The above are all the 3-point tree-level couplings we should count in the tree decomposition of the effective operators. Those not listed above should be considered as loop-induced, which may be more suppressed at low energy. 

Sometimes we also need 4-point couplings, and there are not many of them at tree level.
First, only bosonic fields can be involved, because the wave functions of fermionic fields are always dimensionful. Second, gauge bosons are not allowed, because the gauge invariant field strength must involve derivatives, while the special kinematics required by the gauge couplings are also absent at 4-point. Therefore, the only massless particles that can be involved are scalars, thus we have
\eqs{
    \mc{M}(\varphi,\varphi,\varphi,\varphi) &= 1 \\
    \mc{M}(\varphi,\varphi,\Psi^{(S)},\Psi^{(S)}) &= \vev{{\bf34}}^S[{\bf34}]^S \\
    \mc{M}(\varphi,\Psi^{(S_1)},\Psi^{(S_2)},\Psi^{(S_3)}) &= \frac{1}{(\sqrt{2}M)^{S_1+S_2+S_3}}\left(\vev{{\bf 12}}^{\frac{S_1+S_2-S_3}{2}}\vev{{\bf 23}}^{\frac{-S_1+S_2+S_3}{2}}\vev{{\bf 31}}^{\frac{S_1-S_2+S_3}{2}} \,\times\, c.c.\right)\,.
}
When all the 4 particles are massive with spins $S_1\le S_2\le S_3\le S_4$, there would be generically no unique couplings, but the criteria for the existence of tree-level couplings are $S_1+S_2+S_3+S_4 \in 2\mathbb{Z} $ and $S_4\le S_1+S_2+S_3$.

When all the UV couplings involved in the tree decomposition of an effective operator are tree-level, \ie $d_{\mc{O}}\le 4$, the matching will always give the j-basis operator with the correct quantum numbers. Nevertheless, when higher-dimensional couplings are involved, the resulting operator would have a minimal mass dimension, so that certain j-basis operators at lower dimensions are not accessible through matching. For example, the four fermion operator $(\psi_1\sigma^{\mu\nu}\psi_2)(\psi_3\sigma_{\mu\nu}\psi_4)$ is a $J=1$ operator in the $\{1,2\}\to\{3,4\}$ channel, but a heavy vector resonance in this channel requires dimension-5 couplings with the fermions $\mc{M}(\psi,\psi,V)=\vev{1{\bf 3}}\vev{2{\bf 3}} \simeq \frac{1}{\Lambda}\psi\sigma_{\mu\nu}\psi\partial^\mu V^\nu$ on both sides, resulting in at least dimension-8 effective operators at low energy $\mc{O}\sim 1/(\Lambda^2 M^2)$. Hence the dimension-6 operator cannot be generated in the matching. We take this into account by analysing the mass dimensions of all the UV couplings and setting the minimal dimension for the generated j-basis operators, and the process is named \emph{Dim selection} \cite{Li:2022abx}.

\comment{
We can write down all the allowed tree-level 3-point couplings involving particles up to $J=1$
\eq{
\begin{array}{l|l|r|r|r}
    \# & \text{particles} & \text{operator} & \text{amplitude} & \text{constraint}\\
    \hline
    1& \phi,\phi,\phi & \phi^3 & 1 &\\
    2& \phi,\phi,A & D_\mu\phi D^\mu\phi & x & \text{equal masses} \\
    3& \phi,\phi,V & \phi\partial_\mu\phi V^\mu & \frac{1}{M}\vev{2{\bf3}}[2{\bf3}] & \\
    3'& & \partial_\mu(\phi^2)V^\mu & \text{N/A} & \text{N/A} \\
    4& \psi,\psi,\phi & \bar\psi\psi\phi & \vev{\bf12} & \\
    5& \psi,\psi,A & \bar\psi iD\4 \psi & x\vev{\bf12} & \text{equal masses} \\
    6& \psi,\psi,V & \bar\psi\gamma_\mu\psi V^\mu & \frac{1}{M}\vev{1{\bf3}}[2{\bf3}] & \text{opposite helicities if massless}\\
    7& \phi,V,V & \phi V_\mu V^\mu & \frac{1}{M^2}\vev{\bf 12}[{\bf 12}] & \\
    8& V,V,A & D_\mu V_\nu D^\mu V^\nu & \frac{1}{M}x\vev{\bf 12}[{\bf 12}] & \text{equal masses} \\
    9& & V^\mu V^{\prime\nu}F_{\mu\nu}  & \frac{1}{M^2}\vev{\bf 12}[{\bf 1}3][{\bf 2}3] & \text{unequal masses}\\
    10& V,V,V & \partial_\mu V_\nu V^\mu V^\nu & \frac{1}{M^3}\bra{{\bf2}}p_1|{{\bf2}}]\vev{\bf 13}[{\bf 13}] & \\
    10'& & \partial_\mu V^\mu V_\nu V^\nu & \text{N/A} & \text{N/A}
\end{array}
}
where the scalars $\phi$ can be massive or massless, and $V$ are massive vectors and $A$ is the massless gauge boson. $M$ is the mass of the heavy boson, and $x$ is specifically defined for gauge bosons coupling with an equal-mass current such that $M\tilde\lambda^{\dot\alpha} = x p_1^{\dot\alpha\alpha}\lambda_\alpha$ \cite{Arkani-Hamed:2017jhn}. Note that we only show the amplitudes with helicity $+1$ gauge bosons since $x$ carries $+1$ helicity.

As shown above, the mass dimension of the corresponding operators are not easily recognized from the amplitudes, because extra factors of $1/M$ are present due to the correspondence $\epsilon^\mu \sim \frac{1}{M}\lambda^{(I}\sigma^\mu\tilde\lambda^{J)}$. In addition, extra constraints on the particles are imposed for the couplings involving gauge bosons: the presence of the $x$ factor in $\#2,5,9$ require the equal masses, while the $\#9$ without the $x$ factor requires unequal masses. 
Those external states not listed above, such as scalars or fermions with unequal masses coupling to the gauge boson $A$, are thus considered ``loop-induced''.
We also show the operators that can be removed by field redefinition, such as \#$3'$ and \#$10'$, as explained previously.
}

\comment{
At dimension eight and higher, more operators with high angular momenta are presented in the j-basis. 
It leads to a puzzle: when we classify an operator in terms of whether it could be generated at tree-level, which couplings involving the higher spin particles can be regarded as valid options?
Usually we consider couplings from a renormalizable Lagrangian as tree-level couplings with moderately weak strength, and the others must come from even higher UV completions, either from heavy particles suppressed by large masses, or from loop diagrams suppressed by loop factors. In this way, we can further organize effective operators in terms of their importance at IR even at the same mass dimension. 
{\color{red} Examples ...}
In the case of particles with spin 1 or higher, the UV theory by itself cannot be self-consistent due to the constraint from perturbative unitarity. Even higher energy UV completions are required for such heavy resonances. 
However, we still want some of their couplings to be considered as moderate as ``tree-level'', so that we can reasonably classify the effective operators generated at IR after integrating them out.
For example, we would like the coefficient of the operator $V_\mu\bar\psi\gamma^{\mu}\psi$ involving heavy vectors $V_\mu$ to be ``tree-level''. It seems obvious for this form of the operator since it has mass dimension four. But it is actually equivalent to another form by EOM
\eq{\label{eq:spin-op-ex}
    V_\mu\bar\psi\gamma^{\mu}\psi \ \simeq \ \frac{1}{M^2}F_{\mu\nu} \partial^\mu\bar\psi\gamma^\nu\psi
}
which has dimension six, where $M$ is the vector mass. 
An immediate puzzle is that how should we know from IR whether the mass suppression here should be of order $M$ or some even higher cutoff $\Lambda \gg M$? 
Still, we seem to at least expect their reasonable couplings to gauge bosons given their charges. Such problems lead to a systematic study of the power counting of couplings involving heavy resonances of various spins.

It is especially tricky from the amplitude point of view, since a spin-$j$ particle has to contribute at least $2j$ powers of spinor variables, and get suppressed by masses in the denominators either from its own mass $M$ or some other energy scales. However, for massive particles, we have the freedom to choose from bases consisting of left- or right-handed spinors due to the Dirac equations. For the sake of enumeration of independent amplitudes \cite{Arkani-Hamed:2017jhn}, it is usually convenient to choose a chiral basis with all-left- (or right-) handed spinors. But when it comes to the corresponding fields, the chiral basis is translated directly to field strength tensor fields at higher mass dimensions. For example, you would get the right-hand side of eq.~\eqref{eq:spin-op-ex} instead of the left-hand side. Meanwhile, a \emph{maximally-mixed} basis with $\lfloor j \rfloor$ pairs of left and right spinors and one (for fermions) or none (for bosons) leftovers turns out to be handy regarding the power counting. Each pair of spinors, with symmetric little group indices as usual and suppressed by a factor of mass, is equivalent to the dimensionless polarization vector
\eq{
    \frac{1}{\sqrt{2}m}\bra{\mathbf{i}}\sigma^\mu|\mathbf{i}] \equiv \epsilon_i^\mu\,.
}
which serves as the on-shell solution of the vector field $V^\mu$. This interpretation can be easily generalized to arbitrarily high spins that result in dimension-$1$ bosonic fields $V^{\mu_1\dots\mu_j}$ and dimension-$3/2$ fermionic fields $\Psi^{\mu_1\dots\mu_{\lfloor j \rfloor}}$ as the lowest dimensional field representations of the corresponding heavy particles. 
Therefore, we always do power counting of local amplitudes in the maximally-mixed basis of higher-spin massive particles for consistency.
Specifically, the mass dimension of amplitudes should be counted as the superficial dimension --- the number of spinor brackets --- minus the number of pairs of left- and right-handed external spinor variables.
This ultimately tells us which topology renders the dominant ``tree-level'' generation of j-basis operators.

In general, a spin-$S$ massive particle can be represented by $2S$ spinor variables with totally symmetric little group indices in the amplitude. Suppose there are $l$ angle brackets and $r$ square brackets, it corresponds to a field with certain numbers of spinor indices:
\eq{ 
\Psi_{\{\alpha_1,...,\alpha_l\},\{\dot{\alpha}_1,...,\dot{\alpha}_r\}}\sim |\mathbf{i}\rangle_{\alpha_1}... |\mathbf{i}\rangle_{\alpha_l} [\mathbf{i}|_{\dot{\alpha}_1}...[\mathbf{i}|_{\dot{\alpha}_r} \ ,\quad l+r=2S\ .
}
Due to the Dirac equation $p_i\ket{{\bf i}} = M|{\bf i}]$, fields with different $(l,r)$ can be converted to each other.
Using the Gamma matrices, these fields can be written in the usual forms with Lorentz indices, for instance
\eq{
    V_{\alpha\dot\alpha}\bar\sigma_{\mu}^{\dot\alpha\alpha} \equiv V_\mu \ ,\quad
    F_{\alpha\beta}\sigma_{\mu\nu}^{\alpha\beta} \equiv F_{\mu\nu}\ 
}
are two forms of the spin-1 field.

As an example, let us look at the classical form of Rarita-Schwinger field $\Psi_\zeta^{\mu} \equiv \psi \otimes W^\mu_{\zeta}$ is the direct product of a spin-$1/2$ fermion $\psi$ and a vector $W^\mu_{\zeta}$, 
\eq{
\Psi^{\mu} \sim \left( (\frac{1}{2},0) \oplus (0,\frac{1}{2}) \right) \otimes \left( (1,0) \oplus (\frac{1}{2},\frac{1}{2}) \oplus (0,1) \right)\\ =(\frac{3}{2},0) \oplus (1,\frac{1}{2}) \oplus (\frac{1}{2},1) \oplus (0,\frac{3}{2}) \oplus (\frac{1}{2},0) \oplus (0,\frac{1}{2}). 
}
The spin-$1/2$ part is the same as $\psi$ 
\eq{(\frac{1}{2},0) \sim |\mathbf{i}\rangle, \ (0,\frac{1}{2}) \sim [\mathbf{i}|.} 
We then study the spin-$3/2$ part $\left( (\frac{3}{2},0) \oplus (1,\frac{1}{2}) \oplus (\frac{1}{2},1) \oplus (0,\frac{3}{2}) \right)$ and match the irreps with spinor representation $\Psi_{ \{\alpha_1,..,\alpha_l\}, \{\dot{\alpha}_1,...,\dot{\alpha}_r\} }$
\bea
(\frac{3}{2},0) \sim |\mathbf{i}\rangle^3,\
(0,\frac{3}{2}) \sim [\mathbf{i}|^3,\\
(1,\frac{1}{2}) \sim |\mathbf{i}\rangle^2 [\mathbf{i}|, \
(\frac{1}{2},1) \sim |\mathbf{i}\rangle [\mathbf{i}|^2.\label{eq:rs}
\eea
Eq.~\eqref{eq:rs} should be normalized by a factor $1/(\sqrt{2}m)^{x}$ to match the UV operator that contains the field $\Psi$, and thus the dimension of the operator is
\bea\label{eq:mvpc1}
\mathrm{dim}\{\mbox{operator}\} = N - \sum_i \max(0, \mathrm{ceil}\{s_i - 1\}) + \mathrm{dim}\{\mbox{spinors}\}, 
\eea
where $N$ is the number of particles. 
Therefore the spinor representation of irreps $(1,\frac{1}{2})$ and $(\frac{1}{2},1)$ are 
\bea
(1,\frac{1}{2}) \simeq \lambda \epsilon_\mu =\dfrac{\ket{\bf{i}}[{\bf i}|\sigma_\mu\ket{\bf i}}{\sqrt{2}m_V^{x=1}}, \ (\frac{1}{2},1) \simeq \title{\lambda} \epsilon_\mu =\dfrac{[{\bf i}|\sigma_\mu\ket{\bf i}[\bf{i}|}{\sqrt{2}m_V^{x=1}}. 
\eea

The massive amplitude-operator correspondence is added up to spin-2,
\eq{
\begin{array}{l|rl}
s=\frac{1}{2} & \Psi & \simeq \lambda =|\bf{i}\rangle, \\
\hline
    s=1 & V_\mu & \simeq \epsilon_\mu = \dfrac{[{\bf i}|\sigma_\mu\ket{\bf i}}{\sqrt{2}m_V^{x=1}}, \\
    \hline
    s=\frac{3}{2} & \Psi_\mu & \simeq \lambda \epsilon_\mu =\dfrac{\ket{\bf{i}}[{\bf i}|\sigma_\mu\ket{\bf i}}{\sqrt{2}m_V^{x=1}},\\
    \hline
    s=2 & h_{\mu\nu} & \simeq \epsilon_\mu \epsilon_\nu = \dfrac{[{\bf i}|\sigma_\mu\ket{\bf i}[{\bf i}|\sigma_\nu\ket{\bf i}}{2m_h^{x=2}}, \\
    & (\partial_\mu h_{\nu\rho})\sigma^{\mu\nu}_{\alpha\beta} & \simeq
    \dfrac{{}_\alpha\ket{{\bf i}}[{\bf i}|\sigma_\rho\ket{\bf i}\bra{\bf i}_\beta}{\sqrt{2}m_h^{x=1}}.
\end{array}}

Suppose $l\leq r$ for the UV field $\Psi$, a three-particle amplitude with two SM and one UV particles is 
\bea
A^{(l,r),h_1,h_2}_{\{\alpha_1,...,\alpha_l\},\{\dot{\alpha}_1,...,\dot{\alpha}_r\}} =  (\lambda_1^{a} \lambda_2^{b})_{\{\alpha_1,...,\alpha_l\}} (\tilde{\lambda}_1^{c}\tilde{\lambda}_2^{d})_{\{\dot{\alpha}_1,...,\dot{\alpha}_r\}} [12]^{e} \langle12\rangle^{f},
\eea
where
\bea
\begin{cases}
b=l-a\\
c=a+h_1-h_2+\frac{1}{2}(r-l)\\
d=-a+h_2-h_1+\frac{1}{2}(l+r)\\
e=\frac{1}{2}(\mathrm{dim}+h_1+h_2-3) +\frac{1}{4}(l-5r)\\
f=\frac{1}{2}(\mathrm{dim}-h_1-h_2-3) -\frac{1}{4}(l+3r)
\end{cases}. 
\eea

We shall analyze the treatment with an example of a spin-2 UV that generates an operator within type $D^4 H^2 {H^\dagger}^2$. By j-basis-UV correspondence, we study the j-basis of partition $\{12|34\}$ and obtain a spin-2 UV $(2,1,3,1)$ with order $(\mbox{spin},SU(3),SU(2),U(1))$. Then we apply the spinor bridge method [ref] and obtain the partial amplitudes 
\begin{eqnarray}
A_{12} =\frac{1}{2m_h^2} \langle 2\mathbf{5}\rangle^2 [2\mathbf{5}]^2,\ A_{34} =\frac{1}{2m_h^2} \langle4\mathbf{5}\rangle^2 [4\mathbf{5}]^2, 
\end{eqnarray} 
where the UV is denoted as $\mathbf{5}$. In this case, particle $\mathbf{5}$ is helicity 0, thus carrying a $1/(2m_h^2)$ factor. Then we apply the operators of the above partial amplitudes by amplitude-operator correspondence
\begin{eqnarray}
A_{12}\sim H_i (D^\mu D^\nu H_j) h_{\mu\nu,I} {\tau^I}_k^i \epsilon^{jk},\ A_{34}\sim {H^\dagger}^i (D^\mu D^\nu {H^\dagger}^j) h_{\mu\nu,I} {\tau^I}_i^k \epsilon_{jk}, 
\end{eqnarray}
which are renormalizable. Since the Lorentz structure of partition $\{13|24\}$ is the same as $\{12|34\}$, the UV couplings of $\{13|24\}$ are also renormalizable, but with different gauge structures
\begin{eqnarray}
(2,1,1,0) &\mbox{ - }& H_i (D^\mu D^\nu {H^\dagger}^i) h_{\mu\nu},\ H_i (D^\mu D^\nu {H^\dagger}^i) h_{\mu\nu},\\
(2,1,3,0) &\mbox{ - }& H_i (D^\mu D^\nu {H^\dagger}^j) h_{\mu\nu,I} {\tau^I}^i_j,\ H_i (D^\mu D^\nu {H^\dagger}^j) h_{\mu\nu,I} {\tau^I}^i_j. 
\end{eqnarray}

Similarly we restudied all spin-2 UV that generate dim-8 SMEFT operators. Table~\ref{tab:spin2} lists all the spin-2 UV that generate dim-8 SMEFT operators. To analyze whether the UV couplings are renormalizable, we only need to focus on the Lorentz structures. There are only 3 kinds of amplitudes appearing in the table that can be translated into operators as
\begin{eqnarray}
\frac{1}{2m_h^2} \langle 2\mathbf{5}\rangle^2 [2\mathbf{5}]^2 \sim \phi (D_{\alpha\dot{\alpha}} D_{\beta\dot{\beta}} \phi) h^{\alpha\beta\dot{\alpha}\dot{\beta}} = \phi (D^\mu D^\nu \phi) h_{\mu\nu},
\\
\frac{1}{2m_h^2}\langle1\mathbf{5}\rangle^2 [3\mathbf{5}]^2 \sim {F_L}_{\alpha\beta} {F_R}_{\dot{\alpha}\dot{\beta}} h^{\alpha\beta\dot{\alpha}\dot{\beta}} =2 {F_L}_{\mu\nu} {F_R}_{\rho\sigma} h_{\omega\delta} (g^{\omega\mu} g^{\nu\sigma} g^{\rho\delta} -i\epsilon^{\mu\nu\omega\sigma} g^{\rho\delta} +i\epsilon^{\rho\sigma\delta\nu} g^{\mu\omega} -\epsilon^{\mu\nu\omega\lambda} \epsilon^{\rho\sigma\delta\tau} g_{\lambda\tau}),\\
\frac{1}{2m_h^2} \langle1\mathbf{5}\rangle \langle3\mathbf{5}\rangle [3\mathbf{5}]^2 \sim \psi_\alpha (D_{\beta\dot{\alpha}} {\psi^\dagger}_{\dot{\beta}}) h^{\alpha\beta\dot{\alpha}\dot{\beta}} =(\psi \sigma^\mu D^\nu \psi^\dagger) h_{\mu\nu},
\end{eqnarray}
all of which are renormalizable.

}


\subsection{Subtleties on Field Redefinitions Conversion}

The analysis presented in section~\ref{sec:PLGCP} offers a diagrammatic perspective that allows us to directly discern the potential UV origins responsible for the specific operator. This analysis encompasses crucial information, such as the quantum numbers of new resonances and their interactions with Standard Model (SM) particles. For instance, when examining the operator type $u_{\mathbb{C}}u^\dagger_{\mathbb{C}}LL^\dagger HH^\dagger$, conducting a j-basis analysis on the partition $\{\{L,H^\dagger\},\{u_{_\mathbb{C}},u^\dagger_{_\mathbb{C}}\},\{H,L^\dagger\}\}$ results in the following diagrammatic representation of a UV theory involving three heavy resonances: $\psi (\mathbf{1},\mathbf{1},-1,1/2)$, $\chi (\mathbf{1},\mathbf{2},-1/2,1/2)$, and $A^\mu (\mathbf{1},\mathbf{1},0,1)$, each associated with specific $SU(3)_C$, $SU(2)_L$, $U(1)_Y$, and Spin quantum numbers indicated in the parenthesis:
\begin{align}
\begin{tikzpicture}[baseline=0]
\begin{feynhand}
\vertex[particle] (e1) at (-2,-1) {$H^\dagger$};
\vertex[particle] (e2) at (-2,1) {$L$};
\vertex[particle] (e3) at (-0.5,1) {$H$};
\vertex[particle] (e4) at (0.5,1) {$L^\dagger$};
\vertex[particle] (e5) at (2,1) {$u_{_\mathbb{C}}$};
\vertex[particle] (e6) at (2,-1) {$u^\dagger_{_\mathbb{C}}$};

\vertex[dot] (i1) at (-1.5,0);
\vertex[dot] (i2) at (-0.5,0);
\vertex[dot] (i3) at (0.5,0);
\vertex[dot] (i4) at (1.5,0);

\propag[sca] (e1) to (i1);
\propag[plain] (e2) to (i1);
\propag[sca] (e3) to (i2);
{\setlength{\feynhandlinesize}{1.5pt}
\propag[plain] (i1) to [edge label = $\psi$](i2); }
{\setlength{\feynhandlinesize}{1.5pt} \propag[plain] (i2) to [edge label = $\chi$] (i3);}
\propag[plain] (e4) to (i3);
{\setlength{\feynhandlinesize}{1.0pt} \propag[bos] (i3) to  [edge label = $A$](i4);}
\propag[plain] (i4) to (e5);
\propag[plain] (i4) to (e6);

\end{feynhand}
\end{tikzpicture}
&=\bar{u}_R\gamma^\nu u_R \langle A^\mu  A^\nu \rangle \bar{l}_L\gamma^\mu \langle\chi \bar{\chi}\rangle    H\langle \psi \bar{\psi}\rangle H^\dagger l_L \nonumber\\
&\sim \bar{u}_R\gamma^\nu u_R  \frac{g_{\mu\nu}}{M_{A}}  \bar{l}_L\gamma^\mu  \frac{1}{M_{\chi}}   H  \frac{1}{M_{\psi}}  H^\dagger l_L \nonumber\\
&= \frac{1}{M_\psi M_\chi M^2_A}\bar{u}_R\gamma_\mu u_R\bar{l}_L\gamma^\mu H H^\dagger l_L ,\label{eq:uvdiagram}
\end{align}
where the angle brackets above represent the contraction of the heavy fields in the corresponding amplitude, and we have used the four-component notation for spinor field. The interactions needed in the UV theory are also easy to read out from the above diagram:
\begin{eqnarray}
    {\cal L}_{\rm int}^{\rm UV} \supset( \bar{l}_LH\psi + \bar{\chi}H\psi + \bar{l}_L\slashed{A}\chi  + h.c.)+ \overline{u_R}\slashed{A} u_R .
\end{eqnarray}

In light of the diagrammatic interpretation, it appears that the current approach may not fully encompass UV theories capable of generating the specific type of operator by converting other operators produced during the matching procedure via field redefinition. 
However, we propose that the j-basis analysis described above provides a potential solution for identifying these UV theories. An additional step involves verifying whether the quantum numbers of the resonances align with those of SM particles. 
If there exists a resonance in the candidate UV theory with quantum numbers matching an SM particle, we claim that a UV theory without that resonance is still capable of generating the target operator. This corresponds to a UV theory generating the target operators via field redefinition in removing those \textit{off-shell basis operators}\footnote{The off-shell basis operators are those does not contribute to the local tree-level on-shell amplitude for the scattering states involving external particles that are one-to-one correspondent to the fields in the operator. }.
At the dim-6 level, the field redefinition used to remove those off-shell basis operators is equivalent to replacing the kinetic part in the operators with EOMs of the corresponding fields. 
Diagrammatically it corresponds to the equivalence between the local on-shell amplitude generated by the seemingly non-local diagram with the massless pole being canceled by the insertion of off-shell basis operators and the same amplitude generated by the contact diagram with the converted operators. 
Therefore, replacing a UV resonance propagator with the SM one in the diagram in the j-basis analysis exactly corresponds to the procedure of producing the target operator by converting the operators of other types via EOMs.
For example, in the above $u_{_\mathbb{C}}u^\dagger_{_\mathbb{C}}LL^\dagger HH^\dagger$ example, the quantum numbers of the three resonance $\psi$, $\chi$ and $A^\mu$ are the same as those for SM fields $e^\dagger_{_\mathbb{C}}$, $L$ and $B^\mu$ respectively.
Therefore following the earlier argument, we can replace the field $\psi$ in the diagram with $e^\dagger_{_\mathbb{C}}(e_R)$, and the new UV theory containing only heavy fields $\chi$ and $A^\mu$ are expected to generate the operator of type $u_{_\mathbb{C}}u^\dagger_{_\mathbb{C}}LL^\dagger HH^\dagger$ by converting operator of type $u_{_\mathbb{C}}u^\dagger_{_\mathbb{C}}L^\dagger H e^\dagger_{_\mathbb{C}} D$ with EOM of $e^\dagger_{_\mathbb{C}}$. We can express this procedure diagrammatically as follows:
\begin{align}
&
\begin{tikzpicture}[baseline=0]
\begin{feynhand}
\vertex[particle] (e1) at (-2,-1) {$H^\dagger$};
\vertex[particle] (e2) at (-2,1) {$L$};
\vertex[particle] (e3) at (-0.5,1) {$H$};
\vertex[particle] (e4) at (0.5,1) {$L^\dagger$};
\vertex[particle] (e5) at (2,1) {$u_{_\mathbb{C}}$};
\vertex[particle] (e6) at (2,-1) {$u^\dagger_{_\mathbb{C}}$};

\vertex[dot] (i1) at (-1.5,0);
\vertex[dot] (i2) at (-0.5,0);
\vertex[dot] (i3) at (0.5,0);
\vertex[dot] (i4) at (1.5,0);

\propag[sca] (e1) to (i1);
\propag[plain] (e2) to (i1);
\propag[sca] (e3) to (i2);
{
\propag[plain] (i1) to [edge label = $e^\dagger_{_\mathbb{C}}$](i2); }
{\setlength{\feynhandlinesize}{1.5pt} \propag[plain] (i2) to [edge label = $\chi$] (i3);}
\propag[plain] (e4) to (i3);
{\setlength{\feynhandlinesize}{1.0pt} \propag[bos] (i3) to  [edge label = $A$](i4);}
\propag[plain] (i4) to (e5);
\propag[plain] (i4) to (e6);

\end{feynhand}
\end{tikzpicture}
\sim 
\begin{tikzpicture}[baseline=0]
\begin{feynhand}
\vertex[particle] (e1) at (-2,-1) {$H^\dagger$};
\vertex[particle] (e2) at (-2,1) {$L$};
\vertex[particle] (e3) at (-0.5,1) {$H$};
\vertex[particle] (e4) at (0.3,1) {$L^\dagger$};
\vertex[particle] (e5) at (1,1) {$u_{_\mathbb{C}}$};
\vertex[particle] (e6) at (1,-1) {$u^\dagger_{_\mathbb{C}}$};

\vertex[dot] (i1) at (-1.5,0);
\vertex[ blob,teal] (i3) at (0,0) {};

\propag[sca] (e1) to (i1);
\propag[plain] (e2) to (i1);
\propag[sca] (e3) to (i3);
{
\propag[plain] (i1) to [edge label = $e^\dagger_{_\mathbb{C}}$](i3); }
\propag[plain] (e4) to (i3);
\propag[plain] (i3) to (e5);
\propag[plain] (i3) to (e6);

\end{feynhand}
\end{tikzpicture}\sim 
\begin{tikzpicture}[baseline=0]
\begin{feynhand}
\vertex[particle] (e1) at (-0.5,-1) {$H^\dagger$};
\vertex[particle] (e2) at (-1,0) {$L$};
\vertex[particle] (e3) at (-0.5,1) {$H$};
\vertex[particle] (e4) at (0.5,1) {$L^\dagger$};
\vertex[particle] (e5) at (1,0) {$u_{_\mathbb{C}}$};
\vertex[particle] (e6) at (0.5,-1) {$u^\dagger_{_\mathbb{C}}$};

\vertex[ blob,olive] (i3) at (0,0) {};

\propag[sca] (e1) to (i3);
\propag[plain] (e2) to (i3);
\propag[sca] (e3) to (i3);
\propag[plain] (e4) to (i3);
\propag[plain] (i3) to (e5);
\propag[plain] (i3) to (e6);

\end{feynhand}
\end{tikzpicture}&\\
&=\bar{u}_R\gamma^\nu u_R \langle A^\mu  A^\nu \rangle \bar{l}_L\gamma^\mu \langle\chi \bar{\chi}\rangle    H\langle e_R \bar{e}_R\rangle H^\dagger l_L &\nonumber\\
&\supset \bar{u}_R\gamma^\nu u_R  \frac{1}{M^2_{A}}  \bar{l}_L\gamma^\mu  \frac{\color{blue}\slashed{p}}{M^2_{\chi}}   H  \frac{1}{\color{blue}\slashed{p}}  H^\dagger l_L \Longleftrightarrow {\color{teal} \frac{1}{M^2_\chi M^2_A}\bar{u}_R\gamma^\nu u_R\bar{l}_L\gamma^\mu  H\slashed{D}\langle e_R} \bar{e}_R \rangle l_L H^\dagger &\nonumber\\
&= {\color{olive}\frac{1}{M^2_\chi M^2_A}\bar{u}_R\gamma^\nu u_R\bar{l}_L\gamma^\mu H H^\dagger l_L },& 
\end{align}
where in the second line, the expression on the right-handed side with teal color corresponds to the insertion of the dim-8 operator $O_1=\bar{u}_R\gamma^\nu u_R\bar{l}_L\gamma^\mu  H\slashed{D} e_R$ generated by integrating out $\chi$ and $A$.
This $O_1$ operator contributes a momentum $\slashed{p}$ that cancels with the massless pole of $e_R$ and the resulting amplitude is equal to the one generated by insertion of the operator $\bar{u}_R\gamma^\nu u_R\bar{l}_L\gamma^\mu H H^\dagger l_L$ obtained by the replacement of EOM of $e_R$ in $O_1$. 
The conversions between dim-8 operators with EOMs can always be expressed in such a diagrammatic way, which in turn can be captured by our j-basis analysis, because the external particles in the process fix the possible quantum numbers of the internal lines in the tree diagram, thus by traversing all possible partitions we recursively include all the possible way for generating the operator via conversion with EOMs.

At the dim-8 level, the relation between the field redefinition and the conversion of EOMs is more subtle~\cite{Criado:2018sdb,Banerjee:2022thk}. Apart from the conversion within the dim-8 operators with EOMs of renormalizable Lagrangian, one also needs to consider the dim-8 operators generated by the conversion of dim-6 operators with EOMs of dim-6 Lagrangian and a remaining part that captures the difference between the EOMs replacement and field redefinition.  
We shall show that these two additional effects can also be linked to the diagrammatic origin and thus can be included in the j-basis analysis.
The case of generation of dim-8 operators via the conversion of dim-6 operators with EOMs of dim-6 Lagrangian can be illustrated by replacing the heavy propagator of $\chi$ with the SM lepton doublet field $L$ in the diagram in eq.~\eqref{eq:uvdiagram}:
\begin{align*}
&
\begin{tikzpicture}[baseline=0]
\begin{feynhand}
\vertex[particle] (e1) at (-2,-1) {$H^\dagger$};
\vertex[particle] (e2) at (-2,1) {$L$};
\vertex[particle] (e3) at (-0.5,1) {$H$};
\vertex[particle] (e4) at (0.5,1) {$L^\dagger$};
\vertex[particle] (e5) at (2,1) {$u_{_\mathbb{C}}$};
\vertex[particle] (e6) at (2,-1) {$u^\dagger_{_\mathbb{C}}$};

\vertex[dot] (i1) at (-1.5,0);
\vertex[dot] (i2) at (-0.5,0);
\vertex[dot] (i3) at (0.5,0);
\vertex[dot] (i4) at (1.5,0);

\propag[sca] (e1) to (i1);
\propag[plain] (e2) to (i1);
\propag[sca] (e3) to (i2);
{\setlength{\feynhandlinesize}{1.5pt} 
\propag[plain] (i1) to [edge label = $\psi$](i2); }
{ \propag[plain] (i2) to [edge label = $L$] (i3);}
\propag[plain] (e4) to (i3);
{\setlength{\feynhandlinesize}{1.0pt} \propag[bos] (i3) to  [edge label = $A$](i4);}
\propag[plain] (i4) to (e5);
\propag[plain] (i4) to (e6);

\end{feynhand}
\end{tikzpicture}
\sim 
\begin{tikzpicture}[baseline=0]
\begin{feynhand}
\vertex[particle] (e1) at (-2,-1) {$H^\dagger$};
\vertex[particle] (e2) at (-2,1) {$L$};
\vertex[particle] (e3) at (-1,1) {$H$};
\vertex[particle] (e4) at (-0.5,1) {$L^\dagger$};
\vertex[particle] (e5) at (0.5,1) {$u_{_\mathbb{C}}$};
\vertex[particle] (e6) at (0.5,-1) {$u^\dagger_{_\mathbb{C}}$};

{\setlength{\feynhandblobsize}{5mm}
\vertex[blob, teal] (i1) at (-1.5,0) {};
\vertex[ blob,teal] (i3) at (0,0) {};}

\propag[sca] (e1) to (i1);
\propag[plain] (e2) to (i1);
\propag[sca] (e3) to (i1);
{
\propag[plain] (i1) to [edge label = $L$](i3); }
\propag[plain] (e4) to (i3);
\propag[plain] (i3) to (e5);
\propag[plain] (i3) to (e6);

\end{feynhand}
\end{tikzpicture}= 
\begin{tikzpicture}[baseline=0]
\begin{feynhand}
\vertex[particle] (e1) at (-0.5,-1) {$H^\dagger$};
\vertex[particle] (e2) at (-1,0) {$L$};
\vertex[particle] (e3) at (-0.5,1) {$H$};
\vertex[particle] (e4) at (0.5,1) {$L^\dagger$};
\vertex[particle] (e5) at (1,0) {$u_{_\mathbb{C}}$};
\vertex[particle] (e6) at (0.5,-1) {$u^\dagger_{_\mathbb{C}}$};

\vertex[ blob,olive] (i3) at (0,0) {};

\propag[sca] (e1) to (i3);
\propag[plain] (e2) to (i3);
\propag[sca] (e3) to (i3);
\propag[plain] (e4) to (i3);
\propag[plain] (i3) to (e5);
\propag[plain] (i3) to (e6);

\end{feynhand}
\end{tikzpicture}.&
\end{align*}
In the middle diagram, the left and right blobs in teal color represent the dim-6 operators generated by integrating out $\psi$ and $A$ respectively: $(\slashed{D}\bar{l}_L) HH^\dagger l_L$ and $(\bar{l}_L\gamma^\mu l_L)(\bar{u}_R\gamma_\mu u_R)$. 
The latter contributes to the dim-6 EOM -- $\slashed{D}\bar{l}_L\supset (\bar{u}_R\gamma_\mu u_R)\bar{l}_L\gamma^\mu$, and the equivalence between the amplitudes generated by the middle and the right diagram exactly corresponds to the substitution of the $\slashed{D}\bar{l}_L$ in the operator $(\slashed{D}\bar{l}_L) HH^\dagger l_L$ with EOM of $\bar{l}_L$ at dim-6 resulting in the dim-8 operator $\bar{u}_R\gamma^\mu u_R\bar{l}_L\gamma^\mu H H^\dagger l_L$.

Lastly, the remaining origin for dim-8 operators that comes from the field redefinition and cannot be captured by substitutions of EOMs are those from the second order variation of the renormalizable part of EFT Lagrangian: $1/2 f^\alpha \delta^2{\cal L}_4/(\delta \Phi_\alpha \delta \Phi_\beta)f_\beta$, where $\Phi_\alpha$ is the corresponding SM fields, which to be shifted by $\Phi_\alpha \to \Phi_\alpha + f_\alpha$ to eliminate the dim-6 off-shell basis operators, and the repeated indices $\alpha$ and $\beta$ traversing all the SM fields undergoing redefinition are summed implicitly.   
For the dim-8 operator generated in this way to be tree-level, the corresponding relevant dim-6 operators that were to be removed by the field redefinition must also be generated at the tree level. This requirement reduces dim-6 operators into two classes of operator: $\psi^\dagger\psi H^\dagger H D$ and $H^2H^{\dagger 2}D^2$. 
Removing redundant Green basis operators $|H|^2 H^\dagger D^2 H$ and $\bar{\psi}_2i\overleftrightarrow{\slashed{D}}\psi_1 HH^\dagger$ in these two class generated during the matching amounts to make field redefinition $\psi_1\to \psi_1- \xi\psi_2 HH^\dagger$, $\psi_2\to \psi_2-\xi\psi_1 HH^\dagger$, and $H\to H+\xi' H|H|^2$, where $\xi$ and $\xi'$ are equal to the Wilson coefficient of the corresponding operators generated during matching, and we have omitted gauge contractions in the expressions. 
Diagrammatically, these second-order variation contributions can be further classified into the following three categories:
\begin{align}
 \text{Kinetic term: }   \begin{tikzpicture}[baseline=0]
\begin{feynhand}
\vertex[particle] (e1) at (-2,-1) {$\ $};
\vertex[particle] (e2) at (-2,1) {$\ $};
\vertex[particle] (e3) at (-2,0){$\ $};
\vertex[particle] (e4) at (2,0){$\ $};
\vertex[particle] (e5) at (2,1) {$\ $};
\vertex[particle] (e6) at (2,-1){$\ $};

{\setlength{\feynhandblobsize}{3mm}\vertex[blob,teal] (i2) at (-1,0) {};}
\vertex[crossdot] (i1) at (0,0) {};
{\setlength{\feynhandblobsize}{3mm}\vertex[blob,teal] (i3) at (1,0) {};}

\propag[plain] (e1) to (i2);
\propag[plain] (e2) to (i2);
\propag[plain] (e3) to (i2);
\propag[plain] (i2) to [ edge label = $\Phi^\dagger_\alpha$](i1);
\propag[plain] (i1) to [ edge label = $\Phi_\alpha$](i3);
\propag[plain] (e4) to (i3);
\propag[plain] (i3) to (e5);
\propag[plain] (i3) to (e6);
\draw [decoration={brace}, decorate] (e1.south west) -- (e2.north west)
node [pos=0.5, left] {\(f^\dagger_\alpha\)};
\draw [decoration={brace}, decorate] (e5.north east) -- (e6.south east)
node [pos=0.5, right] {\(f_\alpha\)};
\end{feynhand}
\end{tikzpicture}\longrightarrow
\begin{tikzpicture}[baseline=0]
\begin{feynhand}
\vertex[particle] (e1) at (-2,-1) {$\ $};
\vertex[particle] (e2) at (-2,1) {$\ $};
\vertex[particle] (e3) at (-2,0){$\ $};
\vertex[particle] (e4) at (2,0){$\ $};
\vertex[particle] (e5) at (2,1) {$\ $};
\vertex[particle] (e6) at (2,-1){$\ $};

{\setlength{\feynhandblobsize}{5mm}\vertex[blob,olive] (i1) at (0,0) {};}

\propag[plain] (e1) to (i1);
\propag[plain] (e2) to (i1);
\propag[plain] (e3) to (i1);
\propag[plain] (e4) to (i1);
\propag[plain] (i1) to (e5);
\propag[plain] (i1) to (e6);
\draw [decoration={brace}, decorate] (e1.south west) -- (e2.north west)
node [pos=0.5, left] {\(f^\dagger_\alpha\)};
\draw [decoration={brace}, decorate] (e5.north east) -- (e6.south east)
node [pos=0.5, right] {\(f_\alpha\)};
\end{feynhand}
\end{tikzpicture}
\\
 \text{ Yukawa term: } 
\begin{tikzpicture}[baseline=0]
\begin{feynhand}
\vertex[particle] (e1) at (-2,-1) {$\ $};
\vertex[particle] (e2) at (-2,1) {$\ $};
\vertex[particle] (e3) at (-2,0){$\ $};
\vertex[particle] (e4) at (2,0){$\ $};
\vertex[particle] (e5) at (2,1) {$\ $};
\vertex[particle] (e6) at (2,-1){$\ $};
\vertex[particle] (e7) at (0,1.5){$\Phi_\gamma $};

{\setlength{\feynhandblobsize}{3mm}\vertex[blob,teal] (i2) at (-1,0) {};}
\vertex[dot] (i1) at (0,0) {};
{\setlength{\feynhandblobsize}{3mm}\vertex[blob,teal] (i3) at (1,0) {};}

\propag[plain] (e7) to (i1);
\propag[plain] (e1) to (i2);
\propag[plain] (e2) to (i2);
\propag[plain] (e3) to (i2);
\propag[plain] (i2) to [ edge label = $\Phi_\alpha$](i1);
\propag[plain] (i1) to [ edge label = $\Phi_\beta$](i3);
\propag[plain] (e4) to (i3);
\propag[plain] (i3) to (e5);
\propag[plain] (i3) to (e6);
\draw [decoration={brace}, decorate] (e1.south west) -- (e2.north west)
node [pos=0.5, left] {\(f_\alpha\)};
\draw [decoration={brace}, decorate] (e5.north east) -- (e6.south east)
node [pos=0.5, right] {\(f_\beta\)};
\end{feynhand}
\end{tikzpicture}
\longrightarrow
\begin{tikzpicture}[baseline=0]
\begin{feynhand}
\vertex[particle] (e1) at (-2,-1) {$\ $};
\vertex[particle] (e2) at (-2,1) {$\ $};
\vertex[particle] (e3) at (-2,0){$\ $};
\vertex[particle] (e4) at (2,0){$\ $};
\vertex[particle] (e5) at (2,1) {$\ $};
\vertex[particle] (e6) at (2,-1){$\ $};
\vertex[particle] (e7) at (0,1.5){$\Phi_\gamma $};

{\setlength{\feynhandblobsize}{5mm}\vertex[blob,olive] (i1) at (0,0) {};}
\propag[plain] (e7) to (i1);
\propag[plain] (e1) to (i1);
\propag[plain] (e2) to (i1);
\propag[plain] (e3) to (i1);
\propag[plain] (e4) to (i1);
\propag[plain] (i1) to (e5);
\propag[plain] (i1) to (e6);
\draw [decoration={brace}, decorate] (e1.south west) -- (e2.north west)
node [pos=0.5, left] {\(f_\alpha\)};
\draw [decoration={brace}, decorate] (e5.north east) -- (e6.south east)
node [pos=0.5, right] {\(f_\beta\)};
\end{feynhand}
\end{tikzpicture}
\\
 \text{ Quartic term: } 
\begin{tikzpicture}[baseline=0]
\begin{feynhand}
\vertex[particle] (e1) at (-2,-1) {$\ $};
\vertex[particle] (e2) at (-2,1) {$\ $};
\vertex[particle] (e3) at (-2,0){$\ $};
\vertex[particle] (e4) at (2,0){$\ $};
\vertex[particle] (e5) at (2,1) {$\ $};
\vertex[particle] (e6) at (2,-1){$\ $};
\vertex[particle] (e7) at (0.5,1.5){$\Phi_\gamma $};
\vertex[particle] (e8) at (-0.5,1.5){$\Phi_\delta $};

{\setlength{\feynhandblobsize}{3mm}\vertex[blob,teal] (i2) at (-1,0) {};}
\vertex[dot] (i1) at (0,0) {};
{\setlength{\feynhandblobsize}{3mm}\vertex[blob,teal] (i3) at (1,0) {};}

\propag[plain] (e7) to (i1);
\propag[plain] (e8) to (i1);
\propag[plain] (e1) to (i2);
\propag[plain] (e2) to (i2);
\propag[plain] (e3) to (i2);
\propag[plain] (i2) to [ edge label = $\Phi_\alpha$](i1);
\propag[plain] (i1) to [ edge label = $\Phi_\beta$](i3);
\propag[plain] (e4) to (i3);
\propag[plain] (i3) to (e5);
\propag[plain] (i3) to (e6);
\draw [decoration={brace}, decorate] (e1.south west) -- (e2.north west)
node [pos=0.5, left] {\(f_\alpha\)};
\draw [decoration={brace}, decorate] (e5.north east) -- (e6.south east)
node [pos=0.5, right] {\(f_\alpha\)};
\end{feynhand}
\end{tikzpicture}
\longrightarrow
\begin{tikzpicture}[baseline=0]
\begin{feynhand}
\vertex[particle] (e1) at (-2,-1) {$\ $};
\vertex[particle] (e2) at (-2,1) {$\ $};
\vertex[particle] (e3) at (-2,0){$\ $};
\vertex[particle] (e4) at (2,0){$\ $};
\vertex[particle] (e5) at (2,1) {$\ $};
\vertex[particle] (e6) at (2,-1){$\ $};
\vertex[particle] (e7) at (0.5,1.5){$\Phi_\gamma $};
\vertex[particle] (e8) at (-0.5,1.5){$\Phi_\delta $};

{\setlength{\feynhandblobsize}{5mm}\vertex[blob,olive] (i1) at (0,0) {};}
\propag[plain] (e7) to (i1);
\propag[plain] (e8) to (i1);
\propag[plain] (e1) to (i1);
\propag[plain] (e2) to (i1);
\propag[plain] (e3) to (i1);
\propag[plain] (e4) to (i1);
\propag[plain] (i1) to (e5);
\propag[plain] (i1) to (e6);
\draw [decoration={brace}, decorate] (e1.south west) -- (e2.north west)
node [pos=0.5, left] {\(f_\alpha\)};
\draw [decoration={brace}, decorate] (e5.north east) -- (e6.south east)
node [pos=0.5, right] {\(f_\beta\)};
\end{feynhand}
\end{tikzpicture},
\end{align}
where $\Phi$'s are either SM fermions or Higgs fields, $f$'s are determined by which source particle it stems from. 
The blobs in teal color in the above diagrams correspond to either dim-6 operators $|H|^2 H^\dagger D^2 H$ or $\bar{\psi}_2i\overleftrightarrow{\slashed{D}}\psi_1 HH^\dagger$, and opening further these blobs with finer partition helps to obtain the corresponding UV theories responsible for generating the dim-8 operators in olive blobs on by second-order variation in field redefinition. 
Again this must be included in our j-basis analysis plus identifying the intermediate resonances with the same quantum numbers of SM particles. 
To be more concrete, we illustrate this scenario with an example of the j-basis analysis of operator type $H^2H^{\dagger 3}L e_{_\mathbb{C}}$ with patition $\{\{L, H\}, \{L, H, H^\dagger  \},\{ e_{_\mathbb{C}}, H\}, \{ e_{_\mathbb{C}}, H, H^\dagger \}\}$ in the following diagram:
\begin{align}
&\begin{tikzpicture}[baseline=0]
\begin{feynhand}
\vertex[particle] (e1) at (-2.5,-1) {$H$};
\vertex[particle] (e2) at (-2.5,1) {$L$};
\vertex[particle] (e3) at (-1,1) {$H^\dagger$};
\vertex[particle] (e4) at (1,1) {$H^\dagger$};
\vertex[particle] (e5) at (2.5,1) {$e_{_\mathbb{C}}$};
\vertex[particle] (e6) at (2.5,-1) {$H$};
\vertex[particle] (e7) at (0,1) {$H^\dagger$};

\vertex[dot] (i1) at (-2,0);
\vertex[dot] (i2) at (-1,0);
\vertex[dot] (i3) at (1,0);
\vertex[dot] (i4) at (2,0);
\vertex[dot] (i5) at (0,0);

\propag[sca] (e1) to (i1);
\propag[plain] (e2) to (i1);
\propag[sca] (e7) to (i5);
\propag[sca] (e3) to (i2);
{\setlength{\feynhandlinesize}{1.5pt}
\propag[plain] (i2) to [edge label = $\psi_2$](i5);
\propag[plain] (i5) to [edge label = $\psi_3$] (i3);
\propag[plain] (i1) to [edge label = $\psi_1$](i2); }
\propag[sca] (e4) to (i3);
{\setlength{\feynhandlinesize}{1.5pt} \propag[plain] (i3) to  [edge label = $\psi_4$](i4);}
\propag[plain] (i4) to (e5);
\propag[sca] (i4) to (e6);

\end{feynhand}
\end{tikzpicture}
\quad \begin{pmatrix}
  \psi_1  (\mathbf{1},  \mbf{3} , 1 , 1/2)\\
  \psi_2  (\mathbf{1}, \mbf{2}, -1/2 , 1/2)\\
  \psi_3  (\mathbf{1} , \mbf{1} , 1, 1/2)\\
  \psi_4  (\mathbf{1} , \mbf{2} , 3/2 , 1/2)\\
\end{pmatrix},\  \psi_i(SU(3)_C,SU(2)_L,U(1)_Y,\text{Spin})\nonumber\\
&\quad\quad\quad\quad\quad\quad\quad \Downarrow &\nonumber\\
&\begin{tikzpicture}[baseline=0]
\begin{feynhand}
\vertex[particle] (e1) at (-2.5,-1) {$H$};
\vertex[particle] (e2) at (-2.5,1) {$L$};
\vertex[particle] (e3) at (-1,1) {$H^\dagger$};
\vertex[particle] (e4) at (1,1) {$H^\dagger$};
\vertex[particle] (e5) at (2.5,1) {$e_{_\mathbb{C}}$};
\vertex[particle] (e6) at (2.5,-1) {$H$};
\vertex[particle] (e7) at (0,1) {$H^\dagger$};

\vertex[dot] (i1) at (-2,0);
\vertex[dot] (i2) at (-1,0);
\vertex[dot] (i3) at (1,0);
\vertex[dot] (i4) at (2,0);
\vertex[dot] (i5) at (0,0) {};

\propag[sca] (e1) to (i1);
\propag[plain] (e2) to (i1);
\propag[sca] (e7) to (i5);
\propag[sca] (e3) to (i2);
\propag[plain] (i2) to [edge label = $L$](i5);
\propag[plain] (i5) to [edge label = $e_{_\mathbb{C}}$] (i3);
{\setlength{\feynhandlinesize}{1.5pt}
\propag[plain] (i1) to [edge label = $\psi_1$](i2); }
\propag[sca] (e4) to (i3);
{\setlength{\feynhandlinesize}{1.5pt} \propag[plain] (i3) to  [edge label = $\psi_4$](i4);}
\propag[plain] (i4) to (e5);
\propag[sca] (i4) to (e6);

\end{feynhand}
\end{tikzpicture}
\longrightarrow
\begin{tikzpicture}[baseline=0]
\begin{feynhand}
\vertex[particle] (e1) at (-2,-1) {$H $};
\vertex[particle] (e2) at (-2,1) {$L$};
\vertex[particle] (e3) at (-2,0){$H^\dagger$};
\vertex[particle] (e4) at (2,0){$H^\dagger$};
\vertex[particle] (e5) at (2,1) {$e_{_\mathbb{C}}$};
\vertex[particle] (e6) at (2,-1){$H$};
\vertex[particle] (e7) at (0,1.5){$H^\dagger$};

{\setlength{\feynhandblobsize}{3mm}\vertex[blob,teal] (i2) at (-1,0) {};}
\vertex[dot] (i1) at (0,0) {};
{\setlength{\feynhandblobsize}{3mm}\vertex[blob,teal] (i3) at (1,0) {};}

\propag[plain] (e7) to (i1);
\propag[plain] (e1) to (i2);
\propag[plain] (e2) to (i2);
\propag[plain] (e3) to (i2);
\propag[plain] (i2) to [ edge label = $L$](i1);
\propag[plain] (i1) to [ edge label = $e_{_\mathbb{C}}$](i3);
\propag[plain] (e4) to (i3);
\propag[plain] (i3) to (e5);
\propag[plain] (i3) to (e6);
\end{feynhand}
\end{tikzpicture}.
\end{align}
We find that one of the UV contains four fermion fields $\psi_1$ to $\psi_4$ from our ordinary j-basis analysis, and according to the aforementioned discussion, we identify that $\psi_2$ and $\psi_3$ have the same quantum numbers with the SM fields $L$ and $e_{_\mathbb{C}}$ respectively. 
Therefore we conclude that the UV theories with only $\psi_1$ and $\psi_4$ can also generate dim-8 operator $H^2H^{\dagger 3}L e_{_\mathbb{C}}$ by field redefinition of in removing dim-6 off-shell basis operators $\bar{L}i\overleftrightarrow{\slashed{D}}L HH^\dagger$ and $\bar{e}_Ri\overleftrightarrow{\slashed{D}}e_R HH^\dagger$ when integrating out $\psi_1$ and $\psi_4$, which can actually be easily verified by readers.

To summarize the above discussion:
\begin{itemize}
    \item We have established a diagrammatic correspondence between the UV theory and dim-8 operators, which include two distinct processes for generating a dim-8 operator type: direct integration of heavy resonances and operator conversion through field redefinitions.
    \item Through the j-basis analysis, we can effectively capture all these diagrammatic scenarios by systematically traversing the partitions for a given operator type. With an additional step to identify and drop the resonances in the UV theories that share the same quantum numbers as the Standard Model (SM) particles, we effectively include the UV origins for the other operator types that can be converted to the given operator type with field redefinition.
    \item As a result of this careful analysis, the UV theories that contribute to a specific type of operator, as discovered through our j-basis analysis, are considered complete. We have successfully accounted for all the relevant mechanisms, leading to a comprehensive understanding of the operator's origins.
\end{itemize}
Finally, we comment that people in the phenomenology community like to speak of the UV resonance of certain EFT operators is not meaningful without fixing whole the operator basis at a certain dimension.
Because the concept of ``UV origin" of an operator is ambiguous due to the freedom to convert an operator to a different one with field redefinition.
Due to this ambiguity and the fact that the operators are just intermediate tools to compute the predictions of physical observables, we think that the ``UV origin" of an operator type which fixed the field contents and the number of derivatives is a more reasonable question to investigate.

\section{Complete sets of the UV Resonances for the dim-8 operators}\label{sec:4}

Based on the discussions in previous sections, we summarize here the steps for obtaining the UV origins for a type of operator:
\begin{enumerate}
\item Obtain the independent partitions for the type of operators.
\item For each partition, find out all possible insertion quantum numbers of the resonance using our j-basis analysis. Each set of resonance and associated interactions read out from the corresponding diagram is a candidate UV theory.
\item For each candidate UV theory, we determine whether each interaction is "renormalizable" as discussed in section.~\ref{sec:HSR}. Only retain those UV theories with all interactions that are renormalizable.
\item For each remaining UV theory, find out whether there exist special resonances in the theory that coincide with the quantum numbers of SM particles.  Remove those special resonances in all kinds of ways and keep at least one new resonance in the UV theory. Each way of removing special resonances corresponds to a new UV theory responsible for the operator type under investigation.   
\end{enumerate}

\input{sec4-uvlist}

\section{Conclusion}\label{sec:con}

In this work we investigate the connection between effective operators and their UV resonances in the bottom-up way. We utilize the J-basis technique to investigate all possible tree-level UV origins for the dimension-8 SMEFT operators in the Young Tensor basis.  We list a complete set of 146 (82) possible UV resonances which could generate the dimension-8 SMEFT operators at the tree-level, and we write down all possible UV couplings up to mass dimension 5 (4). Except presenting the above results, two important theoretical issures are also addressed.

First, we clarify the ambiguity in the J-basis/UV resonance correspondence for the UV theories with heavy particles with spin $J\geq 1$, and illustrate that the couplings between the spin-$J$ massive particles and the current of a different angular momentum $J'\neq J$ can be replaced by higher dimensional operators via field redefinition. 
In this way, we can classify the couplings in the UV theories into the ``loop-induced" ones and the authentic tree-level ones. 
Following the line, we enumerate all the 3-pt and 4-pt ``tree-level" couplings in the form of on-shell amplitudes involving massive particles of arbitrary spin-$J$.
The convention also resolves to some extent the issue of whether a non-renormalizable operator can be generated at tree-level in a UV theory.

Second, we tackle the subtlety about the UV theories responsible for the generation of operators via field redefinition. 
In this work, we build a diagrammatic illustration, showing that these UV theories can be included by adding another step after the ordinary J-basis analysis --- recursively identifying and replacing the heavy resonance in the diagram with the SM particles with the same quantum numbers.
In doing so, we can proliferate the UV theories obtained from an ordinary j-basis analysis and thus take into account all the UV theories responsible for the operators in certain operator basis.


Finally, for the readers' convenience, we provide a UV database for dimension 8 operators in the form of \texttt{Mathematica} package, with which one can easily extract the relevant UV theories for generating a specific F-basis operator, including the information of new fields and interactions. Given these information, it provides new strategy on searching for new resonances at the large hadron collider (LHC). Current LHC searches focus on resonances signatures on various new physics models, and various simplified models, which can not be exhaustively searched. In this bottom-up EFT approach, the new resonances are organized by the mass dimension of the effective operators. Truncated to certain mass dimension, such as dimension 8, only limited new resonances contribute to physical processes at the LHC. Thus only searching for such new resonances would be enough. In summary, we believe our results on listing UV resonances would benefit the phenomenology community in the future LHC searches.


\section*{Acknowledgments}

J.H.Y. is supported by the National Science Foundation of China under Grants No. 12022514, No. 12375099 and No. 12047503, and National Key Research and Development Program of China Grant No. 2020YFC2201501, and No. 2021YFA0718304. 
M.-L.X. is supported in part by the Jay Jones Fund in the Department of Physics and Astronomy at Northwestern University, and by the Junior Foundation of Sun Yat-Sen University.
H.-L.L. is supported by the 4.4517.08 IISN-F.N.R.S convention.

\appendix\label{appendix}

\section{UV List for Operators involving 4 particles}
In the following sections, we present the comprehensive results of the j-basis operators of dimension-8 SMEFT for each type of operators, and for each partition. The j-basis operators are expressed in terms of the f-basis operators modulo any pieces that can be converted by the field redefinition and covariant derivative commutator. For those operators with fermion fields, the flavor indices of fermion are labeled in alphabetic order starting with ``$p$, $r$, $s$, $t$,\dots'' depending on their order in the appearance in the operator type.
\begin{longtable}{cc|c}
\caption{UV resonance of type ${B_L}^4$. ``*'' in front of each row indicates topologies with nonrenormalizable vertices. The UV resonances colored in gray are eliminated by UV selection. }\\
\hline\hline\multicolumn{3}{l}{\multirow{2}{*}{\large Type: ${B_L}^4$}}\\
 \multicolumn{3}{l}{} \\
\hline\multicolumn{3}{c}{$\begin{array}{ccl}\mathcal{O}^f_{1}&=&\frac{1}{24}\mathcal{Y}[\yng(4)_{{B_L}}]{{B_L}^{\lambda\mu}}{{B_L}^{\lambda\rho}}{{B_L}^{\mu\nu}}{{B_L}^{\nu\rho}}.\end{array}$}\\
\hline
\multicolumn{2}{c|}{group: $(\text{Spin},SU(3)_c,SU(2)_w,U(1)_y)$} & \multirow{2}{*}{$\mathcal{O}^{j}$}\\
\begin{fmffile}{4}
\begin{fmfgraph*}(40,20)
\fmfpen{thin} \fmfleft{e1,e2} \fmfv{}{g1} \fmfv{}{g2} \fmfright{e4,e3}
\fmf{plain}{e1,g1} \fmf{plain}{g1,e2} \fmf{plain}{g1,g2} \fmf{plain}{e4,g2} \fmf{plain}{g2,e3}
\end{fmfgraph*}
\end{fmffile} &\multicolumn{1}{c|}{ $\{{B_L}_1,{B_L}_2\} \quad \{{B_L}_3,{B_L}_4\}$ } & \\ 
 \hline* & {\color{gray} $(2,1,1,0)$} & {\color{gray} $-64 \mathcal{O}^f_{{1}}$}\\ 
* & $(0,1,1,0)$ & $16 \mathcal{O}^f_{{1}}$\\ 
\hline
\end{longtable}

\begin{longtable}{cc|c}
\caption{UV resonance of type ${B_L}^2{W_L}^2$. ``*'' in front of each row indicates topologies with nonrenormalizable vertices. The UV resonances colored in gray are eliminated by UV selection. }\\
\hline\hline\multicolumn{3}{l}{\multirow{2}{*}{\large Type: ${B_L}^2{W_L}^2$}}\\
 \multicolumn{3}{l}{} \\
\hline\multicolumn{3}{c}{$\begin{array}{ccl}\mathcal{O}^f_{1}&=&\frac{1}{4}\mathcal{Y}[\yng(2)_{{B_L}}]\mathcal{Y}[\yng(2)_{{W_L}}]{{B_L}^{\lambda\rho}}{{B_L}^{\mu\nu}}{{W_L}^{I\lambda\mu}}{{W_L}^{I\nu\rho}},\\
\mathcal{O}^f_{2}&=&\frac{1}{4}\mathcal{Y}[\yng(2)_{{B_L}}]\mathcal{Y}[\yng(2)_{{W_L}}]{{B_L}^{\mu\nu}}{{B_L}^{\mu\nu}}{{W_L}^{I\lambda\rho}}{{W_L}^{I\lambda\rho}}.\end{array}$}\\
\hline
\multicolumn{2}{c|}{group: $(\text{Spin},SU(3)_c,SU(2)_w,U(1)_y)$} & \multirow{2}{*}{$\mathcal{O}^{j}$}\\
\begin{fmffile}{4}
\begin{fmfgraph*}(40,20)
\fmfpen{thin} \fmfleft{e1,e2} \fmfv{}{g1} \fmfv{}{g2} \fmfright{e4,e3}
\fmf{plain}{e1,g1} \fmf{plain}{g1,e2} \fmf{plain}{g1,g2} \fmf{plain}{e4,g2} \fmf{plain}{g2,e3}
\end{fmfgraph*}
\end{fmffile} &\multicolumn{1}{c|}{ $\{{B_L}_1,{B_L}_2\} \quad \{{W_L}_3,{W_L}_4\}$ } & \\ 
 \hline* & {\color{gray} $(2,1,1,0)$} & {\color{gray} $48 \mathcal{O}^f_{{1}} -4 \mathcal{O}^f_{{2}}$}\\ 
* & $(0,1,1,0)$ & $4 \mathcal{O}^f_{{2}}$\\ 
\hline
& \multicolumn{1}{c|}{ $\{{B_L}_1,{W_L}_3\} \quad \{{B_L}_2,{W_L}_4\}$ } & $\mathcal{O}^{j}$\\
 \hline* & {\color{gray} $(2,1,3,0)$} & {\color{gray} $8 \mathcal{O}^f_{{1}} -14 \mathcal{O}^f_{{2}}$}\\ 
* & $(1,1,3,0)$ & $8 \mathcal{O}^f_{{1}} + 2 \mathcal{O}^f_{{2}}$\\ 
* & $(0,1,3,0)$ & $-8 \mathcal{O}^f_{{1}} + 2 \mathcal{O}^f_{{2}}$\\ 
\hline
\end{longtable}

\begin{longtable}{cc|c}
\caption{UV resonance of type ${W_L}^4$. ``*'' in front of each row indicates topologies with nonrenormalizable vertices. The UV resonances colored in gray are eliminated by UV selection. }\\
\hline\hline\multicolumn{3}{l}{\multirow{2}{*}{\large Type: ${W_L}^4$}}\\
 \multicolumn{3}{l}{} \\
\hline\multicolumn{3}{c}{$\begin{array}{ccl}\mathcal{O}^f_{1}&=&\frac{1}{24}\mathcal{Y}[\yng(4)_{{W_L}}]{{W_L}^{I\mu\nu}}{{W_L}^{I\nu\rho}}{{W_L}^{J\lambda\mu}}{{W_L}^{J\lambda\rho}},\\
\mathcal{O}^f_{2}&=&\frac{1}{24}\mathcal{Y}[\yng(4)_{{W_L}}]{{W_L}^{I\lambda\rho}}{{W_L}^{I\mu\nu}}{{W_L}^{J\lambda\rho}}{{W_L}^{J\mu\nu}}.\end{array}$}\\
\hline
\multicolumn{2}{c|}{group: $(\text{Spin},SU(3)_c,SU(2)_w,U(1)_y)$} & \multirow{2}{*}{$\mathcal{O}^{j}$}\\
\begin{fmffile}{4}
\begin{fmfgraph*}(40,20)
\fmfpen{thin} \fmfleft{e1,e2} \fmfv{}{g1} \fmfv{}{g2} \fmfright{e4,e3}
\fmf{plain}{e1,g1} \fmf{plain}{g1,e2} \fmf{plain}{g1,g2} \fmf{plain}{e4,g2} \fmf{plain}{g2,e3}
\end{fmfgraph*}
\end{fmffile} &\multicolumn{1}{c|}{ $\{{W_L}_1,{W_L}_2\} \quad \{{W_L}_3,{W_L}_4\}$ } & \\ 
 \hline* & {\color{gray} $(2,1,5,0)$} & {\color{gray} $352 \mathcal{O}^f_{{1}} + 24 \mathcal{O}^f_{{2}}$}\\ 
* & $(0,1,5,0)$ & $32 \mathcal{O}^f_{{1}} + 24 \mathcal{O}^f_{{2}}$\\ 
* & {\color{gray} $(2,1,1,0)$} & {\color{gray} $-32 \mathcal{O}^f_{{1}} -24 \mathcal{O}^f_{{2}}$}\\ 
* & $(0,1,1,0)$ & $-16 \mathcal{O}^f_{{1}}$\\ 
* & $(1,1,3,0)$ & $32 \mathcal{O}^f_{{1}} + 8 \mathcal{O}^f_{{2}}$\\ 
\hline
\end{longtable}

\begin{longtable}{cc|c}
\caption{UV resonance of type ${B_L}^2{G_L}^2$. ``*'' in front of each row indicates topologies with nonrenormalizable vertices. The UV resonances colored in gray are eliminated by UV selection. }\\
\hline\hline\multicolumn{3}{l}{\multirow{2}{*}{\large Type: ${B_L}^2{G_L}^2$}}\\
 \multicolumn{3}{l}{} \\
\hline\multicolumn{3}{c}{$\begin{array}{ccl}\mathcal{O}^f_{1}&=&\frac{1}{4}\mathcal{Y}[\yng(2)_{{B_L}}]\mathcal{Y}[\yng(2)_{{G_L}}]{{B_L}^{\lambda\rho}}{{B_L}^{\mu\nu}}{{G_L}^{A\lambda\mu}}{{G_L}^{A\nu\rho}},\\
\mathcal{O}^f_{2}&=&\frac{1}{4}\mathcal{Y}[\yng(2)_{{B_L}}]\mathcal{Y}[\yng(2)_{{G_L}}]{{B_L}^{\mu\nu}}{{B_L}^{\mu\nu}}{{G_L}^{A\lambda\rho}}{{G_L}^{A\lambda\rho}}.\end{array}$}\\
\hline
\multicolumn{2}{c|}{group: $(\text{Spin},SU(3)_c,SU(2)_w,U(1)_y)$} & \multirow{2}{*}{$\mathcal{O}^{j}$}\\
\begin{fmffile}{4}
\begin{fmfgraph*}(40,20)
\fmfpen{thin} \fmfleft{e1,e2} \fmfv{}{g1} \fmfv{}{g2} \fmfright{e4,e3}
\fmf{plain}{e1,g1} \fmf{plain}{g1,e2} \fmf{plain}{g1,g2} \fmf{plain}{e4,g2} \fmf{plain}{g2,e3}
\end{fmfgraph*}
\end{fmffile} &\multicolumn{1}{c|}{ $\{{B_L}_1,{B_L}_2\} \quad \{{G_L}_3,{G_L}_4\}$ } & \\ 
 \hline* & {\color{gray} $(2,1,1,0)$} & {\color{gray} $-48 \mathcal{O}^f_{{1}} + 4 \mathcal{O}^f_{{2}}$}\\ 
* & $(0,1,1,0)$ & $-4 \mathcal{O}^f_{{2}}$\\ 
\hline
& \multicolumn{1}{c|}{ $\{{B_L}_1,{G_L}_3\} \quad \{{B_L}_2,{G_L}_4\}$ } & $\mathcal{O}^{j}$\\
 \hline* & {\color{gray} $(2,8,1,0)$} & {\color{gray} $-8 \mathcal{O}^f_{{1}} + 14 \mathcal{O}^f_{{2}}$}\\ 
* & $(1,8,1,0)$ & $-8 \mathcal{O}^f_{{1}} -2 \mathcal{O}^f_{{2}}$\\ 
* & $(0,8,1,0)$ & $8 \mathcal{O}^f_{{1}} -2 \mathcal{O}^f_{{2}}$\\ 
\hline
\end{longtable}

\begin{longtable}{cc|c}
\caption{UV resonance of type ${G_L}^2{W_L}^2$. ``*'' in front of each row indicates topologies with nonrenormalizable vertices. The UV resonances colored in gray are eliminated by UV selection. }\\
\hline\hline\multicolumn{3}{l}{\multirow{2}{*}{\large Type: ${G_L}^2{W_L}^2$}}\\
 \multicolumn{3}{l}{} \\
\hline\multicolumn{3}{c}{$\begin{array}{ccl}\mathcal{O}^f_{1}&=&\frac{1}{4}\mathcal{Y}[\yng(2)_{{G_L}}]\mathcal{Y}[\yng(2)_{{W_L}}]{{G_L}^{A\lambda\rho}}{{G_L}^{A\mu\nu}}{{W_L}^{I\lambda\mu}}{{W_L}^{I\nu\rho}},\\
\mathcal{O}^f_{2}&=&\frac{1}{4}\mathcal{Y}[\yng(2)_{{G_L}}]\mathcal{Y}[\yng(2)_{{W_L}}]{{G_L}^{A\mu\nu}}{{G_L}^{A\mu\nu}}{{W_L}^{I\lambda\rho}}{{W_L}^{I\lambda\rho}}.\end{array}$}\\
\hline
\multicolumn{2}{c|}{group: $(\text{Spin},SU(3)_c,SU(2)_w,U(1)_y)$} & \multirow{2}{*}{$\mathcal{O}^{j}$}\\
\begin{fmffile}{4}
\begin{fmfgraph*}(40,20)
\fmfpen{thin} \fmfleft{e1,e2} \fmfv{}{g1} \fmfv{}{g2} \fmfright{e4,e3}
\fmf{plain}{e1,g1} \fmf{plain}{g1,e2} \fmf{plain}{g1,g2} \fmf{plain}{e4,g2} \fmf{plain}{g2,e3}
\end{fmfgraph*}
\end{fmffile} &\multicolumn{1}{c|}{ $\{{G_L}_1,{G_L}_2\} \quad \{{W_L}_3,{W_L}_4\}$ } & \\ 
 \hline* & {\color{gray} $(2,1,1,0)$} & {\color{gray} $48 \mathcal{O}^f_{{1}} -4 \mathcal{O}^f_{{2}}$}\\ 
* & $(0,1,1,0)$ & $4 \mathcal{O}^f_{{2}}$\\ 
\hline
& \multicolumn{1}{c|}{ $\{{G_L}_1,{W_L}_3\} \quad \{{G_L}_2,{W_L}_4\}$ } & $\mathcal{O}^{j}$\\
 \hline* & {\color{gray} $(2,8,3,0)$} & {\color{gray} $8 \mathcal{O}^f_{{1}} -14 \mathcal{O}^f_{{2}}$}\\ 
* & $(1,8,3,0)$ & $8 \mathcal{O}^f_{{1}} + 2 \mathcal{O}^f_{{2}}$\\ 
* & $(0,8,3,0)$ & $-8 \mathcal{O}^f_{{1}} + 2 \mathcal{O}^f_{{2}}$\\ 
\hline
\end{longtable}

\begin{longtable}{cc|c}
\caption{UV resonance of type ${B_L}{G_L}^3$. ``*'' in front of each row indicates topologies with nonrenormalizable vertices. The UV resonances colored in gray are eliminated by UV selection. }\\
\hline\hline\multicolumn{3}{l}{\multirow{2}{*}{\large Type: ${B_L}{G_L}^3$}}\\
 \multicolumn{3}{l}{} \\
\hline\multicolumn{3}{c}{$\begin{array}{ccl}\mathcal{O}^f_{1}&=&\frac{1}{6}\mathcal{Y}[\yng(3)_{{G_L}}]{{B_L}^{\mu\nu}}d^{ABC}{{G_L}^{A\lambda\rho}}{{G_L}^{B\nu\rho}}{{G_L}^{C\lambda\mu}}.\end{array}$}\\
\hline
\multicolumn{2}{c|}{group: $(\text{Spin},SU(3)_c,SU(2)_w,U(1)_y)$} & \multirow{2}{*}{$\mathcal{O}^{j}$}\\
\begin{fmffile}{4}
\begin{fmfgraph*}(40,20)
\fmfpen{thin} \fmfleft{e1,e2} \fmfv{}{g1} \fmfv{}{g2} \fmfright{e4,e3}
\fmf{plain}{e1,g1} \fmf{plain}{g1,e2} \fmf{plain}{g1,g2} \fmf{plain}{e4,g2} \fmf{plain}{g2,e3}
\end{fmfgraph*}
\end{fmffile} &\multicolumn{1}{c|}{ $\{{B_L}_1,{G_L}_2\} \quad \{{G_L}_3,{G_L}_4\}$ } & \\ 
 \hline* & {\color{gray} $(2,8,1,0)$} & {\color{gray} $-64 \mathcal{O}^f_{{1}}$}\\ 
* & $(0,8,1,0)$ & $16 \mathcal{O}^f_{{1}}$\\ 
\hline
\end{longtable}

\begin{longtable}{cc|c}
\caption{UV resonance of type ${G_L}^4$. ``*'' in front of each row indicates topologies with nonrenormalizable vertices. The UV resonances colored in gray are eliminated by UV selection. }\\
\hline\hline\multicolumn{3}{l}{\multirow{2}{*}{\large Type: ${G_L}^4$}}\\
 \multicolumn{3}{l}{} \\
\hline\multicolumn{3}{c}{$\begin{array}{ccl}\mathcal{O}^f_{1}&=&\frac{1}{24}\mathcal{Y}[\yng(4)_{{G_L}}]d^{ABE}d^{CDE}{{G_L}^{A\mu\nu}}{{G_L}^{B\lambda\rho}}{{G_L}^{C\nu\rho}}{{G_L}^{D\lambda\mu}},\\
\mathcal{O}^f_{2}&=&\frac{1}{24}\mathcal{Y}[\yng(4)_{{G_L}}]d^{ABE}d^{CDE}{{G_L}^{A\mu\nu}}{{G_L}^{B\mu\nu}}{{G_L}^{C\lambda\rho}}{{G_L}^{D\lambda\rho}},\\
\mathcal{O}^f_{3}&=&\frac{1}{24}\mathcal{Y}[\yng(4)_{{G_L}}]f^{ABE}f^{CDE}{{G_L}^{A\mu\nu}}{{G_L}^{B\lambda\rho}}{{G_L}^{C\mu\nu}}{{G_L}^{D\lambda\rho}}.\end{array}$}\\
\hline
\multicolumn{2}{c|}{group: $(\text{Spin},SU(3)_c,SU(2)_w,U(1)_y)$} & \multirow{2}{*}{$\mathcal{O}^{j}$}\\
\begin{fmffile}{4}
\begin{fmfgraph*}(40,20)
\fmfpen{thin} \fmfleft{e1,e2} \fmfv{}{g1} \fmfv{}{g2} \fmfright{e4,e3}
\fmf{plain}{e1,g1} \fmf{plain}{g1,e2} \fmf{plain}{g1,g2} \fmf{plain}{e4,g2} \fmf{plain}{g2,e3}
\end{fmfgraph*}
\end{fmffile} &\multicolumn{1}{c|}{ $\{{G_L}_1,{G_L}_2\} \quad \{{G_L}_3,{G_L}_4\}$ } & \\ 
 \hline* & {\color{gray} $(2,27,1,0)$} & {\color{gray} $522 \mathcal{O}^f_{{1}} -\frac{387}{2} \mathcal{O}^f_{{2}} -125 \mathcal{O}^f_{{3}}$}\\ 
* & $(0,27,1,0)$ & $-90 \mathcal{O}^f_{{1}} + \frac{117}{2} \mathcal{O}^f_{{2}} -25 \mathcal{O}^f_{{3}}$\\
\hline
\end{longtable}

\begin{longtable}{cc|c}
\caption{UV resonance of type ${B_L}DL{L^\dagger}{W_L}$. ``*'' in front of each row indicates topologies with nonrenormalizable vertices. The UV resonances colored in gray are eliminated by UV selection. }\\
\hline\hline\multicolumn{3}{l}{\multirow{2}{*}{\large Type: ${B_L}DL{L^\dagger}{W_L}$}}\\
 \multicolumn{3}{l}{} \\
\hline\multicolumn{3}{c}{$\begin{array}{ccl}\mathcal{O}^f_{{1}}&=& i(({D_\nu}{{L_p}_i}){\sigma_\mu}{{{L^\dagger}_r}^j}){{\tau^I}^i_j}{{B_L}^{\lambda\nu}}{{W_L}^{I\lambda\mu}}.\end{array}$}\\
\hline
\multicolumn{2}{c|}{group: $(\text{Spin},SU(3)_c,SU(2)_w,U(1)_y)$} & \multirow{2}{*}{$\mathcal{O}^{j}$}\\
\begin{fmffile}{4}
\begin{fmfgraph*}(40,20)
\fmfpen{thin} \fmfleft{e1,e2} \fmfv{}{g1} \fmfv{}{g2} \fmfright{e4,e3}
\fmf{plain}{e1,g1} \fmf{plain}{g1,e2} \fmf{plain}{g1,g2} \fmf{plain}{e4,g2} \fmf{plain}{g2,e3}
\end{fmfgraph*}
\end{fmffile} &\multicolumn{1}{c|}{ $\{{B_L}_1,{W_L}_2\} \quad \{L_3,{L^\dagger}_4\}$ } & \\ 
 \hline* & $(1,1,3,0)$ & $-4 \mathcal{O}^f_{1}$\\ 
\hline
& \multicolumn{1}{c|}{ $\{{B_L}_1,L_3\} \quad \{{W_L}_2,{L^\dagger}_4\}$ } & $\mathcal{O}^{j}$\\
 \hline* & $(\frac{3}{2},1,2,-\frac{1}{2})$ & $-4 \mathcal{O}^f_{1}$\\ 
\hline
& \multicolumn{1}{c|}{ $\{{B_L}_1,{L^\dagger}_4\} \quad \{{W_L}_2,L_3\}$ } & $\mathcal{O}^{j}$\\
 \hline* & $(\frac{3}{2},1,2,\frac{1}{2})$ & $-4 \mathcal{O}^f_{1}$\\ 
\hline
\end{longtable}

\begin{longtable}{cc|c}
\caption{UV resonance of type ${B_L}DQ{Q^\dagger}{W_L}$. ``*'' in front of each row indicates topologies with nonrenormalizable vertices. The UV resonances colored in gray are eliminated by UV selection. }\\
\hline\hline\multicolumn{3}{l}{\multirow{2}{*}{\large Type: ${B_L}DQ{Q^\dagger}{W_L}$}}\\
 \multicolumn{3}{l}{} \\
\hline\multicolumn{3}{c}{$\begin{array}{ccl}\mathcal{O}^f_{{1}}&=& i(({D_\nu}{Q_p}^{ai}){\sigma_\mu}{{Q^\dagger}_r}^{aj}){{\tau^I}^i_j}{{B_L}^{\lambda\nu}}{{W_L}^{I\lambda\mu}}.\end{array}$}\\
\hline
\multicolumn{2}{c|}{group: $(\text{Spin},SU(3)_c,SU(2)_w,U(1)_y)$} & \multirow{2}{*}{$\mathcal{O}^{j}$}\\
\begin{fmffile}{4}
\begin{fmfgraph*}(40,20)
\fmfpen{thin} \fmfleft{e1,e2} \fmfv{}{g1} \fmfv{}{g2} \fmfright{e4,e3}
\fmf{plain}{e1,g1} \fmf{plain}{g1,e2} \fmf{plain}{g1,g2} \fmf{plain}{e4,g2} \fmf{plain}{g2,e3}
\end{fmfgraph*}
\end{fmffile} &\multicolumn{1}{c|}{ $\{{B_L}_1,{W_L}_2\} \quad \{Q_3,{Q^\dagger}_4\}$ } & \\ 
 \hline* & $(1,1,3,0)$ & $-4 \mathcal{O}^f_{1}$\\ 
\hline
& \multicolumn{1}{c|}{ $\{{B_L}_1,Q_3\} \quad \{{W_L}_2,{Q^\dagger}_4\}$ } & $\mathcal{O}^{j}$\\
 \hline* & $(\frac{3}{2},3,2,\frac{1}{6})$ & $-4 \mathcal{O}^f_{1}$\\ 
\hline
& \multicolumn{1}{c|}{ $\{{B_L}_1,{Q^\dagger}_4\} \quad \{{W_L}_2,Q_3\}$ } & $\mathcal{O}^{j}$\\
 \hline* & $(\frac{3}{2},\overline{3},2,-\frac{1}{6})$ & $-4 \mathcal{O}^f_{1}$\\ 
\hline
\end{longtable}

\begin{longtable}{cc|c}
\caption{UV resonance of type $DL{L^\dagger}{W_L}^2$. ``*'' in front of each row indicates topologies with nonrenormalizable vertices. The UV resonances colored in gray are eliminated by UV selection. }\\
\hline\hline\multicolumn{3}{l}{\multirow{2}{*}{\large Type: $DL{L^\dagger}{W_L}^2$}}\\
 \multicolumn{3}{l}{} \\
\hline\multicolumn{3}{c}{$\begin{array}{ccl}\mathcal{O}^f_{1}&=&\frac{1}{2}\mathcal{Y}[\yng(2)_{{W_L}}]\epsilon^{IJK}i{{\tau^K}^i_j}{{W_L}^{I\lambda\nu}}{{W_L}^{J\lambda\mu}}(({D_\nu}{{L_p}_i}){\sigma_\mu}{{{L^\dagger}_r}^j}).\end{array}$}\\
\hline
\multicolumn{2}{c|}{group: $(\text{Spin},SU(3)_c,SU(2)_w,U(1)_y)$} & \multirow{2}{*}{$\mathcal{O}^{j}$}\\
\begin{fmffile}{4}
\begin{fmfgraph*}(40,20)
\fmfpen{thin} \fmfleft{e1,e2} \fmfv{}{g1} \fmfv{}{g2} \fmfright{e4,e3}
\fmf{plain}{e1,g1} \fmf{plain}{g1,e2} \fmf{plain}{g1,g2} \fmf{plain}{e4,g2} \fmf{plain}{g2,e3}
\end{fmfgraph*}
\end{fmffile} &\multicolumn{1}{c|}{ $\{{W_L}_1,{W_L}_2\} \quad \{L_3,{L^\dagger}_4\}$ } & \\ 
 \hline* & $(1,1,3,0)$ & $-4 \mathcal{O}^f_{{1}}$\\ 
\hline
& \multicolumn{1}{c|}{ $\{{W_L}_1,L_3\} \quad \{{W_L}_2,{L^\dagger}_4\}$ } & $\mathcal{O}^{j}$\\
 \hline* & $(\frac{3}{2},1,4,-\frac{1}{2})$ & $-4 \mathcal{O}^f_{{1}}$\\ 
* & $(\frac{3}{2},1,2,-\frac{1}{2})$ & $-4 \mathcal{O}^f_{{1}}$\\ 
\hline
\end{longtable}

\begin{longtable}{cc|c}
\caption{UV resonance of type $DQ{Q^\dagger}{W_L}^2$. ``*'' in front of each row indicates topologies with nonrenormalizable vertices. The UV resonances colored in gray are eliminated by UV selection. }\\
\hline\hline\multicolumn{3}{l}{\multirow{2}{*}{\large Type: $DQ{Q^\dagger}{W_L}^2$}}\\
 \multicolumn{3}{l}{} \\
\hline\multicolumn{3}{c}{$\begin{array}{ccl}\mathcal{O}^f_{1}&=&\frac{1}{2}\mathcal{Y}[\yng(2)_{{W_L}}]\epsilon^{IJK}i{{\tau^K}^i_j}{{W_L}^{I\lambda\nu}}{{W_L}^{J\lambda\mu}}(({D_\nu}{Q_p}^{ai}){\sigma_\mu}{{Q^\dagger}_r}^{aj}).\end{array}$}\\
\hline
\multicolumn{2}{c|}{group: $(\text{Spin},SU(3)_c,SU(2)_w,U(1)_y)$} & \multirow{2}{*}{$\mathcal{O}^{j}$}\\
\begin{fmffile}{4}
\begin{fmfgraph*}(40,20)
\fmfpen{thin} \fmfleft{e1,e2} \fmfv{}{g1} \fmfv{}{g2} \fmfright{e4,e3}
\fmf{plain}{e1,g1} \fmf{plain}{g1,e2} \fmf{plain}{g1,g2} \fmf{plain}{e4,g2} \fmf{plain}{g2,e3}
\end{fmfgraph*}
\end{fmffile} &\multicolumn{1}{c|}{ $\{{W_L}_1,{W_L}_2\} \quad \{Q_3,{Q^\dagger}_4\}$ } & \\ 
 \hline* & $(1,1,3,0)$ & $-4 \mathcal{O}^f_{{1}}$\\ 
\hline
& \multicolumn{1}{c|}{ $\{{W_L}_1,Q_3\} \quad \{{W_L}_2,{Q^\dagger}_4\}$ } & $\mathcal{O}^{j}$\\
 \hline* & $(\frac{3}{2},3,4,\frac{1}{6})$ & $-4 \mathcal{O}^f_{{1}}$\\ 
* & $(\frac{3}{2},3,2,\frac{1}{6})$ & $-4 \mathcal{O}^f_{{1}}$\\ 
\hline
\end{longtable}

\begin{longtable}{cc|c}
\caption{UV resonance of type ${B_L}D{d_{\mathbb{C}}}{{d_{\mathbb{C}}}^\dagger}{G_L}$. ``*'' in front of each row indicates topologies with nonrenormalizable vertices. The UV resonances colored in gray are eliminated by UV selection. }\\
\hline\hline\multicolumn{3}{l}{\multirow{2}{*}{\large Type: ${B_L}D{d_{\mathbb{C}}}{{d_{\mathbb{C}}}^\dagger}{G_L}$}}\\
 \multicolumn{3}{l}{} \\
\hline\multicolumn{3}{c}{$\begin{array}{ccl}\mathcal{O}^f_{{1}}&=& i(({D_\nu}{{{d_{\mathbb{C}}}_p}^a}){\sigma_\mu}{{{{d_{\mathbb{C}}}^\dagger}_r}_b}){{\lambda^A}_a^b}{{B_L}^{\lambda\nu}}{{G_L}^{A\lambda\mu}}.\end{array}$}\\
\hline
\multicolumn{2}{c|}{group: $(\text{Spin},SU(3)_c,SU(2)_w,U(1)_y)$} & \multirow{2}{*}{$\mathcal{O}^{j}$}\\
\begin{fmffile}{4}
\begin{fmfgraph*}(40,20)
\fmfpen{thin} \fmfleft{e1,e2} \fmfv{}{g1} \fmfv{}{g2} \fmfright{e4,e3}
\fmf{plain}{e1,g1} \fmf{plain}{g1,e2} \fmf{plain}{g1,g2} \fmf{plain}{e4,g2} \fmf{plain}{g2,e3}
\end{fmfgraph*}
\end{fmffile} &\multicolumn{1}{c|}{ $\{{B_L}_1,{G_L}_2\} \quad \{{d_{\mathbb{C}}}_3,{{d_{\mathbb{C}}}^\dagger}_4\}$ } & \\ 
 \hline* & $(1,8,1,0)$ & $4 \mathcal{O}^f_{1}$\\ 
\hline
& \multicolumn{1}{c|}{ $\{{B_L}_1,{d_{\mathbb{C}}}_3\} \quad \{{G_L}_2,{{d_{\mathbb{C}}}^\dagger}_4\}$ } & $\mathcal{O}^{j}$\\
 \hline* & $(\frac{3}{2},\overline{3},1,\frac{1}{3})$ & $4 \mathcal{O}^f_{1}$\\ 
\hline
& \multicolumn{1}{c|}{ $\{{B_L}_1,{{d_{\mathbb{C}}}^\dagger}_4\} \quad \{{G_L}_2,{d_{\mathbb{C}}}_3\}$ } & $\mathcal{O}^{j}$\\
 \hline* & $(\frac{3}{2},3,1,-\frac{1}{3})$ & $4 \mathcal{O}^f_{1}$\\ 
\hline
\end{longtable}

\begin{longtable}{cc|c}
\caption{UV resonance of type ${B_L}D{G_L}{u_{\mathbb{C}}}{{u_{\mathbb{C}}}^\dagger}$. ``*'' in front of each row indicates topologies with nonrenormalizable vertices. The UV resonances colored in gray are eliminated by UV selection. }\\
\hline\hline\multicolumn{3}{l}{\multirow{2}{*}{\large Type: ${B_L}D{G_L}{u_{\mathbb{C}}}{{u_{\mathbb{C}}}^\dagger}$}}\\
 \multicolumn{3}{l}{} \\
\hline\multicolumn{3}{c}{$\begin{array}{ccl}\mathcal{O}^f_{{1}}&=& i(({D_\nu}{{{u_{\mathbb{C}}}_p}^a}){\sigma_\mu}{{{{u_{\mathbb{C}}}^\dagger}_r}_b}){{\lambda^A}_a^b}{{B_L}^{\lambda\nu}}{{G_L}^{A\lambda\mu}}.\end{array}$}\\
\hline
\multicolumn{2}{c|}{group: $(\text{Spin},SU(3)_c,SU(2)_w,U(1)_y)$} & \multirow{2}{*}{$\mathcal{O}^{j}$}\\
\begin{fmffile}{4}
\begin{fmfgraph*}(40,20)
\fmfpen{thin} \fmfleft{e1,e2} \fmfv{}{g1} \fmfv{}{g2} \fmfright{e4,e3}
\fmf{plain}{e1,g1} \fmf{plain}{g1,e2} \fmf{plain}{g1,g2} \fmf{plain}{e4,g2} \fmf{plain}{g2,e3}
\end{fmfgraph*}
\end{fmffile} &\multicolumn{1}{c|}{ $\{{B_L}_1,{G_L}_2\} \quad \{{u_{\mathbb{C}}}_3,{{u_{\mathbb{C}}}^\dagger}_4\}$ } & \\ 
 \hline* & $(1,8,1,0)$ & $4 \mathcal{O}^f_{1}$\\ 
\hline
& \multicolumn{1}{c|}{ $\{{B_L}_1,{u_{\mathbb{C}}}_3\} \quad \{{G_L}_2,{{u_{\mathbb{C}}}^\dagger}_4\}$ } & $\mathcal{O}^{j}$\\
 \hline* & $(\frac{3}{2},\overline{3},1,-\frac{2}{3})$ & $4 \mathcal{O}^f_{1}$\\ 
\hline
& \multicolumn{1}{c|}{ $\{{B_L}_1,{{u_{\mathbb{C}}}^\dagger}_4\} \quad \{{G_L}_2,{u_{\mathbb{C}}}_3\}$ } & $\mathcal{O}^{j}$\\
 \hline* & $(\frac{3}{2},3,1,\frac{2}{3})$ & $4 \mathcal{O}^f_{1}$\\ 
\hline
\end{longtable}

\begin{longtable}{cc|c}
\caption{UV resonance of type ${B_L}D{G_L}Q{Q^\dagger}$. ``*'' in front of each row indicates topologies with nonrenormalizable vertices. The UV resonances colored in gray are eliminated by UV selection. }\\
\hline\hline\multicolumn{3}{l}{\multirow{2}{*}{\large Type: ${B_L}D{G_L}Q{Q^\dagger}$}}\\
 \multicolumn{3}{l}{} \\
\hline\multicolumn{3}{c}{$\begin{array}{ccl}\mathcal{O}^f_{{1}}&=& i(({D_\nu}{Q_p}^{ai}){\sigma_\mu}{{Q^\dagger}_r}^{bi}){{\lambda^A}_b^a}{{B_L}^{\lambda\nu}}{{G_L}^{A\lambda\mu}}.\end{array}$}\\
\hline
\multicolumn{2}{c|}{group: $(\text{Spin},SU(3)_c,SU(2)_w,U(1)_y)$} & \multirow{2}{*}{$\mathcal{O}^{j}$}\\
\begin{fmffile}{4}
\begin{fmfgraph*}(40,20)
\fmfpen{thin} \fmfleft{e1,e2} \fmfv{}{g1} \fmfv{}{g2} \fmfright{e4,e3}
\fmf{plain}{e1,g1} \fmf{plain}{g1,e2} \fmf{plain}{g1,g2} \fmf{plain}{e4,g2} \fmf{plain}{g2,e3}
\end{fmfgraph*}
\end{fmffile} &\multicolumn{1}{c|}{ $\{{B_L}_1,{G_L}_2\} \quad \{Q_3,{Q^\dagger}_4\}$ } & \\ 
 \hline* & $(1,8,1,0)$ & $-4 \mathcal{O}^f_{1}$\\ 
\hline
& \multicolumn{1}{c|}{ $\{{B_L}_1,Q_3\} \quad \{{G_L}_2,{Q^\dagger}_4\}$ } & $\mathcal{O}^{j}$\\
 \hline* & $(\frac{3}{2},3,2,\frac{1}{6})$ & $-4 \mathcal{O}^f_{1}$\\ 
\hline
& \multicolumn{1}{c|}{ $\{{B_L}_1,{Q^\dagger}_4\} \quad \{{G_L}_2,Q_3\}$ } & $\mathcal{O}^{j}$\\
 \hline* & $(\frac{3}{2},\overline{3},2,-\frac{1}{6})$ & $-4 \mathcal{O}^f_{1}$\\ 
\hline
\end{longtable}

\begin{longtable}{cc|c}
\caption{UV resonance of type $D{G_L}Q{Q^\dagger}{W_L}$. ``*'' in front of each row indicates topologies with nonrenormalizable vertices. The UV resonances colored in gray are eliminated by UV selection. }\\
\hline\hline\multicolumn{3}{l}{\multirow{2}{*}{\large Type: $D{G_L}Q{Q^\dagger}{W_L}$}}\\
 \multicolumn{3}{l}{} \\
\hline\multicolumn{3}{c}{$\begin{array}{ccl}\mathcal{O}^f_{{1}}&=& i(({D_\nu}{Q_p}^{ai}){\sigma_\mu}{{Q^\dagger}_r}^{bj}){{\lambda^A}_b^a}{{\tau^I}^i_j}{{G_L}^{A\lambda\nu}}{{W_L}^{I\lambda\mu}}.\end{array}$}\\
\hline
\multicolumn{2}{c|}{group: $(\text{Spin},SU(3)_c,SU(2)_w,U(1)_y)$} & \multirow{2}{*}{$\mathcal{O}^{j}$}\\
\begin{fmffile}{4}
\begin{fmfgraph*}(40,20)
\fmfpen{thin} \fmfleft{e1,e2} \fmfv{}{g1} \fmfv{}{g2} \fmfright{e4,e3}
\fmf{plain}{e1,g1} \fmf{plain}{g1,e2} \fmf{plain}{g1,g2} \fmf{plain}{e4,g2} \fmf{plain}{g2,e3}
\end{fmfgraph*}
\end{fmffile} &\multicolumn{1}{c|}{ $\{{G_L}_1,{W_L}_2\} \quad \{Q_3,{Q^\dagger}_4\}$ } & \\ 
 \hline* & $(1,8,3,0)$ & $-4 \mathcal{O}^f_{1}$\\ 
\hline
& \multicolumn{1}{c|}{ $\{{G_L}_1,Q_3\} \quad \{{W_L}_2,{Q^\dagger}_4\}$ } & $\mathcal{O}^{j}$\\
 \hline* & $(\frac{3}{2},3,2,\frac{1}{6})$ & $-4 \mathcal{O}^f_{1}$\\ 
\hline
& \multicolumn{1}{c|}{ $\{{G_L}_1,{Q^\dagger}_4\} \quad \{{W_L}_2,Q_3\}$ } & $\mathcal{O}^{j}$\\
 \hline* & $(\frac{3}{2},\overline{3},2,-\frac{1}{6})$ & $-4 \mathcal{O}^f_{1}$\\ 
\hline
\end{longtable}

\begin{longtable}{cc|c}
\caption{UV resonance of type $D{d_{\mathbb{C}}}{{d_{\mathbb{C}}}^\dagger}{G_L}^2$. ``*'' in front of each row indicates topologies with nonrenormalizable vertices. The UV resonances colored in gray are eliminated by UV selection. }\\
\hline\hline\multicolumn{3}{l}{\multirow{2}{*}{\large Type: $D{d_{\mathbb{C}}}{{d_{\mathbb{C}}}^\dagger}{G_L}^2$}}\\
 \multicolumn{3}{l}{} \\
\hline\multicolumn{3}{c}{$\begin{array}{ccl}\mathcal{O}^f_{1}&=&\frac{1}{2}\mathcal{Y}[\yng(2)_{{G_L}}]f^{ABC}{{G_L}^{A\lambda\nu}}{{G_L}^{B\lambda\mu}}i{{\lambda^C}_a^b}(({D_\nu}{{{d_{\mathbb{C}}}_p}^a}){\sigma_\mu}{{{{d_{\mathbb{C}}}^\dagger}_r}_b}).\end{array}$}\\
\hline
\multicolumn{2}{c|}{group: $(\text{Spin},SU(3)_c,SU(2)_w,U(1)_y)$} & \multirow{2}{*}{$\mathcal{O}^{j}$}\\
\begin{fmffile}{4}
\begin{fmfgraph*}(40,20)
\fmfpen{thin} \fmfleft{e1,e2} \fmfv{}{g1} \fmfv{}{g2} \fmfright{e4,e3}
\fmf{plain}{e1,g1} \fmf{plain}{g1,e2} \fmf{plain}{g1,g2} \fmf{plain}{e4,g2} \fmf{plain}{g2,e3}
\end{fmfgraph*}
\end{fmffile} &\multicolumn{1}{c|}{ $\{{G_L}_1,{G_L}_2\} \quad \{{d_{\mathbb{C}}}_3,{{d_{\mathbb{C}}}^\dagger}_4\}$ } & \\ 
 \hline* & $(1,8,1,0)$ & $4 \mathcal{O}^f_{{1}}$\\ 
\hline
& \multicolumn{1}{c|}{ $\{{G_L}_1,{d_{\mathbb{C}}}_3\} \quad \{{G_L}_2,{{d_{\mathbb{C}}}^\dagger}_4\}$ } & $\mathcal{O}^{j}$\\
 \hline* & $(\frac{3}{2},\overline{15},1,\frac{1}{3})$ & $4 \mathcal{O}^f_{{1}}$\\ 
* & $(\frac{3}{2},\overline{6},1,\frac{1}{3})$ & $4 \mathcal{O}^f_{{1}}$\\ 
* & $(\frac{3}{2},\overline{3},1,\frac{1}{3})$ & $4 \mathcal{O}^f_{{1}}$\\ 
\hline
\end{longtable}

\begin{longtable}{cc|c}
\caption{UV resonance of type $D{G_L}^2{u_{\mathbb{C}}}{{u_{\mathbb{C}}}^\dagger}$. ``*'' in front of each row indicates topologies with nonrenormalizable vertices. The UV resonances colored in gray are eliminated by UV selection. }\\
\hline\hline\multicolumn{3}{l}{\multirow{2}{*}{\large Type: $D{G_L}^2{u_{\mathbb{C}}}{{u_{\mathbb{C}}}^\dagger}$}}\\
 \multicolumn{3}{l}{} \\
\hline\multicolumn{3}{c}{$\begin{array}{ccl}\mathcal{O}^f_{1}&=&\frac{1}{2}\mathcal{Y}[\yng(2)_{{G_L}}]f^{ABC}{{G_L}^{A\lambda\nu}}{{G_L}^{B\lambda\mu}}i{{\lambda^C}_a^b}(({D_\nu}{{{u_{\mathbb{C}}}_p}^a}){\sigma_\mu}{{{{u_{\mathbb{C}}}^\dagger}_r}_b}).\end{array}$}\\
\hline
\multicolumn{2}{c|}{group: $(\text{Spin},SU(3)_c,SU(2)_w,U(1)_y)$} & \multirow{2}{*}{$\mathcal{O}^{j}$}\\
\begin{fmffile}{4}
\begin{fmfgraph*}(40,20)
\fmfpen{thin} \fmfleft{e1,e2} \fmfv{}{g1} \fmfv{}{g2} \fmfright{e4,e3}
\fmf{plain}{e1,g1} \fmf{plain}{g1,e2} \fmf{plain}{g1,g2} \fmf{plain}{e4,g2} \fmf{plain}{g2,e3}
\end{fmfgraph*}
\end{fmffile} &\multicolumn{1}{c|}{ $\{{G_L}_1,{G_L}_2\} \quad \{{u_{\mathbb{C}}}_3,{{u_{\mathbb{C}}}^\dagger}_4\}$ } & \\ 
 \hline* & $(1,8,1,0)$ & $4 \mathcal{O}^f_{{1}}$\\ 
\hline
& \multicolumn{1}{c|}{ $\{{G_L}_1,{u_{\mathbb{C}}}_3\} \quad \{{G_L}_2,{{u_{\mathbb{C}}}^\dagger}_4\}$ } & $\mathcal{O}^{j}$\\
 \hline* & $(\frac{3}{2},\overline{15},1,-\frac{2}{3})$ & $4 \mathcal{O}^f_{{1}}$\\ 
* & $(\frac{3}{2},\overline{6},1,-\frac{2}{3})$ & $4 \mathcal{O}^f_{{1}}$\\ 
* & $(\frac{3}{2},\overline{3},1,-\frac{2}{3})$ & $4 \mathcal{O}^f_{{1}}$\\ 
\hline
\end{longtable}

\begin{longtable}{cc|c}
\caption{UV resonance of type $D{G_L}^2Q{Q^\dagger}$. ``*'' in front of each row indicates topologies with nonrenormalizable vertices. The UV resonances colored in gray are eliminated by UV selection. }\\
\hline\hline\multicolumn{3}{l}{\multirow{2}{*}{\large Type: $D{G_L}^2Q{Q^\dagger}$}}\\
 \multicolumn{3}{l}{} \\
\hline\multicolumn{3}{c}{$\begin{array}{ccl}\mathcal{O}^f_{1}&=&\frac{1}{2}\mathcal{Y}[\yng(2)_{{G_L}}]f^{ABC}{{G_L}^{A\lambda\nu}}{{G_L}^{B\lambda\mu}}i{{\lambda^C}_b^a}(({D_\nu}{Q_p}^{ai}){\sigma_\mu}{{Q^\dagger}_r}^{bi}).\end{array}$}\\
\hline
\multicolumn{2}{c|}{group: $(\text{Spin},SU(3)_c,SU(2)_w,U(1)_y)$} & \multirow{2}{*}{$\mathcal{O}^{j}$}\\
\begin{fmffile}{4}
\begin{fmfgraph*}(40,20)
\fmfpen{thin} \fmfleft{e1,e2} \fmfv{}{g1} \fmfv{}{g2} \fmfright{e4,e3}
\fmf{plain}{e1,g1} \fmf{plain}{g1,e2} \fmf{plain}{g1,g2} \fmf{plain}{e4,g2} \fmf{plain}{g2,e3}
\end{fmfgraph*}
\end{fmffile} &\multicolumn{1}{c|}{ $\{{G_L}_1,{G_L}_2\} \quad \{Q_3,{Q^\dagger}_4\}$ } & \\ 
 \hline* & $(1,8,1,0)$ & $-4 \mathcal{O}^f_{{1}}$\\ 
\hline
& \multicolumn{1}{c|}{ $\{{G_L}_1,Q_3\} \quad \{{G_L}_2,{Q^\dagger}_4\}$ } & $\mathcal{O}^{j}$\\
 \hline* & $(\frac{3}{2},15,2,\frac{1}{6})$ & $-4 \mathcal{O}^f_{{1}}$\\ 
* & $(\frac{3}{2},6,2,\frac{1}{6})$ & $-4 \mathcal{O}^f_{{1}}$\\ 
* & $(\frac{3}{2},3,2,\frac{1}{6})$ & $-4 \mathcal{O}^f_{{1}}$\\ 
\hline
\end{longtable}

\begin{longtable}{cc|c}
\caption{UV resonance of type $D^2{d_{\mathbb{C}}}{e_{\mathbb{C}}}{u_{\mathbb{C}}}^2$. ``*'' in front of each row indicates topologies with nonrenormalizable vertices. The UV resonances colored in gray are eliminated by UV selection. }\\
\hline\hline\multicolumn{3}{l}{\multirow{2}{*}{\large Type: $D^2{d_{\mathbb{C}}}{e_{\mathbb{C}}}{u_{\mathbb{C}}}^2$}}\\
 \multicolumn{3}{l}{} \\
\hline\multicolumn{3}{c}{$\begin{array}{ccl}\mathcal{O}^f_{1}&=&\frac{1}{2}\mathcal{Y}[\young(st)](\epsilon_{abc}(({D_\mu}{e_{\mathbb{C}}}_r)({D^\mu}{{{u_{\mathbb{C}}}_t}^c}))({{{d_{\mathbb{C}}}_p}^a}{{{u_{\mathbb{C}}}_s}^b})),\\
\mathcal{O}^f_{2}&=&\frac{1}{2}\mathcal{Y}[\young(s,t)](\epsilon_{abc}(({D_\mu}{{{u_{\mathbb{C}}}_s}^b})({D^\mu}{{{u_{\mathbb{C}}}_t}^c}))({{{d_{\mathbb{C}}}_p}^a}{e_{\mathbb{C}}}_r)),\\
\mathcal{O}^f_{3}&=&\frac{1}{2}\mathcal{Y}[\young(s,t)](\epsilon_{abc}(({D_\mu}{e_{\mathbb{C}}}_r)({D^\mu}{{{u_{\mathbb{C}}}_t}^c}))({{{d_{\mathbb{C}}}_p}^a}{{{u_{\mathbb{C}}}_s}^b})).\end{array}$}\\
\hline
\multicolumn{2}{c|}{group: $(\text{Spin},SU(3)_c,SU(2)_w,U(1)_y)$} & \multirow{2}{*}{$\mathcal{O}^{j}$}\\
\begin{fmffile}{4}
\begin{fmfgraph*}(40,20)
\fmfpen{thin} \fmfleft{e1,e2} \fmfv{}{g1} \fmfv{}{g2} \fmfright{e4,e3}
\fmf{plain}{e1,g1} \fmf{plain}{g1,e2} \fmf{plain}{g1,g2} \fmf{plain}{e4,g2} \fmf{plain}{g2,e3}
\end{fmfgraph*}
\end{fmffile} &\multicolumn{1}{c|}{ $\{{d_{\mathbb{C}}}_1,{e_{\mathbb{C}}}_2\} \quad \{{u_{\mathbb{C}}}_3,{u_{\mathbb{C}}}_4\}$ } & \\ 
 \hline* & $(2,\overline{3},1,\frac{4}{3})$ & $-4 \mathcal{O}^f_{{2}} + 12 \mathcal{O}^f_{{3}}$\\ 
* & $(1,\overline{3},1,\frac{4}{3})$ & $-4 \mathcal{O}^f_{{1}}$\\
\hline
\end{longtable}

\begin{longtable}{cc|c}
\caption{UV resonance of type $D^2{e_{\mathbb{C}}}LQ{u_{\mathbb{C}}}$. ``*'' in front of each row indicates topologies with nonrenormalizable vertices. The UV resonances colored in gray are eliminated by UV selection. }\\
\hline\hline\multicolumn{3}{l}{\multirow{2}{*}{\large Type: $D^2{e_{\mathbb{C}}}LQ{u_{\mathbb{C}}}$}}\\
 \multicolumn{3}{l}{} \\
\hline\multicolumn{3}{c}{$\begin{array}{ccl}\mathcal{O}^f_{{1}}&=& \epsilon^{ij}(({D_\mu}{Q_s}^{aj})({D^\mu}{{{u_{\mathbb{C}}}_t}^a}))({e_{\mathbb{C}}}_p{{L_r}_i}),\\
\mathcal{O}^f_{{2}}&=& \epsilon^{ij}(({D_\mu}{{L_r}_i})({D^\mu}{{{u_{\mathbb{C}}}_t}^a}))({e_{\mathbb{C}}}_p{Q_s}^{aj}),\\
\mathcal{O}^f_{{3}}&=& i\epsilon^{ij}(({D^\mu}{Q_s}^{aj})({D^\nu}{{{u_{\mathbb{C}}}_t}^a}))({e_{\mathbb{C}}}_p\sigma_{\mu\nu}{{L_r}_i}).\end{array}$}\\
\hline
\multicolumn{2}{c|}{group: $(\text{Spin},SU(3)_c,SU(2)_w,U(1)_y)$} & \multirow{2}{*}{$\mathcal{O}^{j}$}\\
\begin{fmffile}{4}
\begin{fmfgraph*}(40,20)
\fmfpen{thin} \fmfleft{e1,e2} \fmfv{}{g1} \fmfv{}{g2} \fmfright{e4,e3}
\fmf{plain}{e1,g1} \fmf{plain}{g1,e2} \fmf{plain}{g1,g2} \fmf{plain}{e4,g2} \fmf{plain}{g2,e3}
\end{fmfgraph*}
\end{fmffile} &\multicolumn{1}{c|}{ $\{{e_{\mathbb{C}}}_1,L_2\} \quad \{Q_3,{u_{\mathbb{C}}}_4\}$ } & \\ 
 \hline* & $(2,1,2,\frac{1}{2})$ & $4 \mathcal{O}^f_{1} -12 \mathcal{O}^f_{2} -6 \mathcal{O}^f_{3}$\\ 
* & $(1,1,2,\frac{1}{2})$ & $-2 \mathcal{O}^f_{3}$\\
\hline
\end{longtable}

\begin{longtable}{cc|c}
\caption{UV resonance of type $D^2{d_{\mathbb{C}}}Q^2{u_{\mathbb{C}}}$. ``*'' in front of each row indicates topologies with nonrenormalizable vertices. The UV resonances colored in gray are eliminated by UV selection. }\\
\hline\hline\multicolumn{3}{l}{\multirow{2}{*}{\large Type: $D^2{d_{\mathbb{C}}}Q^2{u_{\mathbb{C}}}$}}\\
 \multicolumn{3}{l}{} \\
\hline\multicolumn{3}{c}{$\begin{array}{ccl}\mathcal{O}^f_{1}&=&\frac{1}{2}\mathcal{Y}[\young(rs)](\epsilon^{ij}(({D_\mu}{Q_s}^{aj})({D^\mu}{{{u_{\mathbb{C}}}_t}^b}))({{{d_{\mathbb{C}}}_p}^a}{Q_r}^{bi})),\\
\mathcal{O}^f_{2}&=&\frac{1}{2}\mathcal{Y}[\young(rs)](\epsilon^{ij}(({D_\mu}{Q_r}^{bi})({D^\mu}{{{u_{\mathbb{C}}}_t}^b}))({{{d_{\mathbb{C}}}_p}^a}{Q_s}^{aj})),\\
\mathcal{O}^f_{3}&=&\frac{1}{2}\mathcal{Y}[\young(rs)]\epsilon^{ij}i(({D^\mu}{Q_s}^{aj})({D^\nu}{{{u_{\mathbb{C}}}_t}^b}))({{{d_{\mathbb{C}}}_p}^a}\sigma_{\mu\nu}{Q_r}^{bi}),\\
\mathcal{O}^f_{4}&=&\frac{1}{2}\mathcal{Y}[\young(r,s)](\epsilon^{ij}(({D_\mu}{Q_s}^{aj})({D^\mu}{{{u_{\mathbb{C}}}_t}^b}))({{{d_{\mathbb{C}}}_p}^a}{Q_r}^{bi})),\\
\mathcal{O}^f_{5}&=&\frac{1}{2}\mathcal{Y}[\young(r,s)](\epsilon^{ij}(({D_\mu}{Q_r}^{bi})({D^\mu}{{{u_{\mathbb{C}}}_t}^b}))({{{d_{\mathbb{C}}}_p}^a}{Q_s}^{aj})),\\
\mathcal{O}^f_{6}&=&\frac{1}{2}\mathcal{Y}[\young(r,s)]\epsilon^{ij}i(({D^\mu}{Q_s}^{aj})({D^\nu}{{{u_{\mathbb{C}}}_t}^b}))({{{d_{\mathbb{C}}}_p}^a}\sigma_{\mu\nu}{Q_r}^{bi}).\end{array}$}\\
\hline
\multicolumn{2}{c|}{group: $(\text{Spin},SU(3)_c,SU(2)_w,U(1)_y)$} & \multirow{2}{*}{$\mathcal{O}^{j}$}\\
\begin{fmffile}{4}
\begin{fmfgraph*}(40,20)
\fmfpen{thin} \fmfleft{e1,e2} \fmfv{}{g1} \fmfv{}{g2} \fmfright{e4,e3}
\fmf{plain}{e1,g1} \fmf{plain}{g1,e2} \fmf{plain}{g1,g2} \fmf{plain}{e4,g2} \fmf{plain}{g2,e3}
\end{fmfgraph*}
\end{fmffile} &\multicolumn{1}{c|}{ $\{{d_{\mathbb{C}}}_1,Q_2\} \quad \{Q_3,{u_{\mathbb{C}}}_4\}$ } & \\
\hline
\end{longtable}

\begin{longtable}{cc|c}
\caption{UV resonance of type $D^2LQ^3$. ``*'' in front of each row indicates topologies with nonrenormalizable vertices. The UV resonances colored in gray are eliminated by UV selection. }\\
\hline\hline\multicolumn{3}{l}{\multirow{2}{*}{\large Type: $D^2LQ^3$}}\\
 \multicolumn{3}{l}{} \\
\hline\multicolumn{3}{c}{$\begin{array}{ccl}\mathcal{O}^f_{1}&=&\frac{1}{6}\mathcal{Y}[\young(rst)]\epsilon^{abc}\epsilon^{ik}\epsilon^{jl}(({D_\mu}{Q_s}^{bk})({D^\mu}{Q_t}^{cl}))({{L_p}_i}{Q_r}^{aj}),\\
\mathcal{O}^f_{2}&=&\frac{1}{3}\mathcal{Y}[\young(rs,t)]\epsilon^{abc}\epsilon^{ik}\epsilon^{jl}(({D_\mu}{Q_s}^{bk})({D^\mu}{Q_t}^{cl}))({{L_p}_i}{Q_r}^{aj}),\\
\mathcal{O}^f_{3}&=&\frac{1}{3}\mathcal{Y}[\young(rs,t)]\epsilon^{abc}\epsilon^{ij}\epsilon^{kl}(({D_\mu}{Q_s}^{bk})({D^\mu}{Q_t}^{cl}))({{L_p}_i}{Q_r}^{aj}),\\
\mathcal{O}^f_{4}&=&\frac{1}{6}\mathcal{Y}[\young(r,s,t)]\epsilon^{abc}\epsilon^{ik}\epsilon^{jl}(({D_\mu}{Q_s}^{bk})({D^\mu}{Q_t}^{cl}))({{L_p}_i}{Q_r}^{aj}).\end{array}$}\\
\hline
\multicolumn{2}{c|}{group: $(\text{Spin},SU(3)_c,SU(2)_w,U(1)_y)$} & \multirow{2}{*}{$\mathcal{O}^{j}$}\\
\begin{fmffile}{4}
\begin{fmfgraph*}(40,20)
\fmfpen{thin} \fmfleft{e1,e2} \fmfv{}{g1} \fmfv{}{g2} \fmfright{e4,e3}
\fmf{plain}{e1,g1} \fmf{plain}{g1,e2} \fmf{plain}{g1,g2} \fmf{plain}{e4,g2} \fmf{plain}{g2,e3}
\end{fmfgraph*}
\end{fmffile} &\multicolumn{1}{c|}{ $\{L_1,Q_2\} \quad \{Q_3,Q_4\}$ } & \\ 
 \hline* & $(2,3,3,-\frac{1}{3})$ & $-20 \mathcal{O}^f_{{4}} + \mathcal{O}^f_{{2}} (-2 + 2 (s  t)) + \mathcal{O}^f_{{3}} (-8 + 8 (s  t))$\\ 
* & $(1,3,3,-\frac{1}{3})$ & $12 \mathcal{O}^f_{{1}} + \mathcal{O}^f_{{2}} (-6 -6 (s  t))$\\
\hline
\end{longtable}

\begin{longtable}{cc|c}
\caption{UV resonance of type ${B_L}D^2{e_{\mathbb{C}}}{H^\dagger}L$. ``*'' in front of each row indicates topologies with nonrenormalizable vertices. The UV resonances colored in gray are eliminated by UV selection. }\\
\hline\hline\multicolumn{3}{l}{\multirow{2}{*}{\large Type: ${B_L}D^2{e_{\mathbb{C}}}{H^\dagger}L$}}\\
 \multicolumn{3}{l}{} \\
\hline\multicolumn{3}{c}{$\begin{array}{ccl}\mathcal{O}^f_{{1}}&=& i({e_{\mathbb{C}}}_p\sigma_{\mu\nu}({D_\lambda}{{L_r}_i}))({D^\nu}{{H^\dagger}^i}){{B_L}^{\lambda\mu}},\\
\mathcal{O}^f_{{2}}&=& ({e_{\mathbb{C}}}_p({D_\mu}{{L_r}_i}))({D_\nu}{{H^\dagger}^i}){{B_L}^{\mu\nu}}.\end{array}$}\\
\hline
\multicolumn{2}{c|}{group: $(\text{Spin},SU(3)_c,SU(2)_w,U(1)_y)$} & \multirow{2}{*}{$\mathcal{O}^{j}$}\\
\begin{fmffile}{4}
\begin{fmfgraph*}(40,20)
\fmfpen{thin} \fmfleft{e1,e2} \fmfv{}{g1} \fmfv{}{g2} \fmfright{e4,e3}
\fmf{plain}{e1,g1} \fmf{plain}{g1,e2} \fmf{plain}{g1,g2} \fmf{plain}{e4,g2} \fmf{plain}{g2,e3}
\end{fmfgraph*}
\end{fmffile} &\multicolumn{1}{c|}{ $\{{B_L}_1,{e_{\mathbb{C}}}_2\} \quad \{L_3,{H^\dagger}_4\}$ } & \\ 
 \hline* & $(\frac{3}{2},1,1,1)$ & $2 \mathcal{O}^f_{1} + 10 \mathcal{O}^f_{2}$\\ 
* & $(\frac{1}{2},1,1,1)$ & $2 \mathcal{O}^f_{1} -2 \mathcal{O}^f_{2}$\\ 
\hline
& \multicolumn{1}{c|}{ $\{{B_L}_1,L_3\} \quad \{{e_{\mathbb{C}}}_2,{H^\dagger}_4\}$ } & $\mathcal{O}^{j}$\\
 \hline* & $(\frac{3}{2},1,2,-\frac{1}{2})$ & $-2 \mathcal{O}^f_{1} + 10 \mathcal{O}^f_{2}$\\ 
* & $(\frac{1}{2},1,2,-\frac{1}{2})$ & $2 \mathcal{O}^f_{1} + 2 \mathcal{O}^f_{2}$\\ 
\hline
& \multicolumn{1}{c|}{ $\{{B_L}_1,{H^\dagger}_4\} \quad \{{e_{\mathbb{C}}}_2,L_3\}$ } & $\mathcal{O}^{j}$\\
 \hline* & $(2,1,2,-\frac{1}{2})$ & $4 \mathcal{O}^f_{1}$\\ 
* & $(1,1,2,-\frac{1}{2})$ & $4 \mathcal{O}^f_{2}$\\ 
\hline
\end{longtable}

\begin{longtable}{cc|c}
\caption{UV resonance of type ${B_L}D^2{d_{\mathbb{C}}}{H^\dagger}Q$. ``*'' in front of each row indicates topologies with nonrenormalizable vertices. The UV resonances colored in gray are eliminated by UV selection. }\\
\hline\hline\multicolumn{3}{l}{\multirow{2}{*}{\large Type: ${B_L}D^2{d_{\mathbb{C}}}{H^\dagger}Q$}}\\
 \multicolumn{3}{l}{} \\
\hline\multicolumn{3}{c}{$\begin{array}{ccl}\mathcal{O}^f_{{1}}&=& i({{{d_{\mathbb{C}}}_p}^a}\sigma_{\mu\nu}({D_\lambda}{Q_r}^{ai}))({D^\nu}{{H^\dagger}^i}){{B_L}^{\lambda\mu}},\\
\mathcal{O}^f_{{2}}&=& ({{{d_{\mathbb{C}}}_p}^a}({D_\mu}{Q_r}^{ai}))({D_\nu}{{H^\dagger}^i}){{B_L}^{\mu\nu}}.\end{array}$}\\
\hline
\multicolumn{2}{c|}{group: $(\text{Spin},SU(3)_c,SU(2)_w,U(1)_y)$} & \multirow{2}{*}{$\mathcal{O}^{j}$}\\
\begin{fmffile}{4}
\begin{fmfgraph*}(40,20)
\fmfpen{thin} \fmfleft{e1,e2} \fmfv{}{g1} \fmfv{}{g2} \fmfright{e4,e3}
\fmf{plain}{e1,g1} \fmf{plain}{g1,e2} \fmf{plain}{g1,g2} \fmf{plain}{e4,g2} \fmf{plain}{g2,e3}
\end{fmfgraph*}
\end{fmffile} &\multicolumn{1}{c|}{ $\{{B_L}_1,{d_{\mathbb{C}}}_2\} \quad \{Q_3,{H^\dagger}_4\}$ } & \\ 
 \hline* & $(\frac{3}{2},\overline{3},1,\frac{1}{3})$ & $2 \mathcal{O}^f_{1} + 10 \mathcal{O}^f_{2}$\\ 
* & $(\frac{1}{2},\overline{3},1,\frac{1}{3})$ & $2 \mathcal{O}^f_{1} -2 \mathcal{O}^f_{2}$\\ 
\hline
& \multicolumn{1}{c|}{ $\{{B_L}_1,Q_3\} \quad \{{d_{\mathbb{C}}}_2,{H^\dagger}_4\}$ } & $\mathcal{O}^{j}$\\
 \hline* & $(\frac{3}{2},3,2,\frac{1}{6})$ & $-2 \mathcal{O}^f_{1} + 10 \mathcal{O}^f_{2}$\\ 
* & $(\frac{1}{2},3,2,\frac{1}{6})$ & $2 \mathcal{O}^f_{1} + 2 \mathcal{O}^f_{2}$\\ 
\hline
& \multicolumn{1}{c|}{ $\{{B_L}_1,{H^\dagger}_4\} \quad \{{d_{\mathbb{C}}}_2,Q_3\}$ } & $\mathcal{O}^{j}$\\
 \hline* & $(2,1,2,-\frac{1}{2})$ & $4 \mathcal{O}^f_{1}$\\ 
* & $(1,1,2,-\frac{1}{2})$ & $4 \mathcal{O}^f_{2}$\\ 
\hline
\end{longtable}

\begin{longtable}{cc|c}
\caption{UV resonance of type ${B_L}D^2HQ{u_{\mathbb{C}}}$. ``*'' in front of each row indicates topologies with nonrenormalizable vertices. The UV resonances colored in gray are eliminated by UV selection. }\\
\hline\hline\multicolumn{3}{l}{\multirow{2}{*}{\large Type: ${B_L}D^2HQ{u_{\mathbb{C}}}$}}\\
 \multicolumn{3}{l}{} \\
\hline\multicolumn{3}{c}{$\begin{array}{ccl}\mathcal{O}^f_{{1}}&=& i\epsilon^{ij}({Q_p}^{ai}\sigma_{\mu\nu}({D_\lambda}{{{u_{\mathbb{C}}}_r}^a}))({D^\nu}{H_j}){{B_L}^{\lambda\mu}},\\
\mathcal{O}^f_{{2}}&=& \epsilon^{ij}({Q_p}^{ai}({D_\mu}{{{u_{\mathbb{C}}}_r}^a}))({D_\nu}{H_j}){{B_L}^{\mu\nu}}.\end{array}$}\\
\hline
\multicolumn{2}{c|}{group: $(\text{Spin},SU(3)_c,SU(2)_w,U(1)_y)$} & \multirow{2}{*}{$\mathcal{O}^{j}$}\\
\begin{fmffile}{4}
\begin{fmfgraph*}(40,20)
\fmfpen{thin} \fmfleft{e1,e2} \fmfv{}{g1} \fmfv{}{g2} \fmfright{e4,e3}
\fmf{plain}{e1,g1} \fmf{plain}{g1,e2} \fmf{plain}{g1,g2} \fmf{plain}{e4,g2} \fmf{plain}{g2,e3}
\end{fmfgraph*}
\end{fmffile} &\multicolumn{1}{c|}{ $\{{B_L}_1,Q_2\} \quad \{{u_{\mathbb{C}}}_3,H_4\}$ } & \\ 
 \hline* & $(\frac{3}{2},3,2,\frac{1}{6})$ & $2 \mathcal{O}^f_{1} + 10 \mathcal{O}^f_{2}$\\ 
* & $(\frac{1}{2},3,2,\frac{1}{6})$ & $2 \mathcal{O}^f_{1} -2 \mathcal{O}^f_{2}$\\ 
\hline
& \multicolumn{1}{c|}{ $\{{B_L}_1,{u_{\mathbb{C}}}_3\} \quad \{Q_2,H_4\}$ } & $\mathcal{O}^{j}$\\
 \hline* & $(\frac{3}{2},\overline{3},1,-\frac{2}{3})$ & $-2 \mathcal{O}^f_{1} + 10 \mathcal{O}^f_{2}$\\ 
* & $(\frac{1}{2},\overline{3},1,-\frac{2}{3})$ & $2 \mathcal{O}^f_{1} + 2 \mathcal{O}^f_{2}$\\ 
\hline
& \multicolumn{1}{c|}{ $\{{B_L}_1,H_4\} \quad \{Q_2,{u_{\mathbb{C}}}_3\}$ } & $\mathcal{O}^{j}$\\
 \hline* & $(2,1,2,\frac{1}{2})$ & $4 \mathcal{O}^f_{1}$\\ 
* & $(1,1,2,\frac{1}{2})$ & $4 \mathcal{O}^f_{2}$\\ 
\hline
\end{longtable}

\begin{longtable}{cc|c}
\caption{UV resonance of type $D^2{e_{\mathbb{C}}}{H^\dagger}L{W_L}$. ``*'' in front of each row indicates topologies with nonrenormalizable vertices. The UV resonances colored in gray are eliminated by UV selection. }\\
\hline\hline\multicolumn{3}{l}{\multirow{2}{*}{\large Type: $D^2{e_{\mathbb{C}}}{H^\dagger}L{W_L}$}}\\
 \multicolumn{3}{l}{} \\
\hline\multicolumn{3}{c}{$\begin{array}{ccl}\mathcal{O}^f_{{1}}&=& i({e_{\mathbb{C}}}_p\sigma_{\mu\nu}({D_\lambda}{{L_r}_i}))({D^\nu}{{H^\dagger}^j}){{\tau^I}^i_j}{{W_L}^{I\lambda\mu}},\\
\mathcal{O}^f_{{2}}&=& ({e_{\mathbb{C}}}_p({D_\mu}{{L_r}_i}))({D_\nu}{{H^\dagger}^j}){{\tau^I}^i_j}{{W_L}^{I\mu\nu}}.\end{array}$}\\
\hline
\multicolumn{2}{c|}{group: $(\text{Spin},SU(3)_c,SU(2)_w,U(1)_y)$} & \multirow{2}{*}{$\mathcal{O}^{j}$}\\
\begin{fmffile}{4}
\begin{fmfgraph*}(40,20)
\fmfpen{thin} \fmfleft{e1,e2} \fmfv{}{g1} \fmfv{}{g2} \fmfright{e4,e3}
\fmf{plain}{e1,g1} \fmf{plain}{g1,e2} \fmf{plain}{g1,g2} \fmf{plain}{e4,g2} \fmf{plain}{g2,e3}
\end{fmfgraph*}
\end{fmffile} &\multicolumn{1}{c|}{ $\{{W_L}_1,{e_{\mathbb{C}}}_2\} \quad \{L_3,{H^\dagger}_4\}$ } & \\ 
 \hline* & $(\frac{3}{2},1,3,-1)$ & $2 \mathcal{O}^f_{1} + 10 \mathcal{O}^f_{2}$\\ 
* & $(\frac{1}{2},1,3,-1)$ & $2 \mathcal{O}^f_{1} -2 \mathcal{O}^f_{2}$\\ 
\hline
& \multicolumn{1}{c|}{ $\{{W_L}_1,L_3\} \quad \{{e_{\mathbb{C}}}_2,{H^\dagger}_4\}$ } & $\mathcal{O}^{j}$\\
 \hline* & $(\frac{3}{2},1,2,\frac{1}{2})$ & $-2 \mathcal{O}^f_{1} + 10 \mathcal{O}^f_{2}$\\ 
* & $(\frac{1}{2},1,2,\frac{1}{2})$ & $2 \mathcal{O}^f_{1} + 2 \mathcal{O}^f_{2}$\\ 
\hline
& \multicolumn{1}{c|}{ $\{{W_L}_1,{H^\dagger}_4\} \quad \{{e_{\mathbb{C}}}_2,L_3\}$ } & $\mathcal{O}^{j}$\\
 \hline* & $(2,1,2,\frac{1}{2})$ & $4 \mathcal{O}^f_{1}$\\ 
* & $(1,1,2,\frac{1}{2})$ & $4 \mathcal{O}^f_{2}$\\ 
\hline
\end{longtable}

\begin{longtable}{cc|c}
\caption{UV resonance of type $D^2{d_{\mathbb{C}}}{H^\dagger}Q{W_L}$. ``*'' in front of each row indicates topologies with nonrenormalizable vertices. The UV resonances colored in gray are eliminated by UV selection. }\\
\hline\hline\multicolumn{3}{l}{\multirow{2}{*}{\large Type: $D^2{d_{\mathbb{C}}}{H^\dagger}Q{W_L}$}}\\
 \multicolumn{3}{l}{} \\
\hline\multicolumn{3}{c}{$\begin{array}{ccl}\mathcal{O}^f_{{1}}&=& i({{{d_{\mathbb{C}}}_p}^a}\sigma_{\mu\nu}({D_\lambda}{Q_r}^{ai}))({D^\nu}{{H^\dagger}^j}){{\tau^I}^i_j}{{W_L}^{I\lambda\mu}},\\
\mathcal{O}^f_{{2}}&=& ({{{d_{\mathbb{C}}}_p}^a}({D_\mu}{Q_r}^{ai}))({D_\nu}{{H^\dagger}^j}){{\tau^I}^i_j}{{W_L}^{I\mu\nu}}.\end{array}$}\\
\hline
\multicolumn{2}{c|}{group: $(\text{Spin},SU(3)_c,SU(2)_w,U(1)_y)$} & \multirow{2}{*}{$\mathcal{O}^{j}$}\\
\begin{fmffile}{4}
\begin{fmfgraph*}(40,20)
\fmfpen{thin} \fmfleft{e1,e2} \fmfv{}{g1} \fmfv{}{g2} \fmfright{e4,e3}
\fmf{plain}{e1,g1} \fmf{plain}{g1,e2} \fmf{plain}{g1,g2} \fmf{plain}{e4,g2} \fmf{plain}{g2,e3}
\end{fmfgraph*}
\end{fmffile} &\multicolumn{1}{c|}{ $\{{W_L}_1,{d_{\mathbb{C}}}_2\} \quad \{Q_3,{H^\dagger}_4\}$ } & \\ 
 \hline* & $(\frac{3}{2},3,3,-\frac{1}{3})$ & $2 \mathcal{O}^f_{1} + 10 \mathcal{O}^f_{2}$\\ 
* & $(\frac{1}{2},3,3,-\frac{1}{3})$ & $2 \mathcal{O}^f_{1} -2 \mathcal{O}^f_{2}$\\ 
\hline
& \multicolumn{1}{c|}{ $\{{W_L}_1,Q_3\} \quad \{{d_{\mathbb{C}}}_2,{H^\dagger}_4\}$ } & $\mathcal{O}^{j}$\\
 \hline* & $(\frac{3}{2},\overline{3},2,-\frac{1}{6})$ & $-2 \mathcal{O}^f_{1} + 10 \mathcal{O}^f_{2}$\\ 
* & $(\frac{1}{2},\overline{3},2,-\frac{1}{6})$ & $2 \mathcal{O}^f_{1} + 2 \mathcal{O}^f_{2}$\\ 
\hline
& \multicolumn{1}{c|}{ $\{{W_L}_1,{H^\dagger}_4\} \quad \{{d_{\mathbb{C}}}_2,Q_3\}$ } & $\mathcal{O}^{j}$\\
 \hline* & $(2,1,2,\frac{1}{2})$ & $4 \mathcal{O}^f_{1}$\\ 
* & $(1,1,2,\frac{1}{2})$ & $4 \mathcal{O}^f_{2}$\\ 
\hline
\end{longtable}

\begin{longtable}{cc|c}
\caption{UV resonance of type $D^2HQ{u_{\mathbb{C}}}{W_L}$. ``*'' in front of each row indicates topologies with nonrenormalizable vertices. The UV resonances colored in gray are eliminated by UV selection. }\\
\hline\hline\multicolumn{3}{l}{\multirow{2}{*}{\large Type: $D^2HQ{u_{\mathbb{C}}}{W_L}$}}\\
 \multicolumn{3}{l}{} \\
\hline\multicolumn{3}{c}{$\begin{array}{ccl}\mathcal{O}^f_{{1}}&=& i\epsilon^{jk}({Q_p}^{ai}\sigma_{\mu\nu}({D_\lambda}{{{u_{\mathbb{C}}}_r}^a}))({D^\nu}{H_j}){{\tau^I}^i_k}{{W_L}^{I\lambda\mu}},\\
\mathcal{O}^f_{{2}}&=& \epsilon^{jk}({Q_p}^{ai}({D_\mu}{{{u_{\mathbb{C}}}_r}^a}))({D_\nu}{H_j}){{\tau^I}^i_k}{{W_L}^{I\mu\nu}}.\end{array}$}\\
\hline
\multicolumn{2}{c|}{group: $(\text{Spin},SU(3)_c,SU(2)_w,U(1)_y)$} & \multirow{2}{*}{$\mathcal{O}^{j}$}\\
\begin{fmffile}{4}
\begin{fmfgraph*}(40,20)
\fmfpen{thin} \fmfleft{e1,e2} \fmfv{}{g1} \fmfv{}{g2} \fmfright{e4,e3}
\fmf{plain}{e1,g1} \fmf{plain}{g1,e2} \fmf{plain}{g1,g2} \fmf{plain}{e4,g2} \fmf{plain}{g2,e3}
\end{fmfgraph*}
\end{fmffile} &\multicolumn{1}{c|}{ $\{{W_L}_1,Q_2\} \quad \{{u_{\mathbb{C}}}_3,H_4\}$ } & \\ 
 \hline* & $(\frac{3}{2},\overline{3},2,-\frac{1}{6})$ & $2 \mathcal{O}^f_{1} + 10 \mathcal{O}^f_{2}$\\ 
* & $(\frac{1}{2},\overline{3},2,-\frac{1}{6})$ & $2 \mathcal{O}^f_{1} -2 \mathcal{O}^f_{2}$\\ 
\hline
& \multicolumn{1}{c|}{ $\{{W_L}_1,{u_{\mathbb{C}}}_3\} \quad \{Q_2,H_4\}$ } & $\mathcal{O}^{j}$\\
 \hline* & $(\frac{3}{2},3,3,\frac{2}{3})$ & $-2 \mathcal{O}^f_{1} + 10 \mathcal{O}^f_{2}$\\ 
* & $(\frac{1}{2},3,3,\frac{2}{3})$ & $2 \mathcal{O}^f_{1} + 2 \mathcal{O}^f_{2}$\\ 
\hline
& \multicolumn{1}{c|}{ $\{{W_L}_1,H_4\} \quad \{Q_2,{u_{\mathbb{C}}}_3\}$ } & $\mathcal{O}^{j}$\\
 \hline* & $(2,1,2,-\frac{1}{2})$ & $4 \mathcal{O}^f_{1}$\\ 
* & $(1,1,2,-\frac{1}{2})$ & $4 \mathcal{O}^f_{2}$\\ 
\hline
\end{longtable}

\begin{longtable}{cc|c}
\caption{UV resonance of type $D^2{d_{\mathbb{C}}}{G_L}{H^\dagger}Q$. ``*'' in front of each row indicates topologies with nonrenormalizable vertices. The UV resonances colored in gray are eliminated by UV selection. }\\
\hline\hline\multicolumn{3}{l}{\multirow{2}{*}{\large Type: $D^2{d_{\mathbb{C}}}{G_L}{H^\dagger}Q$}}\\
 \multicolumn{3}{l}{} \\
\hline\multicolumn{3}{c}{$\begin{array}{ccl}\mathcal{O}^f_{{1}}&=& i({{{d_{\mathbb{C}}}_p}^a}\sigma_{\mu\nu}({D_\lambda}{Q_r}^{bi}))({D^\nu}{{H^\dagger}^i}){{\lambda^A}_a^b}{{G_L}^{A\lambda\mu}},\\
\mathcal{O}^f_{{2}}&=& ({{{d_{\mathbb{C}}}_p}^a}({D_\mu}{Q_r}^{bi}))({D_\nu}{{H^\dagger}^i}){{\lambda^A}_a^b}{{G_L}^{A\mu\nu}}.\end{array}$}\\
\hline
\multicolumn{2}{c|}{group: $(\text{Spin},SU(3)_c,SU(2)_w,U(1)_y)$} & \multirow{2}{*}{$\mathcal{O}^{j}$}\\
\begin{fmffile}{4}
\begin{fmfgraph*}(40,20)
\fmfpen{thin} \fmfleft{e1,e2} \fmfv{}{g1} \fmfv{}{g2} \fmfright{e4,e3}
\fmf{plain}{e1,g1} \fmf{plain}{g1,e2} \fmf{plain}{g1,g2} \fmf{plain}{e4,g2} \fmf{plain}{g2,e3}
\end{fmfgraph*}
\end{fmffile} &\multicolumn{1}{c|}{ $\{{G_L}_1,{d_{\mathbb{C}}}_2\} \quad \{Q_3,{H^\dagger}_4\}$ } & \\ 
 \hline* & $(\frac{3}{2},\overline{3},1,\frac{1}{3})$ & $2 \mathcal{O}^f_{1} + 10 \mathcal{O}^f_{2}$\\ 
* & $(\frac{1}{2},\overline{3},1,\frac{1}{3})$ & $2 \mathcal{O}^f_{1} -2 \mathcal{O}^f_{2}$\\ 
\hline
& \multicolumn{1}{c|}{ $\{{G_L}_1,Q_3\} \quad \{{d_{\mathbb{C}}}_2,{H^\dagger}_4\}$ } & $\mathcal{O}^{j}$\\
 \hline* & $(\frac{3}{2},\overline{3},2,-\frac{1}{6})$ & $-2 \mathcal{O}^f_{1} + 10 \mathcal{O}^f_{2}$\\ 
* & $(\frac{1}{2},\overline{3},2,-\frac{1}{6})$ & $2 \mathcal{O}^f_{1} + 2 \mathcal{O}^f_{2}$\\ 
\hline
& \multicolumn{1}{c|}{ $\{{G_L}_1,{H^\dagger}_4\} \quad \{{d_{\mathbb{C}}}_2,Q_3\}$ } & $\mathcal{O}^{j}$\\
 \hline* & $(2,8,2,\frac{1}{2})$ & $4 \mathcal{O}^f_{1}$\\ 
* & $(1,8,2,\frac{1}{2})$ & $4 \mathcal{O}^f_{2}$\\ 
\hline
\end{longtable}

\begin{longtable}{cc|c}
\caption{UV resonance of type $D^2{G_L}HQ{u_{\mathbb{C}}}$. ``*'' in front of each row indicates topologies with nonrenormalizable vertices. The UV resonances colored in gray are eliminated by UV selection. }\\
\hline\hline\multicolumn{3}{l}{\multirow{2}{*}{\large Type: $D^2{G_L}HQ{u_{\mathbb{C}}}$}}\\
 \multicolumn{3}{l}{} \\
\hline\multicolumn{3}{c}{$\begin{array}{ccl}\mathcal{O}^f_{{1}}&=& i\epsilon^{ij}({Q_p}^{ai}\sigma_{\mu\nu}({D_\lambda}{{{u_{\mathbb{C}}}_r}^b}))({D^\nu}{H_j}){{\lambda^A}_b^a}{{G_L}^{A\lambda\mu}},\\
\mathcal{O}^f_{{2}}&=& \epsilon^{ij}({Q_p}^{ai}({D_\mu}{{{u_{\mathbb{C}}}_r}^b}))({D_\nu}{H_j}){{\lambda^A}_b^a}{{G_L}^{A\mu\nu}}.\end{array}$}\\
\hline
\multicolumn{2}{c|}{group: $(\text{Spin},SU(3)_c,SU(2)_w,U(1)_y)$} & \multirow{2}{*}{$\mathcal{O}^{j}$}\\
\begin{fmffile}{4}
\begin{fmfgraph*}(40,20)
\fmfpen{thin} \fmfleft{e1,e2} \fmfv{}{g1} \fmfv{}{g2} \fmfright{e4,e3}
\fmf{plain}{e1,g1} \fmf{plain}{g1,e2} \fmf{plain}{g1,g2} \fmf{plain}{e4,g2} \fmf{plain}{g2,e3}
\end{fmfgraph*}
\end{fmffile} &\multicolumn{1}{c|}{ $\{{G_L}_1,Q_2\} \quad \{{u_{\mathbb{C}}}_3,H_4\}$ } & \\ 
 \hline* & $(\frac{3}{2},3,2,\frac{1}{6})$ & $2 \mathcal{O}^f_{1} + 10 \mathcal{O}^f_{2}$\\ 
* & $(\frac{1}{2},3,2,\frac{1}{6})$ & $2 \mathcal{O}^f_{1} -2 \mathcal{O}^f_{2}$\\ 
\hline
& \multicolumn{1}{c|}{ $\{{G_L}_1,{u_{\mathbb{C}}}_3\} \quad \{Q_2,H_4\}$ } & $\mathcal{O}^{j}$\\
 \hline* & $(\frac{3}{2},\overline{3},1,-\frac{2}{3})$ & $-2 \mathcal{O}^f_{1} + 10 \mathcal{O}^f_{2}$\\ 
* & $(\frac{1}{2},\overline{3},1,-\frac{2}{3})$ & $2 \mathcal{O}^f_{1} + 2 \mathcal{O}^f_{2}$\\ 
\hline
& \multicolumn{1}{c|}{ $\{{G_L}_1,H_4\} \quad \{Q_2,{u_{\mathbb{C}}}_3\}$ } & $\mathcal{O}^{j}$\\
 \hline* & $(2,8,2,\frac{1}{2})$ & $4 \mathcal{O}^f_{1}$\\ 
* & $(1,8,2,\frac{1}{2})$ & $4 \mathcal{O}^f_{2}$\\ 
\hline
\end{longtable}

\begin{longtable}{cc|c}
\caption{UV resonance of type ${B_L}^2D^2H{H^\dagger}$. ``*'' in front of each row indicates topologies with nonrenormalizable vertices. The UV resonances colored in gray are eliminated by UV selection. }\\
\hline\hline\multicolumn{3}{l}{\multirow{2}{*}{\large Type: ${B_L}^2D^2H{H^\dagger}$}}\\
 \multicolumn{3}{l}{} \\
\hline\multicolumn{3}{c}{$\begin{array}{ccl}\mathcal{O}^f_{1}&=&\frac{1}{2}\mathcal{Y}[\yng(2)_{{B_L}}]{{B_L}^{\lambda\nu}}{{B_L}^{\nu\lambda}}({D_\mu}{H_i})({D^\mu}{{H^\dagger}^i}).\end{array}$}\\
\hline
\multicolumn{2}{c|}{group: $(\text{Spin},SU(3)_c,SU(2)_w,U(1)_y)$} & \multirow{2}{*}{$\mathcal{O}^{j}$}\\
\begin{fmffile}{4}
\begin{fmfgraph*}(40,20)
\fmfpen{thin} \fmfleft{e1,e2} \fmfv{}{g1} \fmfv{}{g2} \fmfright{e4,e3}
\fmf{plain}{e1,g1} \fmf{plain}{g1,e2} \fmf{plain}{g1,g2} \fmf{plain}{e4,g2} \fmf{plain}{g2,e3}
\end{fmfgraph*}
\end{fmffile} &\multicolumn{1}{c|}{ $\{{B_L}_1,{B_L}_2\} \quad \{H_3,{H^\dagger}_4\}$ } & \\ 
 \hline* & $(0,1,1,0)$ & $-4 \mathcal{O}^f_{{1}}$\\ 
\hline
& \multicolumn{1}{c|}{ $\{{B_L}_1,H_3\} \quad \{{B_L}_2,{H^\dagger}_4\}$ } & $\mathcal{O}^{j}$\\
 \hline* & $(2,1,2,\frac{1}{2})$ & $10 \mathcal{O}^f_{{1}}$\\ 
* & $(1,1,2,\frac{1}{2})$ & $-2 \mathcal{O}^f_{{1}}$\\ 
\hline
\end{longtable}%

\section{UV List for Operators involving 5 particles}

\input{dim8-5}

\section{UV List for Operators involving 6 particles}

\begin{longtable}{cccc|c}
\caption{UV resonance of type ${e_{\mathbb{C}}}^2{H^\dagger}^2L^2$. ``*'' in front of each row indicates topologies with nonrenormalizable vertices. The UV resonances colored in gray are eliminated by UV selection. }\\
\hline\hline\multicolumn{5}{l}{\multirow{2}{*}{\large Type: ${e_{\mathbb{C}}}^2{H^\dagger}^2L^2$}}\\
 \multicolumn{5}{l}{} \\
\hline\multicolumn{5}{c}{$\begin{array}{ccl}\mathcal{O}^f_{1}&=&\frac{1}{8}\mathcal{Y}[\yng(2)_{{H^\dagger}}]\mathcal{Y}[\young(pr)]\mathcal{Y}[\young(st)]{{H^\dagger}^i}{{H^\dagger}^j}({e_{\mathbb{C}}}_p{e_{\mathbb{C}}}_r)({{L_s}_i}{{L_t}_j}),\\
\mathcal{O}^f_{2}&=&\frac{1}{8}\mathcal{Y}[\yng(2)_{{H^\dagger}}]\mathcal{Y}[\young(p,r)]\mathcal{Y}[\young(s,t)]{{H^\dagger}^i}{{H^\dagger}^j}({e_{\mathbb{C}}}_p{{L_s}_i})({e_{\mathbb{C}}}_r{{L_t}_j}).\end{array}$}\\
\hline
\multicolumn{4}{c|}{group: $(\text{Spin},SU(3)_c,SU(2)_w,U(1)_y)$} & \multirow{2}{*}{$\mathcal{O}^{j}$}\\
\begin{fmffile}{6-1}
\begin{fmfgraph*}(30,30)
\fmfpen{thin} \fmfsurroundn{e}{6} \fmfvn{}{g}{4} 
\fmf{plain}{e1,g1} \fmf{plain}{g1,e2} \fmf{plain}{g1,g4} \fmf{plain}{e3,g2} \fmf{plain}{g2,e4} \fmf{plain}{g2,g4} \fmf{plain}{e5,g3} \fmf{plain}{g3,g4} \fmf{plain}{g3,e6}
\end{fmfgraph*}
\end{fmffile} &\multicolumn{3}{c|}{ $\{{e_{\mathbb{C}}}_1,{e_{\mathbb{C}}}_2\} \quad \{L_3,L_4\} \quad \{{H^\dagger}_5,{H^\dagger}_6\}$ } & \\ 
 \hline* & $(1,1,1,2)$ & $(0,1,3,-1)$ & $(1,1,3,-1)$ & $4 \mathcal{O}^f_{{2}}$\\ 
& $(0,1,1,2)$ & $(0,1,3,-1)$ & $(0,1,3,-1)$ & $2 \mathcal{O}^f_{{1}}$\\
\hline
\end{longtable}

\begin{longtable}{cccc|c}
\caption{UV resonance of type ${d_{\mathbb{C}}}{e_{\mathbb{C}}}H{H^\dagger}{u_{\mathbb{C}}}^2$. ``*'' in front of each row indicates topologies with nonrenormalizable vertices. The UV resonances colored in gray are eliminated by UV selection. }\\
\hline\hline\multicolumn{5}{l}{\multirow{2}{*}{\large Type: ${d_{\mathbb{C}}}{e_{\mathbb{C}}}H{H^\dagger}{u_{\mathbb{C}}}^2$}}\\
 \multicolumn{5}{l}{} \\
\hline\multicolumn{5}{c}{$\begin{array}{ccl}\mathcal{O}^f_{1}&=&\frac{1}{2}\mathcal{Y}[\young(st)]\epsilon_{abc}{{H^\dagger}^i}{H_i}({{{d_{\mathbb{C}}}_p}^a}{{{u_{\mathbb{C}}}_s}^b})({e_{\mathbb{C}}}_r{{{u_{\mathbb{C}}}_t}^c}),\\
\mathcal{O}^f_{2}&=&\frac{1}{2}\mathcal{Y}[\young(s,t)]\epsilon_{abc}{{H^\dagger}^i}{H_i}({{{d_{\mathbb{C}}}_p}^a}{e_{\mathbb{C}}}_r)({{{u_{\mathbb{C}}}_s}^b}{{{u_{\mathbb{C}}}_t}^c}).\end{array}$}\\
\hline
\multicolumn{4}{c|}{group: $(\text{Spin},SU(3)_c,SU(2)_w,U(1)_y)$} & \multirow{2}{*}{$\mathcal{O}^{j}$}\\
\begin{fmffile}{6-1}
\begin{fmfgraph*}(30,30)
\fmfpen{thin} \fmfsurroundn{e}{6} \fmfvn{}{g}{4} 
\fmf{plain}{e1,g1} \fmf{plain}{g1,e2} \fmf{plain}{g1,g4} \fmf{plain}{e3,g2} \fmf{plain}{g2,e4} \fmf{plain}{g2,g4} \fmf{plain}{e5,g3} \fmf{plain}{g3,g4} \fmf{plain}{g3,e6}
\end{fmfgraph*}
\end{fmffile} &\multicolumn{3}{c|}{ $\{{d_{\mathbb{C}}}_1,{e_{\mathbb{C}}}_2\} \quad \{{u_{\mathbb{C}}}_3,{u_{\mathbb{C}}}_4\} \quad \{H_5,{H^\dagger}_6\}$ } & \\ 
 \hline* & $(1,\overline{3},1,\frac{4}{3})$ & $(0,1,1,0)$ & $(1,3,1,-\frac{4}{3})$ & $2 \mathcal{O}^f_{{1}}$\\ 
& $(0,\overline{3},1,\frac{4}{3})$ & $(0,1,1,0)$ & $(0,3,1,-\frac{4}{3})$ & $\mathcal{O}^f_{{2}}$\\
\hline
\end{longtable}

\begin{longtable}{cccc|c}
\caption{UV resonance of type ${d_{\mathbb{C}}}{e_{\mathbb{C}}}{H^\dagger}^2LQ$. ``*'' in front of each row indicates topologies with nonrenormalizable vertices. The UV resonances colored in gray are eliminated by UV selection. }\\
\hline\hline\multicolumn{5}{l}{\multirow{2}{*}{\large Type: ${d_{\mathbb{C}}}{e_{\mathbb{C}}}{H^\dagger}^2LQ$}}\\
 \multicolumn{5}{l}{} \\
\hline\multicolumn{5}{c}{$\begin{array}{ccl}\mathcal{O}^f_{1}&=&\frac{1}{2}\mathcal{Y}[\yng(2)_{{H^\dagger}}]{{H^\dagger}^i}{{H^\dagger}^j}({{{d_{\mathbb{C}}}_p}^a}{e_{\mathbb{C}}}_r)({{L_s}_i}{Q_t}^{aj}),\\
\mathcal{O}^f_{2}&=&\frac{1}{2}\mathcal{Y}[\yng(2)_{{H^\dagger}}]{{H^\dagger}^i}{{H^\dagger}^j}({{{d_{\mathbb{C}}}_p}^a}{{L_s}_i})({e_{\mathbb{C}}}_r{Q_t}^{aj}).\end{array}$}\\
\hline
\multicolumn{4}{c|}{group: $(\text{Spin},SU(3)_c,SU(2)_w,U(1)_y)$} & \multirow{2}{*}{$\mathcal{O}^{j}$}\\
\begin{fmffile}{6-1}
\begin{fmfgraph*}(30,30)
\fmfpen{thin} \fmfsurroundn{e}{6} \fmfvn{}{g}{4} 
\fmf{plain}{e1,g1} \fmf{plain}{g1,e2} \fmf{plain}{g1,g4} \fmf{plain}{e3,g2} \fmf{plain}{g2,e4} \fmf{plain}{g2,g4} \fmf{plain}{e5,g3} \fmf{plain}{g3,g4} \fmf{plain}{g3,e6}
\end{fmfgraph*}
\end{fmffile} &\multicolumn{3}{c|}{ $\{{d_{\mathbb{C}}}_1,{e_{\mathbb{C}}}_2\} \quad \{L_3,Q_4\} \quad \{{H^\dagger}_5,{H^\dagger}_6\}$ } & \\ 
 \hline* & $(1,\overline{3},1,\frac{4}{3})$ & $(0,1,3,-1)$ & $(1,3,3,-\frac{1}{3})$ & $2 \mathcal{O}^f_{{1}} + 4 \mathcal{O}^f_{{2}}$\\ 
& $(0,\overline{3},1,\frac{4}{3})$ & $(0,1,3,-1)$ & $(0,3,3,-\frac{1}{3})$ & $2 \mathcal{O}^f_{{1}}$\\
\hline
\end{longtable}

\begin{longtable}{cccc|c}
\caption{UV resonance of type ${e_{\mathbb{C}}}H{H^\dagger}LQ{u_{\mathbb{C}}}$. ``*'' in front of each row indicates topologies with nonrenormalizable vertices. The UV resonances colored in gray are eliminated by UV selection. }\\
\hline\hline\multicolumn{5}{l}{\multirow{2}{*}{\large Type: ${e_{\mathbb{C}}}H{H^\dagger}LQ{u_{\mathbb{C}}}$}}\\
 \multicolumn{5}{l}{} \\
\hline\multicolumn{5}{c}{$\begin{array}{ccl}\mathcal{O}^f_{{1}}&=& \epsilon^{kj}({e_{\mathbb{C}}}_p{{L_r}_i})({Q_s}^{aj}{{{u_{\mathbb{C}}}_t}^a}){H_k}{{H^\dagger}^i},\\
\mathcal{O}^f_{{2}}&=& \epsilon^{kj}({e_{\mathbb{C}}}_p{Q_s}^{aj})({{L_r}_i}{{{u_{\mathbb{C}}}_t}^a}){H_k}{{H^\dagger}^i},\\
\mathcal{O}^f_{{3}}&=& \epsilon^{ij}({e_{\mathbb{C}}}_p{{L_r}_i})({Q_s}^{aj}{{{u_{\mathbb{C}}}_t}^a}){H_k}{{H^\dagger}^k},\\
\mathcal{O}^f_{{4}}&=& \epsilon^{ij}({e_{\mathbb{C}}}_p{Q_s}^{aj})({{L_r}_i}{{{u_{\mathbb{C}}}_t}^a}){H_k}{{H^\dagger}^k}.\end{array}$}\\
\hline
\multicolumn{4}{c|}{group: $(\text{Spin},SU(3)_c,SU(2)_w,U(1)_y)$} & \multirow{2}{*}{$\mathcal{O}^{j}$}\\
\begin{fmffile}{6-1}
\begin{fmfgraph*}(30,30)
\fmfpen{thin} \fmfsurroundn{e}{6} \fmfvn{}{g}{4} 
\fmf{plain}{e1,g1} \fmf{plain}{g1,e2} \fmf{plain}{g1,g4} \fmf{plain}{e3,g2} \fmf{plain}{g2,e4} \fmf{plain}{g2,g4} \fmf{plain}{e5,g3} \fmf{plain}{g3,g4} \fmf{plain}{g3,e6}
\end{fmfgraph*}
\end{fmffile} &\multicolumn{3}{c|}{ $\{{e_{\mathbb{C}}}_1,L_2\} \quad \{Q_3,{u_{\mathbb{C}}}_4\} \quad \{H_5,{H^\dagger}_6\}$ } & \\ 
 \hline* & $(1,1,2,\frac{1}{2})$ & $(0,1,3,0)$ & $(1,1,2,-\frac{1}{2})$ & $2 \mathcal{O}^f_{1} + 4 \mathcal{O}^f_{2} -\mathcal{O}^f_{3} -2 \mathcal{O}^f_{4}$\\ 
& $(0,1,2,\frac{1}{2})$ & $(0,1,3,0)$ & $(0,1,2,-\frac{1}{2})$ & $2 \mathcal{O}^f_{1} -\mathcal{O}^f_{3}$\\
\hline
\end{longtable}

\begin{longtable}{cccc|c}
\caption{UV resonance of type ${d_{\mathbb{C}}}^2{H^\dagger}^2Q^2$. ``*'' in front of each row indicates topologies with nonrenormalizable vertices. The UV resonances colored in gray are eliminated by UV selection. }\\
\hline\hline\multicolumn{5}{l}{\multirow{2}{*}{\large Type: ${d_{\mathbb{C}}}^2{H^\dagger}^2Q^2$}}\\
 \multicolumn{5}{l}{} \\
\hline\multicolumn{5}{c}{$\begin{array}{ccl}\mathcal{O}^f_{1}&=&\frac{1}{8}\mathcal{Y}[\yng(2)_{{H^\dagger}}]\mathcal{Y}[\young(pr)]\mathcal{Y}[\young(st)]{{H^\dagger}^i}{{H^\dagger}^j}({{{d_{\mathbb{C}}}_p}^a}{{{d_{\mathbb{C}}}_r}^b})({Q_s}^{ai}{Q_t}^{bj}),\\
\mathcal{O}^f_{2}&=&\frac{1}{8}\mathcal{Y}[\yng(2)_{{H^\dagger}}]\mathcal{Y}[\young(pr)]\mathcal{Y}[\young(st)]{{H^\dagger}^i}{{H^\dagger}^j}({{{d_{\mathbb{C}}}_p}^a}{Q_s}^{ai})({{{d_{\mathbb{C}}}_r}^b}{Q_t}^{bj}),\\
\mathcal{O}^f_{3}&=&\frac{1}{8}\mathcal{Y}[\yng(2)_{{H^\dagger}}]\mathcal{Y}[\young(p,r)]\mathcal{Y}[\young(s,t)]{{H^\dagger}^i}{{H^\dagger}^j}({{{d_{\mathbb{C}}}_p}^a}{{{d_{\mathbb{C}}}_r}^b})({Q_s}^{ai}{Q_t}^{bj}),\\
\mathcal{O}^f_{4}&=&\frac{1}{8}\mathcal{Y}[\yng(2)_{{H^\dagger}}]\mathcal{Y}[\young(p,r)]\mathcal{Y}[\young(s,t)]{{H^\dagger}^i}{{H^\dagger}^j}({{{d_{\mathbb{C}}}_p}^a}{Q_s}^{ai})({{{d_{\mathbb{C}}}_r}^b}{Q_t}^{bj}).\end{array}$}\\
\hline
\multicolumn{4}{c|}{group: $(\text{Spin},SU(3)_c,SU(2)_w,U(1)_y)$} & \multirow{2}{*}{$\mathcal{O}^{j}$}\\
\begin{fmffile}{6-1}
\begin{fmfgraph*}(30,30)
\fmfpen{thin} \fmfsurroundn{e}{6} \fmfvn{}{g}{4} 
\fmf{plain}{e1,g1} \fmf{plain}{g1,e2} \fmf{plain}{g1,g4} \fmf{plain}{e3,g2} \fmf{plain}{g2,e4} \fmf{plain}{g2,g4} \fmf{plain}{e5,g3} \fmf{plain}{g3,g4} \fmf{plain}{g3,e6}
\end{fmfgraph*}
\end{fmffile} &\multicolumn{3}{c|}{ $\{{d_{\mathbb{C}}}_1,{d_{\mathbb{C}}}_2\} \quad \{Q_3,Q_4\} \quad \{{H^\dagger}_5,{H^\dagger}_6\}$ } & \\ 
 \hline* & $(1,6,1,\frac{2}{3})$ & $(0,1,3,-1)$ & $(1,\overline{6},3,\frac{1}{3})$ & $4 \mathcal{O}^f_{{3}} + 8 \mathcal{O}^f_{{4}}$\\
\hline
\end{longtable}

\begin{landscape}
\begin{longtable}{cccc|c}
\caption{UV resonance of type ${d_{\mathbb{C}}}H{H^\dagger}Q^2{u_{\mathbb{C}}}$. ``*'' in front of each row indicates topologies with nonrenormalizable vertices. The UV resonances colored in gray are eliminated by UV selection. }\\
\hline\hline\multicolumn{5}{l}{\multirow{2}{*}{\large Type: ${d_{\mathbb{C}}}H{H^\dagger}Q^2{u_{\mathbb{C}}}$}}\\
 \multicolumn{5}{l}{} \\
\hline\multicolumn{5}{c}{$\begin{array}{ccl}\mathcal{O}^f_{1}&=&\frac{1}{2}\mathcal{Y}[\young(rs)]\epsilon^{kj}{{H^\dagger}^i}{H_k}({Q_s}^{aj}{{{u_{\mathbb{C}}}_t}^b})({{{d_{\mathbb{C}}}_p}^a}{Q_r}^{bi}),\\
\mathcal{O}^f_{2}&=&\frac{1}{2}\mathcal{Y}[\young(rs)]\epsilon^{kj}{{H^\dagger}^i}{H_k}({Q_r}^{bi}{{{u_{\mathbb{C}}}_t}^b})({{{d_{\mathbb{C}}}_p}^a}{Q_s}^{aj}),\\
\mathcal{O}^f_{3}&=&\frac{1}{2}\mathcal{Y}[\young(rs)]\epsilon^{ij}{{H^\dagger}^k}{H_k}({Q_s}^{aj}{{{u_{\mathbb{C}}}_t}^b})({{{d_{\mathbb{C}}}_p}^a}{Q_r}^{bi}),\\
\mathcal{O}^f_{4}&=&\frac{1}{2}\mathcal{Y}[\young(rs)]\epsilon^{ij}{{H^\dagger}^k}{H_k}({Q_r}^{bi}{{{u_{\mathbb{C}}}_t}^b})({{{d_{\mathbb{C}}}_p}^a}{Q_s}^{aj}),\\
\mathcal{O}^f_{5}&=&\frac{1}{2}\mathcal{Y}[\young(r,s)]\epsilon^{kj}{{H^\dagger}^i}{H_k}({Q_s}^{aj}{{{u_{\mathbb{C}}}_t}^b})({{{d_{\mathbb{C}}}_p}^a}{Q_r}^{bi}),\\
\mathcal{O}^f_{6}&=&\frac{1}{2}\mathcal{Y}[\young(r,s)]\epsilon^{kj}{{H^\dagger}^i}{H_k}({Q_r}^{bi}{{{u_{\mathbb{C}}}_t}^b})({{{d_{\mathbb{C}}}_p}^a}{Q_s}^{aj}),\\
\mathcal{O}^f_{7}&=&\frac{1}{2}\mathcal{Y}[\young(r,s)]\epsilon^{ij}{{H^\dagger}^k}{H_k}({Q_s}^{aj}{{{u_{\mathbb{C}}}_t}^b})({{{d_{\mathbb{C}}}_p}^a}{Q_r}^{bi}),\\
\mathcal{O}^f_{8}&=&\frac{1}{2}\mathcal{Y}[\young(r,s)]\epsilon^{ij}{{H^\dagger}^k}{H_k}({Q_r}^{bi}{{{u_{\mathbb{C}}}_t}^b})({{{d_{\mathbb{C}}}_p}^a}{Q_s}^{aj}).\end{array}$}\\
\hline
\multicolumn{4}{c|}{group: $(\text{Spin},SU(3)_c,SU(2)_w,U(1)_y)$} & \multirow{2}{*}{$\mathcal{O}^{j}$}\\
\hline
\end{longtable}
\end{landscape}

\begin{longtable}{cccc|c}
\caption{UV resonance of type $H^2Q^2{u_{\mathbb{C}}}^2$. ``*'' in front of each row indicates topologies with nonrenormalizable vertices. The UV resonances colored in gray are eliminated by UV selection. }\\
\hline\hline\multicolumn{5}{l}{\multirow{2}{*}{\large Type: $H^2Q^2{u_{\mathbb{C}}}^2$}}\\
 \multicolumn{5}{l}{} \\
\hline\multicolumn{5}{c}{$\begin{array}{ccl}\mathcal{O}^f_{1}&=&\frac{1}{8}\mathcal{Y}[\yng(2)_{H}]\mathcal{Y}[\young(pr)]\mathcal{Y}[\young(st)]\epsilon^{ik}\epsilon^{jl}{H_k}{H_l}({{{u_{\mathbb{C}}}_s}^a}{{{u_{\mathbb{C}}}_t}^b})({Q_p}^{ai}{Q_r}^{bj}),\\
\mathcal{O}^f_{2}&=&\frac{1}{8}\mathcal{Y}[\yng(2)_{H}]\mathcal{Y}[\young(pr)]\mathcal{Y}[\young(st)]\epsilon^{ik}\epsilon^{jl}{H_k}{H_l}({Q_p}^{ai}{{{u_{\mathbb{C}}}_s}^a})({Q_r}^{bj}{{{u_{\mathbb{C}}}_t}^b}),\\
\mathcal{O}^f_{3}&=&\frac{1}{8}\mathcal{Y}[\yng(2)_{H}]\mathcal{Y}[\young(p,r)]\mathcal{Y}[\young(s,t)]\epsilon^{ik}\epsilon^{jl}{H_k}{H_l}({{{u_{\mathbb{C}}}_s}^a}{{{u_{\mathbb{C}}}_t}^b})({Q_p}^{ai}{Q_r}^{bj}),\\
\mathcal{O}^f_{4}&=&\frac{1}{8}\mathcal{Y}[\yng(2)_{H}]\mathcal{Y}[\young(p,r)]\mathcal{Y}[\young(s,t)]\epsilon^{ik}\epsilon^{jl}{H_k}{H_l}({Q_p}^{ai}{{{u_{\mathbb{C}}}_s}^a})({Q_r}^{bj}{{{u_{\mathbb{C}}}_t}^b}).\end{array}$}\\
\hline
\multicolumn{4}{c|}{group: $(\text{Spin},SU(3)_c,SU(2)_w,U(1)_y)$} & \multirow{2}{*}{$\mathcal{O}^{j}$}\\
\begin{fmffile}{6-1}
\begin{fmfgraph*}(30,30)
\fmfpen{thin} \fmfsurroundn{e}{6} \fmfvn{}{g}{4} 
\fmf{plain}{e1,g1} \fmf{plain}{g1,e2} \fmf{plain}{g1,g4} \fmf{plain}{e3,g2} \fmf{plain}{g2,e4} \fmf{plain}{g2,g4} \fmf{plain}{e5,g3} \fmf{plain}{g3,g4} \fmf{plain}{g3,e6}
\end{fmfgraph*}
\end{fmffile} &\multicolumn{3}{c|}{ $\{Q_1,Q_2\} \quad \{{u_{\mathbb{C}}}_3,{u_{\mathbb{C}}}_4\} \quad \{H_5,H_6\}$ } & \\ 
 \hline* & $(0,1,3,1)$ & $(1,\overline{6},3,\frac{1}{3})$ & $(1,6,1,-\frac{4}{3})$ & $4 \mathcal{O}^f_{{3}} + 8 \mathcal{O}^f_{{4}}$\\
\hline
\end{longtable}

\begin{landscape}
\begin{longtable}{cccc|c}
\caption{UV resonance of type $H{H^\dagger}LQ^3$. ``*'' in front of each row indicates topologies with nonrenormalizable vertices. The UV resonances colored in gray are eliminated by UV selection. }\\
\hline\hline\multicolumn{5}{l}{\multirow{2}{*}{\large Type: $H{H^\dagger}LQ^3$}}\\
 \multicolumn{5}{l}{} \\
\hline\multicolumn{5}{c}{$\begin{array}{ccl}\mathcal{O}^f_{1}&=&\frac{1}{6}\mathcal{Y}[\young(rst)]\epsilon^{abc}\epsilon^{kl}\epsilon^{mj}{{H^\dagger}^i}{H_m}({{L_p}_i}{Q_r}^{aj})({Q_s}^{bk}{Q_t}^{cl}),\\
\mathcal{O}^f_{2}&=&\frac{1}{6}\mathcal{Y}[\young(rst)]\epsilon^{abc}\epsilon^{ki}\epsilon^{mj}{{H^\dagger}^l}{H_m}({{L_p}_i}{Q_s}^{bk})({Q_r}^{aj}{Q_t}^{cl}),\\
\mathcal{O}^f_{3}&=&\frac{1}{3}\mathcal{Y}[\young(rs,t)]\epsilon^{abc}\epsilon^{kl}\epsilon^{mj}{{H^\dagger}^i}{H_m}({{L_p}_i}{Q_r}^{aj})({Q_s}^{bk}{Q_t}^{cl}),\\
\mathcal{O}^f_{4}&=&\frac{1}{3}\mathcal{Y}[\young(rs,t)]\epsilon^{abc}\epsilon^{ki}\epsilon^{mj}{{H^\dagger}^l}{H_m}({{L_p}_i}{Q_r}^{aj})({Q_s}^{bk}{Q_t}^{cl}),\\
\mathcal{O}^f_{5}&=&\frac{1}{3}\mathcal{Y}[\young(rs,t)]\epsilon^{abc}\epsilon^{ki}\epsilon^{mj}{{H^\dagger}^l}{H_m}({{L_p}_i}{Q_s}^{bk})({Q_r}^{aj}{Q_t}^{cl}),\\
\mathcal{O}^f_{6}&=&\frac{1}{6}\mathcal{Y}[\young(r,s,t)]\epsilon^{abc}\epsilon^{kl}\epsilon^{mj}{{H^\dagger}^i}{H_m}({{L_p}_i}{Q_s}^{bk})({Q_r}^{aj}{Q_t}^{cl}),\\
\mathcal{O}^f_{7}&=&\frac{1}{6}\mathcal{Y}[\young(r,s,t)]\epsilon^{abc}\epsilon^{ki}\epsilon^{mj}{{H^\dagger}^l}{H_m}({{L_p}_i}{Q_r}^{aj})({Q_s}^{bk}{Q_t}^{cl}).\end{array}$}\\
\hline
\multicolumn{4}{c|}{group: $(\text{Spin},SU(3)_c,SU(2)_w,U(1)_y)$} & \multirow{2}{*}{$\mathcal{O}^{j}$}\\
\hline
\end{longtable}
\end{landscape}

\begin{longtable}{cccc|c}
\caption{UV resonance of type ${B_L}{e_{\mathbb{C}}}H{H^\dagger}^2L$. ``*'' in front of each row indicates topologies with nonrenormalizable vertices. The UV resonances colored in gray are eliminated by UV selection. }\\
\hline\hline\multicolumn{5}{l}{\multirow{2}{*}{\large Type: ${B_L}{e_{\mathbb{C}}}H{H^\dagger}^2L$}}\\
 \multicolumn{5}{l}{} \\
\hline\multicolumn{5}{c}{$\begin{array}{ccl}\mathcal{O}^f_{1}&=&\frac{1}{2}\mathcal{Y}[\yng(2)_{{H^\dagger}}]{{B_L}^{\mu\nu}}{{H^\dagger}^i}{{H^\dagger}^j}{H_j}i({e_{\mathbb{C}}}_p\sigma_{\mu\nu}{{L_r}_i}).\end{array}$}\\
\hline
\multicolumn{4}{c|}{group: $(\text{Spin},SU(3)_c,SU(2)_w,U(1)_y)$} & \multirow{2}{*}{$\mathcal{O}^{j}$}\\
\begin{fmffile}{6-1}
\begin{fmfgraph*}(30,30)
\fmfpen{thin} \fmfsurroundn{e}{6} \fmfvn{}{g}{4} 
\fmf{plain}{e1,g1} \fmf{plain}{g1,e2} \fmf{plain}{g1,g4} \fmf{plain}{e3,g2} \fmf{plain}{g2,e4} \fmf{plain}{g2,g4} \fmf{plain}{e5,g3} \fmf{plain}{g3,g4} \fmf{plain}{g3,e6}
\end{fmfgraph*}
\end{fmffile} &\multicolumn{3}{c|}{ $\{{B_L}_1,{e_{\mathbb{C}}}_2\} \quad \{L_3,H_4\} \quad \{{H^\dagger}_5,{H^\dagger}_6\}$ } & \\ 
 \hline* & $(\frac{1}{2},1,1,1)$ & $(0,1,3,-1)$ & $(\frac{1}{2},1,3,0)$ & $-\mathcal{O}^f_{{1}}$\\ 
\hline
& \multicolumn{3}{c|}{ $\{{B_L}_1,{e_{\mathbb{C}}}_2\} \quad \{L_3,{H^\dagger}_5\} \quad \{H_4,{H^\dagger}_6\}$ } & $\mathcal{O}^{j}$\\
\hline
\end{longtable}

\begin{longtable}{cccc|c}
\caption{UV resonance of type ${B_L}{d_{\mathbb{C}}}H{H^\dagger}^2Q$. ``*'' in front of each row indicates topologies with nonrenormalizable vertices. The UV resonances colored in gray are eliminated by UV selection. }\\
\hline\hline\multicolumn{5}{l}{\multirow{2}{*}{\large Type: ${B_L}{d_{\mathbb{C}}}H{H^\dagger}^2Q$}}\\
 \multicolumn{5}{l}{} \\
\hline\multicolumn{5}{c}{$\begin{array}{ccl}\mathcal{O}^f_{1}&=&\frac{1}{2}\mathcal{Y}[\yng(2)_{{H^\dagger}}]{{B_L}^{\mu\nu}}{{H^\dagger}^i}{{H^\dagger}^j}{H_j}i({{{d_{\mathbb{C}}}_p}^a}\sigma_{\mu\nu}{Q_r}^{ai}).\end{array}$}\\
\hline
\multicolumn{4}{c|}{group: $(\text{Spin},SU(3)_c,SU(2)_w,U(1)_y)$} & \multirow{2}{*}{$\mathcal{O}^{j}$}\\
\begin{fmffile}{6-1}
\begin{fmfgraph*}(30,30)
\fmfpen{thin} \fmfsurroundn{e}{6} \fmfvn{}{g}{4} 
\fmf{plain}{e1,g1} \fmf{plain}{g1,e2} \fmf{plain}{g1,g4} \fmf{plain}{e3,g2} \fmf{plain}{g2,e4} \fmf{plain}{g2,g4} \fmf{plain}{e5,g3} \fmf{plain}{g3,g4} \fmf{plain}{g3,e6}
\end{fmfgraph*}
\end{fmffile} &\multicolumn{3}{c|}{ $\{{B_L}_1,{d_{\mathbb{C}}}_2\} \quad \{Q_3,H_4\} \quad \{{H^\dagger}_5,{H^\dagger}_6\}$ } & \\ 
 \hline* & $(\frac{1}{2},\overline{3},1,\frac{1}{3})$ & $(0,1,3,-1)$ & $(\frac{1}{2},3,3,\frac{2}{3})$ & $-\mathcal{O}^f_{{1}}$\\ 
\hline
& \multicolumn{3}{c|}{ $\{{B_L}_1,{d_{\mathbb{C}}}_2\} \quad \{Q_3,{H^\dagger}_5\} \quad \{H_4,{H^\dagger}_6\}$ } & $\mathcal{O}^{j}$\\
\hline
\end{longtable}

\begin{longtable}{cccc|c}
\caption{UV resonance of type ${B_L}H^2{H^\dagger}Q{u_{\mathbb{C}}}$. ``*'' in front of each row indicates topologies with nonrenormalizable vertices. The UV resonances colored in gray are eliminated by UV selection. }\\
\hline\hline\multicolumn{5}{l}{\multirow{2}{*}{\large Type: ${B_L}H^2{H^\dagger}Q{u_{\mathbb{C}}}$}}\\
 \multicolumn{5}{l}{} \\
\hline\multicolumn{5}{c}{$\begin{array}{ccl}\mathcal{O}^f_{1}&=&\frac{1}{2}\mathcal{Y}[\yng(2)_{H}]{{B_L}^{\mu\nu}}\epsilon^{ij}{{H^\dagger}^k}{H_j}{H_k}i({Q_p}^{ai}\sigma_{\mu\nu}{{{u_{\mathbb{C}}}_r}^a}).\end{array}$}\\
\hline
\multicolumn{4}{c|}{group: $(\text{Spin},SU(3)_c,SU(2)_w,U(1)_y)$} & \multirow{2}{*}{$\mathcal{O}^{j}$}\\
\begin{fmffile}{6-1}
\begin{fmfgraph*}(30,30)
\fmfpen{thin} \fmfsurroundn{e}{6} \fmfvn{}{g}{4} 
\fmf{plain}{e1,g1} \fmf{plain}{g1,e2} \fmf{plain}{g1,g4} \fmf{plain}{e3,g2} \fmf{plain}{g2,e4} \fmf{plain}{g2,g4} \fmf{plain}{e5,g3} \fmf{plain}{g3,g4} \fmf{plain}{g3,e6}
\end{fmfgraph*}
\end{fmffile} &\multicolumn{3}{c|}{ $\{{B_L}_1,Q_2\} \quad \{{u_{\mathbb{C}}}_3,H_4\} \quad \{H_5,{H^\dagger}_6\}$ } & \\ 
 \hline* & $(\frac{1}{2},3,2,\frac{1}{6})$ & $(0,1,3,0)$ & $(\frac{1}{2},\overline{3},2,-\frac{1}{6})$ & $\frac{1}{2} \mathcal{O}^f_{{1}}$\\ 
* & $(\frac{1}{2},3,2,\frac{1}{6})$ & $(0,1,1,0)$ & $(\frac{1}{2},\overline{3},2,-\frac{1}{6})$ & $-\frac{1}{2} \mathcal{O}^f_{{1}}$\\
\hline
\end{longtable}

\begin{longtable}{cccc|c}
\caption{UV resonance of type ${e_{\mathbb{C}}}H{H^\dagger}^2L{W_L}$. ``*'' in front of each row indicates topologies with nonrenormalizable vertices. The UV resonances colored in gray are eliminated by UV selection. }\\
\hline\hline\multicolumn{5}{l}{\multirow{2}{*}{\large Type: ${e_{\mathbb{C}}}H{H^\dagger}^2L{W_L}$}}\\
 \multicolumn{5}{l}{} \\
\hline\multicolumn{5}{c}{$\begin{array}{ccl}\mathcal{O}^f_{1}&=&\frac{1}{2}\mathcal{Y}[\yng(2)_{{H^\dagger}}]{{H^\dagger}^i}{{H^\dagger}^k}{H_j}i{{\tau^I}^j_k}{{W_L}^{I\mu\nu}}({e_{\mathbb{C}}}_p\sigma_{\mu\nu}{{L_r}_i}),\\
\mathcal{O}^f_{2}&=&\frac{1}{2}\mathcal{Y}[\yng(2)_{{H^\dagger}}]{{H^\dagger}^j}{{H^\dagger}^k}{H_j}i{{\tau^I}^i_k}{{W_L}^{I\mu\nu}}({e_{\mathbb{C}}}_p\sigma_{\mu\nu}{{L_r}_i}).\end{array}$}\\
\hline
\multicolumn{4}{c|}{group: $(\text{Spin},SU(3)_c,SU(2)_w,U(1)_y)$} & \multirow{2}{*}{$\mathcal{O}^{j}$}\\
\begin{fmffile}{6-1}
\begin{fmfgraph*}(30,30)
\fmfpen{thin} \fmfsurroundn{e}{6} \fmfvn{}{g}{4} 
\fmf{plain}{e1,g1} \fmf{plain}{g1,e2} \fmf{plain}{g1,g4} \fmf{plain}{e3,g2} \fmf{plain}{g2,e4} \fmf{plain}{g2,g4} \fmf{plain}{e5,g3} \fmf{plain}{g3,g4} \fmf{plain}{g3,e6}
\end{fmfgraph*}
\end{fmffile} &\multicolumn{3}{c|}{ $\{{W_L}_1,{e_{\mathbb{C}}}_2\} \quad \{L_3,H_4\} \quad \{{H^\dagger}_5,{H^\dagger}_6\}$ } & \\ 
 \hline* & $(0,1,3,-1)$ & $(\frac{1}{2},1,3,0)$ & $(\frac{1}{2},1,3,1)$ & $-\frac{1}{2} \mathcal{O}^f_{{1}} -\frac{1}{2} \mathcal{O}^f_{{2}}$\\ 
* & $(0,1,3,-1)$ & $(\frac{1}{2},1,1,0)$ & $(\frac{1}{2},1,3,1)$ & $\frac{1}{2} \mathcal{O}^f_{{1}} -\frac{1}{2} \mathcal{O}^f_{{2}}$\\
\hline
\end{longtable}

\begin{longtable}{cccc|c}
\caption{UV resonance of type ${d_{\mathbb{C}}}H{H^\dagger}^2Q{W_L}$. ``*'' in front of each row indicates topologies with nonrenormalizable vertices. The UV resonances colored in gray are eliminated by UV selection. }\\
\hline\hline\multicolumn{5}{l}{\multirow{2}{*}{\large Type: ${d_{\mathbb{C}}}H{H^\dagger}^2Q{W_L}$}}\\
 \multicolumn{5}{l}{} \\
\hline\multicolumn{5}{c}{$\begin{array}{ccl}\mathcal{O}^f_{1}&=&\frac{1}{2}\mathcal{Y}[\yng(2)_{{H^\dagger}}]{{H^\dagger}^i}{{H^\dagger}^k}{H_j}i{{\tau^I}^j_k}{{W_L}^{I\mu\nu}}({{{d_{\mathbb{C}}}_p}^a}\sigma_{\mu\nu}{Q_r}^{ai}),\\
\mathcal{O}^f_{2}&=&\frac{1}{2}\mathcal{Y}[\yng(2)_{{H^\dagger}}]{{H^\dagger}^j}{{H^\dagger}^k}{H_j}i{{\tau^I}^i_k}{{W_L}^{I\mu\nu}}({{{d_{\mathbb{C}}}_p}^a}\sigma_{\mu\nu}{Q_r}^{ai}).\end{array}$}\\
\hline
\multicolumn{4}{c|}{group: $(\text{Spin},SU(3)_c,SU(2)_w,U(1)_y)$} & \multirow{2}{*}{$\mathcal{O}^{j}$}\\
\begin{fmffile}{6-1}
\begin{fmfgraph*}(30,30)
\fmfpen{thin} \fmfsurroundn{e}{6} \fmfvn{}{g}{4} 
\fmf{plain}{e1,g1} \fmf{plain}{g1,e2} \fmf{plain}{g1,g4} \fmf{plain}{e3,g2} \fmf{plain}{g2,e4} \fmf{plain}{g2,g4} \fmf{plain}{e5,g3} \fmf{plain}{g3,g4} \fmf{plain}{g3,e6}
\end{fmfgraph*}
\end{fmffile} &\multicolumn{3}{c|}{ $\{{W_L}_1,{d_{\mathbb{C}}}_2\} \quad \{Q_3,H_4\} \quad \{{H^\dagger}_5,{H^\dagger}_6\}$ } & \\ 
 \hline* & $(0,1,3,-1)$ & $(\frac{1}{2},3,3,\frac{2}{3})$ & $(\frac{1}{2},\overline{3},3,\frac{1}{3})$ & $-\frac{1}{2} \mathcal{O}^f_{{1}} -\frac{1}{2} \mathcal{O}^f_{{2}}$\\ 
* & $(0,1,3,-1)$ & $(\frac{1}{2},3,1,\frac{2}{3})$ & $(\frac{1}{2},\overline{3},3,\frac{1}{3})$ & $\frac{1}{2} \mathcal{O}^f_{{1}} -\frac{1}{2} \mathcal{O}^f_{{2}}$\\
\hline
\end{longtable}

\begin{longtable}{cccc|c}
\caption{UV resonance of type $H^2{H^\dagger}Q{u_{\mathbb{C}}}{W_L}$. ``*'' in front of each row indicates topologies with nonrenormalizable vertices. The UV resonances colored in gray are eliminated by UV selection. }\\
\hline\hline\multicolumn{5}{l}{\multirow{2}{*}{\large Type: $H^2{H^\dagger}Q{u_{\mathbb{C}}}{W_L}$}}\\
 \multicolumn{5}{l}{} \\
\hline\multicolumn{5}{c}{$\begin{array}{ccl}\mathcal{O}^f_{1}&=&\frac{1}{2}\mathcal{Y}[\yng(2)_{H}]\epsilon^{im}{{H^\dagger}^k}{H_j}{H_k}i{{\tau^I}^j_m}{{W_L}^{I\mu\nu}}({Q_p}^{ai}\sigma_{\mu\nu}{{{u_{\mathbb{C}}}_r}^a}),\\
\mathcal{O}^f_{2}&=&\frac{1}{2}\mathcal{Y}[\yng(2)_{H}]\epsilon^{ik}{{H^\dagger}^l}{H_j}{H_k}i{{\tau^I}^j_l}{{W_L}^{I\mu\nu}}({Q_p}^{ai}\sigma_{\mu\nu}{{{u_{\mathbb{C}}}_r}^a}).\end{array}$}\\
\hline
\multicolumn{4}{c|}{group: $(\text{Spin},SU(3)_c,SU(2)_w,U(1)_y)$} & \multirow{2}{*}{$\mathcal{O}^{j}$}\\
\begin{fmffile}{6-1}
\begin{fmfgraph*}(30,30)
\fmfpen{thin} \fmfsurroundn{e}{6} \fmfvn{}{g}{4} 
\fmf{plain}{e1,g1} \fmf{plain}{g1,e2} \fmf{plain}{g1,g4} \fmf{plain}{e3,g2} \fmf{plain}{g2,e4} \fmf{plain}{g2,g4} \fmf{plain}{e5,g3} \fmf{plain}{g3,g4} \fmf{plain}{g3,e6}
\end{fmfgraph*}
\end{fmffile} &\multicolumn{3}{c|}{ $\{{W_L}_1,Q_2\} \quad \{{u_{\mathbb{C}}}_3,H_4\} \quad \{H_5,{H^\dagger}_6\}$ } & \\ 
 \hline* & $(0,1,3,0)$ & $(\frac{1}{2},\overline{3},2,-\frac{1}{6})$ & $(\frac{1}{2},3,4,\frac{1}{6})$ & $-\mathcal{O}^f_{{1}} + 3 \mathcal{O}^f_{{2}}$\\ 
* & $(0,1,3,0)$ & $(\frac{1}{2},\overline{3},2,-\frac{1}{6})$ & $(\frac{1}{2},3,2,\frac{1}{6})$ & $\frac{1}{2} \mathcal{O}^f_{{1}}$\\
\hline
\end{longtable}

\begin{longtable}{cccc|c}
\caption{UV resonance of type ${d_{\mathbb{C}}}{G_L}H{H^\dagger}^2Q$. ``*'' in front of each row indicates topologies with nonrenormalizable vertices. The UV resonances colored in gray are eliminated by UV selection. }\\
\hline\hline\multicolumn{5}{l}{\multirow{2}{*}{\large Type: ${d_{\mathbb{C}}}{G_L}H{H^\dagger}^2Q$}}\\
 \multicolumn{5}{l}{} \\
\hline\multicolumn{5}{c}{$\begin{array}{ccl}\mathcal{O}^f_{1}&=&\frac{1}{2}\mathcal{Y}[\yng(2)_{{H^\dagger}}]{{G_L}^{A\mu\nu}}{{H^\dagger}^i}{{H^\dagger}^j}{H_j}i{{\lambda^A}_a^b}({{{d_{\mathbb{C}}}_p}^a}\sigma_{\mu\nu}{Q_r}^{bi}).\end{array}$}\\
\hline
\multicolumn{4}{c|}{group: $(\text{Spin},SU(3)_c,SU(2)_w,U(1)_y)$} & \multirow{2}{*}{$\mathcal{O}^{j}$}\\
\begin{fmffile}{6-1}
\begin{fmfgraph*}(30,30)
\fmfpen{thin} \fmfsurroundn{e}{6} \fmfvn{}{g}{4} 
\fmf{plain}{e1,g1} \fmf{plain}{g1,e2} \fmf{plain}{g1,g4} \fmf{plain}{e3,g2} \fmf{plain}{g2,e4} \fmf{plain}{g2,g4} \fmf{plain}{e5,g3} \fmf{plain}{g3,g4} \fmf{plain}{g3,e6}
\end{fmfgraph*}
\end{fmffile} &\multicolumn{3}{c|}{ $\{{G_L}_1,{d_{\mathbb{C}}}_2\} \quad \{Q_3,H_4\} \quad \{{H^\dagger}_5,{H^\dagger}_6\}$ } & \\ 
 \hline* & $(\frac{1}{2},\overline{3},1,\frac{1}{3})$ & $(0,1,3,-1)$ & $(\frac{1}{2},3,3,\frac{2}{3})$ & $-\mathcal{O}^f_{{1}}$\\ 
\hline
& \multicolumn{3}{c|}{ $\{{G_L}_1,{d_{\mathbb{C}}}_2\} \quad \{Q_3,{H^\dagger}_5\} \quad \{H_4,{H^\dagger}_6\}$ } & $\mathcal{O}^{j}$\\
\hline
\end{longtable}

\begin{longtable}{cccc|c}
\caption{UV resonance of type ${G_L}H^2{H^\dagger}Q{u_{\mathbb{C}}}$. ``*'' in front of each row indicates topologies with nonrenormalizable vertices. The UV resonances colored in gray are eliminated by UV selection. }\\
\hline\hline\multicolumn{5}{l}{\multirow{2}{*}{\large Type: ${G_L}H^2{H^\dagger}Q{u_{\mathbb{C}}}$}}\\
 \multicolumn{5}{l}{} \\
\hline\multicolumn{5}{c}{$\begin{array}{ccl}\mathcal{O}^f_{1}&=&\frac{1}{2}\mathcal{Y}[\yng(2)_{H}]\epsilon^{ij}{{G_L}^{A\mu\nu}}{{H^\dagger}^k}{H_j}{H_k}i{{\lambda^A}_b^a}({Q_p}^{ai}\sigma_{\mu\nu}{{{u_{\mathbb{C}}}_r}^b}).\end{array}$}\\
\hline
\multicolumn{4}{c|}{group: $(\text{Spin},SU(3)_c,SU(2)_w,U(1)_y)$} & \multirow{2}{*}{$\mathcal{O}^{j}$}\\
\begin{fmffile}{6-1}
\begin{fmfgraph*}(30,30)
\fmfpen{thin} \fmfsurroundn{e}{6} \fmfvn{}{g}{4} 
\fmf{plain}{e1,g1} \fmf{plain}{g1,e2} \fmf{plain}{g1,g4} \fmf{plain}{e3,g2} \fmf{plain}{g2,e4} \fmf{plain}{g2,g4} \fmf{plain}{e5,g3} \fmf{plain}{g3,g4} \fmf{plain}{g3,e6}
\end{fmfgraph*}
\end{fmffile} &\multicolumn{3}{c|}{ $\{{G_L}_1,Q_2\} \quad \{{u_{\mathbb{C}}}_3,H_4\} \quad \{H_5,{H^\dagger}_6\}$ } & \\ 
 \hline* & $(\frac{1}{2},3,2,\frac{1}{6})$ & $(0,1,3,0)$ & $(\frac{1}{2},\overline{3},2,-\frac{1}{6})$ & $\frac{1}{2} \mathcal{O}^f_{{1}}$\\ 
* & $(\frac{1}{2},3,2,\frac{1}{6})$ & $(0,1,1,0)$ & $(\frac{1}{2},\overline{3},2,-\frac{1}{6})$ & $-\frac{1}{2} \mathcal{O}^f_{{1}}$\\
\hline
\end{longtable}

\begin{longtable}{cccc|c}
\caption{UV resonance of type ${B_L}^2H^2{H^\dagger}^2$. ``*'' in front of each row indicates topologies with nonrenormalizable vertices. The UV resonances colored in gray are eliminated by UV selection. }\\
\hline\hline\multicolumn{5}{l}{\multirow{2}{*}{\large Type: ${B_L}^2H^2{H^\dagger}^2$}}\\
 \multicolumn{5}{l}{} \\
\hline\multicolumn{5}{c}{$\begin{array}{ccl}\mathcal{O}^f_{1}&=&\frac{1}{8}\mathcal{Y}[\yng(2)_{{B_L}}]\mathcal{Y}[\yng(2)_{H}]\mathcal{Y}[\yng(2)_{{H^\dagger}}]{{B_L}^{\mu\nu}}{{B_L}^{\mu\nu}}{{H^\dagger}^i}{{H^\dagger}^j}{H_i}{H_j}.\end{array}$}\\
\hline
\multicolumn{4}{c|}{group: $(\text{Spin},SU(3)_c,SU(2)_w,U(1)_y)$} & \multirow{2}{*}{$\mathcal{O}^{j}$}\\
\begin{fmffile}{6-1}
\begin{fmfgraph*}(30,30)
\fmfpen{thin} \fmfsurroundn{e}{6} \fmfvn{}{g}{4} 
\fmf{plain}{e1,g1} \fmf{plain}{g1,e2} \fmf{plain}{g1,g4} \fmf{plain}{e3,g2} \fmf{plain}{g2,e4} \fmf{plain}{g2,g4} \fmf{plain}{e5,g3} \fmf{plain}{g3,g4} \fmf{plain}{g3,e6}
\end{fmfgraph*}
\end{fmffile} &\multicolumn{3}{c|}{ $\{{B_L}_1,{B_L}_2\} \quad \{H_3,H_4\} \quad \{{H^\dagger}_5,{H^\dagger}_6\}$ } & \\ 
 \hline* & $(0,1,1,0)$ & $(0,1,3,1)$ & $(0,1,3,-1)$ & $4 \mathcal{O}^f_{{1}}$\\ 
\hline
& \multicolumn{3}{c|}{ $\{{B_L}_1,{B_L}_2\} \quad \{H_3,{H^\dagger}_5\} \quad \{H_4,{H^\dagger}_6\}$ } & $\mathcal{O}^{j}$\\
\hline
\end{longtable}

\begin{longtable}{cccc|c}
\caption{UV resonance of type ${B_L}H^2{H^\dagger}^2{W_L}$. ``*'' in front of each row indicates topologies with nonrenormalizable vertices. The UV resonances colored in gray are eliminated by UV selection. }\\
\hline\hline\multicolumn{5}{l}{\multirow{2}{*}{\large Type: ${B_L}H^2{H^\dagger}^2{W_L}$}}\\
 \multicolumn{5}{l}{} \\
\hline\multicolumn{5}{c}{$\begin{array}{ccl}\mathcal{O}^f_{1}&=&\frac{1}{4}\mathcal{Y}[\yng(2)_{H}]\mathcal{Y}[\yng(2)_{{H^\dagger}}]{{B_L}^{\mu\nu}}{{H^\dagger}^i}{{H^\dagger}^k}{H_i}{H_j}{{\tau^I}^j_k}{{W_L}^{I\mu\nu}}.\end{array}$}\\
\hline
\multicolumn{4}{c|}{group: $(\text{Spin},SU(3)_c,SU(2)_w,U(1)_y)$} & \multirow{2}{*}{$\mathcal{O}^{j}$}\\
\begin{fmffile}{6-1}
\begin{fmfgraph*}(30,30)
\fmfpen{thin} \fmfsurroundn{e}{6} \fmfvn{}{g}{4} 
\fmf{plain}{e1,g1} \fmf{plain}{g1,e2} \fmf{plain}{g1,g4} \fmf{plain}{e3,g2} \fmf{plain}{g2,e4} \fmf{plain}{g2,g4} \fmf{plain}{e5,g3} \fmf{plain}{g3,g4} \fmf{plain}{g3,e6}
\end{fmfgraph*}
\end{fmffile} &\multicolumn{3}{c|}{ $\{{B_L}_1,{W_L}_2\} \quad \{H_3,H_4\} \quad \{{H^\dagger}_5,{H^\dagger}_6\}$ } & \\ 
 \hline* & $(0,1,3,0)$ & $(0,1,3,1)$ & $(0,1,3,-1)$ & $4 \mathcal{O}^f_{{1}}$\\ 
\hline
& \multicolumn{3}{c|}{ $\{{B_L}_1,{W_L}_2\} \quad \{H_3,{H^\dagger}_5\} \quad \{H_4,{H^\dagger}_6\}$ } & $\mathcal{O}^{j}$\\
\hline
\end{longtable}

\begin{longtable}{cccc|c}
\caption{UV resonance of type $H^2{H^\dagger}^2{W_L}^2$. ``*'' in front of each row indicates topologies with nonrenormalizable vertices. The UV resonances colored in gray are eliminated by UV selection. }\\
\hline\hline\multicolumn{5}{l}{\multirow{2}{*}{\large Type: $H^2{H^\dagger}^2{W_L}^2$}}\\
 \multicolumn{5}{l}{} \\
\hline\multicolumn{5}{c}{$\begin{array}{ccl}\mathcal{O}^f_{1}&=&\frac{1}{8}\mathcal{Y}[\yng(2)_{H}]\mathcal{Y}[\yng(2)_{{H^\dagger}}]\mathcal{Y}[\yng(2)_{{W_L}}]{{H^\dagger}^k}{{H^\dagger}^l}{H_i}{H_j}{{\tau^I}^i_k}{{\tau^J}^j_l}{{W_L}^{I\mu\nu}}{{W_L}^{J\mu\nu}},\\
\mathcal{O}^f_{2}&=&\frac{1}{8}\mathcal{Y}[\yng(2)_{H}]\mathcal{Y}[\yng(2)_{{H^\dagger}}]\mathcal{Y}[\yng(2)_{{W_L}}]{{H^\dagger}^i}{{H^\dagger}^j}{H_i}{H_j}{{W_L}^{I\mu\nu}}{{W_L}^{I\mu\nu}}.\end{array}$}\\
\hline
\multicolumn{4}{c|}{group: $(\text{Spin},SU(3)_c,SU(2)_w,U(1)_y)$} & \multirow{2}{*}{$\mathcal{O}^{j}$}\\
\begin{fmffile}{6-1}
\begin{fmfgraph*}(30,30)
\fmfpen{thin} \fmfsurroundn{e}{6} \fmfvn{}{g}{4} 
\fmf{plain}{e1,g1} \fmf{plain}{g1,e2} \fmf{plain}{g1,g4} \fmf{plain}{e3,g2} \fmf{plain}{g2,e4} \fmf{plain}{g2,g4} \fmf{plain}{e5,g3} \fmf{plain}{g3,g4} \fmf{plain}{g3,e6}
\end{fmfgraph*}
\end{fmffile} &\multicolumn{3}{c|}{ $\{{W_L}_1,{W_L}_2\} \quad \{H_3,H_4\} \quad \{{H^\dagger}_5,{H^\dagger}_6\}$ } & \\ 
 \hline* & $(0,1,3,1)$ & $(0,1,3,-1)$ & $(0,1,5,0)$ & $-12 \mathcal{O}^f_{{1}} + 4 \mathcal{O}^f_{{2}}$\\ 
* & $(0,1,3,1)$ & $(0,1,3,-1)$ & $(0,1,1,0)$ & $4 \mathcal{O}^f_{{2}}$\\
\hline
\end{longtable}

\begin{longtable}{cccc|c}
\caption{UV resonance of type ${G_L}^2H^2{H^\dagger}^2$. ``*'' in front of each row indicates topologies with nonrenormalizable vertices. The UV resonances colored in gray are eliminated by UV selection. }\\
\hline\hline\multicolumn{5}{l}{\multirow{2}{*}{\large Type: ${G_L}^2H^2{H^\dagger}^2$}}\\
 \multicolumn{5}{l}{} \\
\hline\multicolumn{5}{c}{$\begin{array}{ccl}\mathcal{O}^f_{1}&=&\frac{1}{8}\mathcal{Y}[\yng(2)_{{G_L}}]\mathcal{Y}[\yng(2)_{H}]\mathcal{Y}[\yng(2)_{{H^\dagger}}]{{G_L}^{A\mu\nu}}{{G_L}^{A\mu\nu}}{{H^\dagger}^i}{{H^\dagger}^j}{H_i}{H_j}.\end{array}$}\\
\hline
\multicolumn{4}{c|}{group: $(\text{Spin},SU(3)_c,SU(2)_w,U(1)_y)$} & \multirow{2}{*}{$\mathcal{O}^{j}$}\\
\begin{fmffile}{6-1}
\begin{fmfgraph*}(30,30)
\fmfpen{thin} \fmfsurroundn{e}{6} \fmfvn{}{g}{4} 
\fmf{plain}{e1,g1} \fmf{plain}{g1,e2} \fmf{plain}{g1,g4} \fmf{plain}{e3,g2} \fmf{plain}{g2,e4} \fmf{plain}{g2,g4} \fmf{plain}{e5,g3} \fmf{plain}{g3,g4} \fmf{plain}{g3,e6}
\end{fmfgraph*}
\end{fmffile} &\multicolumn{3}{c|}{ $\{{G_L}_1,{G_L}_2\} \quad \{H_3,H_4\} \quad \{{H^\dagger}_5,{H^\dagger}_6\}$ } & \\ 
 \hline* & $(0,1,1,0)$ & $(0,1,3,1)$ & $(0,1,3,-1)$ & $4 \mathcal{O}^f_{{1}}$\\ 
\hline
& \multicolumn{3}{c|}{ $\{{G_L}_1,{G_L}_2\} \quad \{H_3,{H^\dagger}_5\} \quad \{H_4,{H^\dagger}_6\}$ } & $\mathcal{O}^{j}$\\
\hline
\end{longtable}

\begin{longtable}{cccc|c}
\caption{UV resonance of type ${e_{\mathbb{C}}}^2{{e_{\mathbb{C}}}^\dagger}^2H{H^\dagger}$. ``*'' in front of each row indicates topologies with nonrenormalizable vertices. The UV resonances colored in gray are eliminated by UV selection. }\\
\hline\hline\multicolumn{5}{l}{\multirow{2}{*}{\large Type: ${e_{\mathbb{C}}}^2{{e_{\mathbb{C}}}^\dagger}^2H{H^\dagger}$}}\\
 \multicolumn{5}{l}{} \\
\hline\multicolumn{5}{c}{$\begin{array}{ccl}\mathcal{O}^f_{{1}}&=& ({e_{\mathbb{C}}}_p{e_{\mathbb{C}}}_r)({{e_{\mathbb{C}}}^\dagger}_s{{e_{\mathbb{C}}}^\dagger}_t){H_i}{{H^\dagger}^i}.\end{array}$}\\
\hline
\multicolumn{4}{c|}{group: $(\text{Spin},SU(3)_c,SU(2)_w,U(1)_y)$} & \multirow{2}{*}{$\mathcal{O}^{j}$}\\
\begin{fmffile}{6-1}
\begin{fmfgraph*}(30,30)
\fmfpen{thin} \fmfsurroundn{e}{6} \fmfvn{}{g}{4} 
\fmf{plain}{e1,g1} \fmf{plain}{g1,e2} \fmf{plain}{g1,g4} \fmf{plain}{e3,g2} \fmf{plain}{g2,e4} \fmf{plain}{g2,g4} \fmf{plain}{e5,g3} \fmf{plain}{g3,g4} \fmf{plain}{g3,e6}
\end{fmfgraph*}
\end{fmffile} &\multicolumn{3}{c|}{ $\{{e_{\mathbb{C}}}_1,{e_{\mathbb{C}}}_2\} \quad \{H_3,{H^\dagger}_4\} \quad \{{{e_{\mathbb{C}}}^\dagger}_5,{{e_{\mathbb{C}}}^\dagger}_6\}$ } & \\ 
 \hline& $(0,1,1,2)$ & $(0,1,1,-2)$ & $(0,1,1,0)$ & $\mathcal{O}^f_{1}$\\ 
\hline
& \multicolumn{3}{c|}{ $\{{e_{\mathbb{C}}}_1,{e_{\mathbb{C}}}_2\} \quad \{H_3,{{e_{\mathbb{C}}}^\dagger}_5\} \quad \{{H^\dagger}_4,{{e_{\mathbb{C}}}^\dagger}_6\}$ } & $\mathcal{O}^{j}$\\
\hline
\end{longtable}

\begin{longtable}{cccc|c}
\caption{UV resonance of type ${e_{\mathbb{C}}}{H^\dagger}^2L{Q^\dagger}{{u_{\mathbb{C}}}^\dagger}$. ``*'' in front of each row indicates topologies with nonrenormalizable vertices. The UV resonances colored in gray are eliminated by UV selection. }\\
\hline\hline\multicolumn{5}{l}{\multirow{2}{*}{\large Type: ${e_{\mathbb{C}}}{H^\dagger}^2L{Q^\dagger}{{u_{\mathbb{C}}}^\dagger}$}}\\
 \multicolumn{5}{l}{} \\
\hline\multicolumn{5}{c}{$\begin{array}{ccl}\mathcal{O}^f_{1}&=&\frac{1}{2}\mathcal{Y}[\yng(2)_{{H^\dagger}}]\epsilon_{lj}{{H^\dagger}^i}{{H^\dagger}^j}({e_{\mathbb{C}}}_p{{L_r}_i})({{Q^\dagger}_s}^{al}{{{{u_{\mathbb{C}}}^\dagger}_t}_a}).\end{array}$}\\
\hline
\multicolumn{4}{c|}{group: $(\text{Spin},SU(3)_c,SU(2)_w,U(1)_y)$} & \multirow{2}{*}{$\mathcal{O}^{j}$}\\
\begin{fmffile}{6-1}
\begin{fmfgraph*}(30,30)
\fmfpen{thin} \fmfsurroundn{e}{6} \fmfvn{}{g}{4} 
\fmf{plain}{e1,g1} \fmf{plain}{g1,e2} \fmf{plain}{g1,g4} \fmf{plain}{e3,g2} \fmf{plain}{g2,e4} \fmf{plain}{g2,g4} \fmf{plain}{e5,g3} \fmf{plain}{g3,g4} \fmf{plain}{g3,e6}
\end{fmfgraph*}
\end{fmffile} &\multicolumn{3}{c|}{ $\{{e_{\mathbb{C}}}_1,L_2\} \quad \{{H^\dagger}_3,{H^\dagger}_4\} \quad \{{Q^\dagger}_5,{{u_{\mathbb{C}}}^\dagger}_6\}$ } & \\ 
 \hline& $(0,1,2,\frac{1}{2})$ & $(0,1,3,-1)$ & $(0,1,2,\frac{1}{2})$ & $-2 \mathcal{O}^f_{{1}}$\\ 
\hline
& \multicolumn{3}{c|}{ $\{{e_{\mathbb{C}}}_1,L_2\} \quad \{{H^\dagger}_3,{Q^\dagger}_5\} \quad \{{H^\dagger}_4,{{u_{\mathbb{C}}}^\dagger}_6\}$ } & $\mathcal{O}^{j}$\\
\hline
\end{longtable}

\begin{longtable}{cccc|c}
\caption{UV resonance of type ${d_{\mathbb{C}}}{H^\dagger}^2L{L^\dagger}{{u_{\mathbb{C}}}^\dagger}$. ``*'' in front of each row indicates topologies with nonrenormalizable vertices. The UV resonances colored in gray are eliminated by UV selection. }\\
\hline\hline\multicolumn{5}{l}{\multirow{2}{*}{\large Type: ${d_{\mathbb{C}}}{H^\dagger}^2L{L^\dagger}{{u_{\mathbb{C}}}^\dagger}$}}\\
 \multicolumn{5}{l}{} \\
\hline\multicolumn{5}{c}{$\begin{array}{ccl}\mathcal{O}^f_{1}&=&\frac{1}{2}\mathcal{Y}[\yng(2)_{{H^\dagger}}]\epsilon_{lj}{{H^\dagger}^i}{{H^\dagger}^j}({{{d_{\mathbb{C}}}_p}^a}{{L_r}_i})({{{L^\dagger}_s}^l}{{{{u_{\mathbb{C}}}^\dagger}_t}_a}).\end{array}$}\\
\hline
\multicolumn{4}{c|}{group: $(\text{Spin},SU(3)_c,SU(2)_w,U(1)_y)$} & \multirow{2}{*}{$\mathcal{O}^{j}$}\\
\begin{fmffile}{6-1}
\begin{fmfgraph*}(30,30)
\fmfpen{thin} \fmfsurroundn{e}{6} \fmfvn{}{g}{4} 
\fmf{plain}{e1,g1} \fmf{plain}{g1,e2} \fmf{plain}{g1,g4} \fmf{plain}{e3,g2} \fmf{plain}{g2,e4} \fmf{plain}{g2,g4} \fmf{plain}{e5,g3} \fmf{plain}{g3,g4} \fmf{plain}{g3,e6}
\end{fmfgraph*}
\end{fmffile} &\multicolumn{3}{c|}{ $\{{d_{\mathbb{C}}}_1,L_2\} \quad \{{H^\dagger}_3,{H^\dagger}_4\} \quad \{{L^\dagger}_5,{{u_{\mathbb{C}}}^\dagger}_6\}$ } & \\ 
 \hline& $(0,\overline{3},2,-\frac{1}{6})$ & $(0,1,3,-1)$ & $(0,3,2,\frac{7}{6})$ & $-2 \mathcal{O}^f_{{1}}$\\ 
\hline
& \multicolumn{3}{c|}{ $\{{d_{\mathbb{C}}}_1,L_2\} \quad \{{H^\dagger}_3,{L^\dagger}_5\} \quad \{{H^\dagger}_4,{{u_{\mathbb{C}}}^\dagger}_6\}$ } & $\mathcal{O}^{j}$\\
\hline
\end{longtable}

\begin{longtable}{cccc|c}
\caption{UV resonance of type ${e_{\mathbb{C}}}{{e_{\mathbb{C}}}^\dagger}H{H^\dagger}L{L^\dagger}$. ``*'' in front of each row indicates topologies with nonrenormalizable vertices. The UV resonances colored in gray are eliminated by UV selection. }\\
\hline\hline\multicolumn{5}{l}{\multirow{2}{*}{\large Type: ${e_{\mathbb{C}}}{{e_{\mathbb{C}}}^\dagger}H{H^\dagger}L{L^\dagger}$}}\\
 \multicolumn{5}{l}{} \\
\hline\multicolumn{5}{c}{$\begin{array}{ccl}\mathcal{O}^f_{{1}}&=& ({e_{\mathbb{C}}}_p{{L_r}_i})({{e_{\mathbb{C}}}^\dagger}_s{{{L^\dagger}_t}^j}){H_j}{{H^\dagger}^i},\\
\mathcal{O}^f_{{2}}&=& ({e_{\mathbb{C}}}_p{{L_r}_i})({{e_{\mathbb{C}}}^\dagger}_s{{{L^\dagger}_t}^i}){H_j}{{H^\dagger}^j}.\end{array}$}\\
\hline
\multicolumn{4}{c|}{group: $(\text{Spin},SU(3)_c,SU(2)_w,U(1)_y)$} & \multirow{2}{*}{$\mathcal{O}^{j}$}\\
\begin{fmffile}{6-1}
\begin{fmfgraph*}(30,30)
\fmfpen{thin} \fmfsurroundn{e}{6} \fmfvn{}{g}{4} 
\fmf{plain}{e1,g1} \fmf{plain}{g1,e2} \fmf{plain}{g1,g4} \fmf{plain}{e3,g2} \fmf{plain}{g2,e4} \fmf{plain}{g2,g4} \fmf{plain}{e5,g3} \fmf{plain}{g3,g4} \fmf{plain}{g3,e6}
\end{fmfgraph*}
\end{fmffile} &\multicolumn{3}{c|}{ $\{{e_{\mathbb{C}}}_1,L_2\} \quad \{H_3,{H^\dagger}_4\} \quad \{{{e_{\mathbb{C}}}^\dagger}_5,{L^\dagger}_6\}$ } & \\ 
 \hline& $(0,1,2,\frac{1}{2})$ & $(0,1,2,-\frac{1}{2})$ & $(0,1,3,0)$ & $-2 \mathcal{O}^f_{1} + \mathcal{O}^f_{2}$\\ 
& $(0,1,2,\frac{1}{2})$ & $(0,1,2,-\frac{1}{2})$ & $(0,1,1,0)$ & $\mathcal{O}^f_{2}$\\
\hline
\end{longtable}

\begin{longtable}{cccc|c}
\caption{UV resonance of type ${{d_{\mathbb{C}}}^\dagger}{e_{\mathbb{C}}}H{H^\dagger}L{Q^\dagger}$. ``*'' in front of each row indicates topologies with nonrenormalizable vertices. The UV resonances colored in gray are eliminated by UV selection. }\\
\hline\hline\multicolumn{5}{l}{\multirow{2}{*}{\large Type: ${{d_{\mathbb{C}}}^\dagger}{e_{\mathbb{C}}}H{H^\dagger}L{Q^\dagger}$}}\\
 \multicolumn{5}{l}{} \\
\hline\multicolumn{5}{c}{$\begin{array}{ccl}\mathcal{O}^f_{{1}}&=& ({e_{\mathbb{C}}}_p{{L_r}_i})({{{{d_{\mathbb{C}}}^\dagger}_s}_a}{{Q^\dagger}_t}^{aj}){H_j}{{H^\dagger}^i},\\
\mathcal{O}^f_{{2}}&=& ({e_{\mathbb{C}}}_p{{L_r}_i})({{{{d_{\mathbb{C}}}^\dagger}_s}_a}{{Q^\dagger}_t}^{ai}){H_j}{{H^\dagger}^j}.\end{array}$}\\
\hline
\multicolumn{4}{c|}{group: $(\text{Spin},SU(3)_c,SU(2)_w,U(1)_y)$} & \multirow{2}{*}{$\mathcal{O}^{j}$}\\
\begin{fmffile}{6-1}
\begin{fmfgraph*}(30,30)
\fmfpen{thin} \fmfsurroundn{e}{6} \fmfvn{}{g}{4} 
\fmf{plain}{e1,g1} \fmf{plain}{g1,e2} \fmf{plain}{g1,g4} \fmf{plain}{e3,g2} \fmf{plain}{g2,e4} \fmf{plain}{g2,g4} \fmf{plain}{e5,g3} \fmf{plain}{g3,g4} \fmf{plain}{g3,e6}
\end{fmfgraph*}
\end{fmffile} &\multicolumn{3}{c|}{ $\{{e_{\mathbb{C}}}_1,L_2\} \quad \{H_3,{H^\dagger}_4\} \quad \{{{d_{\mathbb{C}}}^\dagger}_5,{Q^\dagger}_6\}$ } & \\ 
 \hline& $(0,1,2,-\frac{1}{2})$ & $(0,1,2,\frac{1}{2})$ & $(0,1,3,0)$ & $-2 \mathcal{O}^f_{1} + \mathcal{O}^f_{2}$\\ 
& $(0,1,2,-\frac{1}{2})$ & $(0,1,2,\frac{1}{2})$ & $(0,1,1,0)$ & $\mathcal{O}^f_{2}$\\
\hline
\end{longtable}

\begin{longtable}{cccc|c}
\caption{UV resonance of type ${d_{\mathbb{C}}}{{d_{\mathbb{C}}}^\dagger}H{H^\dagger}L{L^\dagger}$. ``*'' in front of each row indicates topologies with nonrenormalizable vertices. The UV resonances colored in gray are eliminated by UV selection. }\\
\hline\hline\multicolumn{5}{l}{\multirow{2}{*}{\large Type: ${d_{\mathbb{C}}}{{d_{\mathbb{C}}}^\dagger}H{H^\dagger}L{L^\dagger}$}}\\
 \multicolumn{5}{l}{} \\
\hline\multicolumn{5}{c}{$\begin{array}{ccl}\mathcal{O}^f_{{1}}&=& ({{{d_{\mathbb{C}}}_p}^a}{{L_r}_i})({{{{d_{\mathbb{C}}}^\dagger}_s}_a}{{{L^\dagger}_t}^j}){H_j}{{H^\dagger}^i},\\
\mathcal{O}^f_{{2}}&=& ({{{d_{\mathbb{C}}}_p}^a}{{L_r}_i})({{{{d_{\mathbb{C}}}^\dagger}_s}_a}{{{L^\dagger}_t}^i}){H_j}{{H^\dagger}^j}.\end{array}$}\\
\hline
\multicolumn{4}{c|}{group: $(\text{Spin},SU(3)_c,SU(2)_w,U(1)_y)$} & \multirow{2}{*}{$\mathcal{O}^{j}$}\\
\begin{fmffile}{6-1}
\begin{fmfgraph*}(30,30)
\fmfpen{thin} \fmfsurroundn{e}{6} \fmfvn{}{g}{4} 
\fmf{plain}{e1,g1} \fmf{plain}{g1,e2} \fmf{plain}{g1,g4} \fmf{plain}{e3,g2} \fmf{plain}{g2,e4} \fmf{plain}{g2,g4} \fmf{plain}{e5,g3} \fmf{plain}{g3,g4} \fmf{plain}{g3,e6}
\end{fmfgraph*}
\end{fmffile} &\multicolumn{3}{c|}{ $\{{d_{\mathbb{C}}}_1,L_2\} \quad \{H_3,{H^\dagger}_4\} \quad \{{{d_{\mathbb{C}}}^\dagger}_5,{L^\dagger}_6\}$ } & \\ 
 \hline& $(0,\overline{3},2,-\frac{1}{6})$ & $(0,3,2,\frac{1}{6})$ & $(0,1,3,0)$ & $-2 \mathcal{O}^f_{1} + \mathcal{O}^f_{2}$\\ 
& $(0,\overline{3},2,-\frac{1}{6})$ & $(0,3,2,\frac{1}{6})$ & $(0,1,1,0)$ & $\mathcal{O}^f_{2}$\\
\hline
\end{longtable}

\begin{longtable}{cccc|c}
\caption{UV resonance of type $H{H^\dagger}L{L^\dagger}{u_{\mathbb{C}}}{{u_{\mathbb{C}}}^\dagger}$. ``*'' in front of each row indicates topologies with nonrenormalizable vertices. The UV resonances colored in gray are eliminated by UV selection. }\\
\hline\hline\multicolumn{5}{l}{\multirow{2}{*}{\large Type: $H{H^\dagger}L{L^\dagger}{u_{\mathbb{C}}}{{u_{\mathbb{C}}}^\dagger}$}}\\
 \multicolumn{5}{l}{} \\
\hline\multicolumn{5}{c}{$\begin{array}{ccl}\mathcal{O}^f_{{1}}&=& ({{L_p}_i}{{{u_{\mathbb{C}}}_r}^a})({{{L^\dagger}_s}^j}{{{{u_{\mathbb{C}}}^\dagger}_t}_a}){H_j}{{H^\dagger}^i},\\
\mathcal{O}^f_{{2}}&=& ({{L_p}_i}{{{u_{\mathbb{C}}}_r}^a})({{{L^\dagger}_s}^i}{{{{u_{\mathbb{C}}}^\dagger}_t}_a}){H_j}{{H^\dagger}^j}.\end{array}$}\\
\hline
\multicolumn{4}{c|}{group: $(\text{Spin},SU(3)_c,SU(2)_w,U(1)_y)$} & \multirow{2}{*}{$\mathcal{O}^{j}$}\\
\begin{fmffile}{6-1}
\begin{fmfgraph*}(30,30)
\fmfpen{thin} \fmfsurroundn{e}{6} \fmfvn{}{g}{4} 
\fmf{plain}{e1,g1} \fmf{plain}{g1,e2} \fmf{plain}{g1,g4} \fmf{plain}{e3,g2} \fmf{plain}{g2,e4} \fmf{plain}{g2,g4} \fmf{plain}{e5,g3} \fmf{plain}{g3,g4} \fmf{plain}{g3,e6}
\end{fmfgraph*}
\end{fmffile} &\multicolumn{3}{c|}{ $\{L_1,{u_{\mathbb{C}}}_2\} \quad \{H_3,{H^\dagger}_4\} \quad \{{L^\dagger}_5,{{u_{\mathbb{C}}}^\dagger}_6\}$ } & \\ 
 \hline& $(0,1,3,0)$ & $(0,\overline{3},2,-\frac{7}{6})$ & $(0,3,2,\frac{7}{6})$ & $-2 \mathcal{O}^f_{1} + \mathcal{O}^f_{2}$\\ 
& $(0,1,1,0)$ & $(0,\overline{3},2,-\frac{7}{6})$ & $(0,3,2,\frac{7}{6})$ & $\mathcal{O}^f_{2}$\\
\hline
\end{longtable}

\begin{longtable}{cccc|c}
\caption{UV resonance of type ${e_{\mathbb{C}}}{{e_{\mathbb{C}}}^\dagger}H{H^\dagger}Q{Q^\dagger}$. ``*'' in front of each row indicates topologies with nonrenormalizable vertices. The UV resonances colored in gray are eliminated by UV selection. }\\
\hline\hline\multicolumn{5}{l}{\multirow{2}{*}{\large Type: ${e_{\mathbb{C}}}{{e_{\mathbb{C}}}^\dagger}H{H^\dagger}Q{Q^\dagger}$}}\\
 \multicolumn{5}{l}{} \\
\hline\multicolumn{5}{c}{$\begin{array}{ccl}\mathcal{O}^f_{{1}}&=& ({e_{\mathbb{C}}}_p{Q_r}^{ai})({{e_{\mathbb{C}}}^\dagger}_s{{Q^\dagger}_t}^{aj}){H_j}{{H^\dagger}^i},\\
\mathcal{O}^f_{{2}}&=& ({e_{\mathbb{C}}}_p{Q_r}^{ai})({{e_{\mathbb{C}}}^\dagger}_s{{Q^\dagger}_t}^{ai}){H_j}{{H^\dagger}^j}.\end{array}$}\\
\hline
\multicolumn{4}{c|}{group: $(\text{Spin},SU(3)_c,SU(2)_w,U(1)_y)$} & \multirow{2}{*}{$\mathcal{O}^{j}$}\\
\begin{fmffile}{6-1}
\begin{fmfgraph*}(30,30)
\fmfpen{thin} \fmfsurroundn{e}{6} \fmfvn{}{g}{4} 
\fmf{plain}{e1,g1} \fmf{plain}{g1,e2} \fmf{plain}{g1,g4} \fmf{plain}{e3,g2} \fmf{plain}{g2,e4} \fmf{plain}{g2,g4} \fmf{plain}{e5,g3} \fmf{plain}{g3,g4} \fmf{plain}{g3,e6}
\end{fmfgraph*}
\end{fmffile} &\multicolumn{3}{c|}{ $\{{e_{\mathbb{C}}}_1,Q_2\} \quad \{H_3,{H^\dagger}_4\} \quad \{{{e_{\mathbb{C}}}^\dagger}_5,{Q^\dagger}_6\}$ } & \\ 
 \hline& $(0,3,2,\frac{7}{6})$ & $(0,\overline{3},2,-\frac{7}{6})$ & $(0,1,3,0)$ & $-2 \mathcal{O}^f_{1} + \mathcal{O}^f_{2}$\\ 
& $(0,3,2,\frac{7}{6})$ & $(0,\overline{3},2,-\frac{7}{6})$ & $(0,1,1,0)$ & $\mathcal{O}^f_{2}$\\
\hline
\end{longtable}
\input{dim8-6-2}

\section{UV List for Operators involving 7 particles}

\begin{longtable}{ccccc|c}
\caption{UV resonance of type ${e_{\mathbb{C}}}H^2{H^\dagger}^3L$. ``*'' in front of each row indicates topologies with nonrenormalizable vertices. The UV resonances colored in gray are eliminated by UV selection. }\\
\hline\hline\multicolumn{6}{l}{\multirow{2}{*}{\large Type: ${e_{\mathbb{C}}}H^2{H^\dagger}^3L$}}\\
 \multicolumn{6}{l}{} \\
\hline\multicolumn{6}{c}{$\begin{array}{ccl}\mathcal{O}^f_{1}&=&\frac{1}{12}\mathcal{Y}[\yng(2)_{H}]\mathcal{Y}[\yng(3)_{{H^\dagger}}]{{H^\dagger}^i}{{H^\dagger}^j}{{H^\dagger}^k}{H_j}{H_k}({e_{\mathbb{C}}}_p{{L_r}_i}).\end{array}$}\\
\hline
\multicolumn{5}{c|}{group: $(\text{Spin},SU(3)_c,SU(2)_w,U(1)_y)$} & \multirow{2}{*}{$\mathcal{O}^{j}$}\\
\begin{fmffile}{7-1}
\begin{fmfgraph*}(30,30)
\fmfpen{thin} \fmfsurroundn{e}{7} \fmfvn{}{g}{5} 
\fmf{plain}{e1,g1} \fmf{plain}{g1,e2} \fmf{plain}{g1,g2} \fmf{plain}{e3,g2} \fmf{plain}{g2,g3} \fmf{plain}{g3,g4} \fmf{plain}{e4,g3} \fmf{plain}{g4,g5} \fmf{plain}{e6,g5} \fmf{plain}{e5,g4} \fmf{plain}{g5,e7}
\end{fmfgraph*}
\end{fmffile} &\multicolumn{4}{c|}{ $\{{e_{\mathbb{C}}}_1,L_2\} \quad \{{e_{\mathbb{C}}}_1,L_2,H_3\} \quad \{H_4\} \quad \{{H^\dagger}_5,{H^\dagger}_6,{H^\dagger}_7\} \quad \{{H^\dagger}_5,{H^\dagger}_6\}$ } & \\ 
 \hline& $(0,1,2,\frac{1}{2})$ & $(0,1,3,-1)$ & $(0,1,3,1)$ & $(0,1,4,-\frac{3}{2})$ & $-3 \mathcal{O}^f_{{1}}$\\ 
\hline
& \multicolumn{4}{c|}{ $\{{e_{\mathbb{C}}}_1,L_2\} \quad \{{e_{\mathbb{C}}}_1,L_2,H_3\} \quad \{H_4\} \quad \{{H^\dagger}_5,{H^\dagger}_6,{H^\dagger}_7\} \quad \{{H^\dagger}_5,{H^\dagger}_7\}$ } & $\mathcal{O}^{j}$\\
\hline
\end{longtable}

\begin{longtable}{ccccc|c}
\caption{UV resonance of type ${d_{\mathbb{C}}}H^2{H^\dagger}^3Q$. ``*'' in front of each row indicates topologies with nonrenormalizable vertices. The UV resonances colored in gray are eliminated by UV selection. }\\
\hline\hline\multicolumn{6}{l}{\multirow{2}{*}{\large Type: ${d_{\mathbb{C}}}H^2{H^\dagger}^3Q$}}\\
 \multicolumn{6}{l}{} \\
\hline\multicolumn{6}{c}{$\begin{array}{ccl}\mathcal{O}^f_{1}&=&\frac{1}{12}\mathcal{Y}[\yng(2)_{H}]\mathcal{Y}[\yng(3)_{{H^\dagger}}]{{H^\dagger}^i}{{H^\dagger}^j}{{H^\dagger}^k}{H_j}{H_k}({{{d_{\mathbb{C}}}_p}^a}{Q_r}^{ai}).\end{array}$}\\
\hline
\multicolumn{5}{c|}{group: $(\text{Spin},SU(3)_c,SU(2)_w,U(1)_y)$} & \multirow{2}{*}{$\mathcal{O}^{j}$}\\
\begin{fmffile}{7-1}
\begin{fmfgraph*}(30,30)
\fmfpen{thin} \fmfsurroundn{e}{7} \fmfvn{}{g}{5} 
\fmf{plain}{e1,g1} \fmf{plain}{g1,e2} \fmf{plain}{g1,g2} \fmf{plain}{e3,g2} \fmf{plain}{g2,g3} \fmf{plain}{g3,g4} \fmf{plain}{e4,g3} \fmf{plain}{g4,g5} \fmf{plain}{e6,g5} \fmf{plain}{e5,g4} \fmf{plain}{g5,e7}
\end{fmfgraph*}
\end{fmffile} &\multicolumn{4}{c|}{ $\{{d_{\mathbb{C}}}_1,Q_2\} \quad \{{d_{\mathbb{C}}}_1,Q_2,H_3\} \quad \{H_4\} \quad \{{H^\dagger}_5,{H^\dagger}_6,{H^\dagger}_7\} \quad \{{H^\dagger}_5,{H^\dagger}_6\}$ } & \\ 
 \hline& $(0,1,2,\frac{1}{2})$ & $(0,1,3,-1)$ & $(0,1,3,1)$ & $(0,1,4,-\frac{3}{2})$ & $-3 \mathcal{O}^f_{{1}}$\\ 
\hline
& \multicolumn{4}{c|}{ $\{{d_{\mathbb{C}}}_1,Q_2\} \quad \{{d_{\mathbb{C}}}_1,Q_2,H_3\} \quad \{H_4\} \quad \{{H^\dagger}_5,{H^\dagger}_6,{H^\dagger}_7\} \quad \{{H^\dagger}_5,{H^\dagger}_7\}$ } & $\mathcal{O}^{j}$\\
\hline
\end{longtable}

\begin{longtable}{ccccc|c}
\caption{UV resonance of type $H^3{H^\dagger}^2Q{u_{\mathbb{C}}}$. ``*'' in front of each row indicates topologies with nonrenormalizable vertices. The UV resonances colored in gray are eliminated by UV selection. }\\
\hline\hline\multicolumn{6}{l}{\multirow{2}{*}{\large Type: $H^3{H^\dagger}^2Q{u_{\mathbb{C}}}$}}\\
 \multicolumn{6}{l}{} \\
\hline\multicolumn{6}{c}{$\begin{array}{ccl}\mathcal{O}^f_{1}&=&\frac{1}{12}\mathcal{Y}[\yng(2)_{{H^\dagger}}]\mathcal{Y}[\yng(3)_{H}]\epsilon^{ik}{{H^\dagger}^j}{{H^\dagger}^l}{H_j}{H_k}{H_l}({Q_p}^{ai}{{{u_{\mathbb{C}}}_r}^a}).\end{array}$}\\
\hline
\multicolumn{5}{c|}{group: $(\text{Spin},SU(3)_c,SU(2)_w,U(1)_y)$} & \multirow{2}{*}{$\mathcal{O}^{j}$}\\
\begin{fmffile}{7-1}
\begin{fmfgraph*}(30,30)
\fmfpen{thin} \fmfsurroundn{e}{7} \fmfvn{}{g}{5} 
\fmf{plain}{e1,g1} \fmf{plain}{g1,e2} \fmf{plain}{g1,g2} \fmf{plain}{e3,g2} \fmf{plain}{g2,g3} \fmf{plain}{g3,g4} \fmf{plain}{e4,g3} \fmf{plain}{g4,g5} \fmf{plain}{e6,g5} \fmf{plain}{e5,g4} \fmf{plain}{g5,e7}
\end{fmfgraph*}
\end{fmffile} &\multicolumn{4}{c|}{ $\{Q_1,{u_{\mathbb{C}}}_2\} \quad \{Q_1,{u_{\mathbb{C}}}_2,H_3\} \quad \{H_5\} \quad \{H_4,{H^\dagger}_6,{H^\dagger}_7\} \quad \{{H^\dagger}_6,{H^\dagger}_7\}$ } & \\ 
 \hline& $(0,1,3,-1)$ & $(0,1,2,-\frac{1}{2})$ & $(0,1,4,-\frac{1}{2})$ & $(0,1,3,0)$ & $\mathcal{O}^f_{{1}}$\\ 
& $(0,1,3,-1)$ & $(0,1,2,-\frac{1}{2})$ & $(0,1,2,-\frac{1}{2})$ & $(0,1,3,0)$ & $-2 \mathcal{O}^f_{{1}}$\\
\hline
\end{longtable}

\section{UV List for Operators involving 8 particles}

\begin{landscape}
%

\end{landscape}

\bibliographystyle{JHEP}
\bibliography{reference}

\providecommand{\href}[2]{#2}\begingroup\raggedright\begin{thebibliography}{10}

\bibitem{Isidori:2023pyp}
G.~Isidori, F.~Wilsch, and D.~Wyler, {\it {The Standard Model effective field
  theory at work}},  \href{http://arxiv.org/abs/2303.16922}{{\tt
  arXiv:2303.16922}}.

\bibitem{Falkowski:2023hsg}
A.~Falkowski, {\it {Lectures on SMEFT}},  {\em Eur. Phys. J. C} {\bf 83}
  (2023), no.~7 656.

\bibitem{Weinberg:1979sa}
S.~Weinberg, {\it {Baryon and Lepton Nonconserving Processes}},  {\em Phys.
  Rev. Lett.} {\bf 43} (1979) 1566--1570.

\bibitem{Buchmuller:1985jz}
W.~Buchmuller and D.~Wyler, {\it {Effective Lagrangian Analysis of New
  Interactions and Flavor Conservation}},  {\em Nucl. Phys. B} {\bf 268} (1986)
  621--653.

\bibitem{Grzadkowski:2010es}
B.~Grzadkowski, M.~Iskrzynski, M.~Misiak, and J.~Rosiek, {\it {Dimension-Six
  Terms in the Standard Model Lagrangian}},  {\em JHEP} {\bf 10} (2010) 085,
  [\href{http://arxiv.org/abs/1008.4884}{{\tt arXiv:1008.4884}}].

\bibitem{Lehman:2014jma}
L.~Lehman, {\it {Extending the Standard Model Effective Field Theory with the
  Complete Set of Dimension-7 Operators}},  {\em Phys. Rev. D} {\bf 90} (2014),
  no.~12 125023, [\href{http://arxiv.org/abs/1410.4193}{{\tt
  arXiv:1410.4193}}].

\bibitem{Liao:2016hru}
Y.~Liao and X.-D. Ma, {\it {Renormalization Group Evolution of Dimension-seven
  Baryon- and Lepton-number-violating Operators}},  {\em JHEP} {\bf 11} (2016)
  043, [\href{http://arxiv.org/abs/1607.07309}{{\tt arXiv:1607.07309}}].

\bibitem{Li:2020gnx}
H.-L. Li, Z.~Ren, J.~Shu, M.-L. Xiao, J.-H. Yu, and Y.-H. Zheng, {\it {Complete
  set of dimension-eight operators in the standard model effective field
  theory}},  {\em Phys. Rev. D} {\bf 104} (2021), no.~1 015026,
  [\href{http://arxiv.org/abs/2005.00008}{{\tt arXiv:2005.00008}}].

\bibitem{Murphy:2020rsh}
C.~W. Murphy, {\it {Dimension-8 operators in the Standard Model Eective Field
  Theory}},  {\em JHEP} {\bf 10} (2020) 174,
  [\href{http://arxiv.org/abs/2005.00059}{{\tt arXiv:2005.00059}}].

\bibitem{Li:2020xlh}
H.-L. Li, Z.~Ren, M.-L. Xiao, J.-H. Yu, and Y.-H. Zheng, {\it {Complete set of
  dimension-nine operators in the standard model effective field theory}},
  {\em Phys. Rev. D} {\bf 104} (2021), no.~1 015025,
  [\href{http://arxiv.org/abs/2007.07899}{{\tt arXiv:2007.07899}}].

\bibitem{Liao:2020jmn}
Y.~Liao and X.-D. Ma, {\it {An explicit construction of the dimension-9
  operator basis in the standard model effective field theory}},  {\em JHEP}
  {\bf 11} (2020) 152, [\href{http://arxiv.org/abs/2007.08125}{{\tt
  arXiv:2007.08125}}].

\bibitem{Harlander:2023psl}
R.~V. Harlander, T.~Kempkens, and M.~C. Schaaf, {\it {Standard model effective
  field theory up to mass dimension 12}},  {\em Phys. Rev. D} {\bf 108} (2023),
  no.~5 055020, [\href{http://arxiv.org/abs/2305.06832}{{\tt
  arXiv:2305.06832}}].

\bibitem{Li:2022tec}
H.-L. Li, Z.~Ren, M.-L. Xiao, J.-H. Yu, and Y.-H. Zheng, {\it {Operators for
  generic effective field theory at any dimension: on-shell amplitude basis
  construction}},  {\em JHEP} {\bf 04} (2022) 140,
  [\href{http://arxiv.org/abs/2201.04639}{{\tt arXiv:2201.04639}}].

\bibitem{Ellis:2020unq}
J.~Ellis, M.~Madigan, K.~Mimasu, V.~Sanz, and T.~You, {\it {Top, Higgs, Diboson
  and Electroweak Fit to the Standard Model Effective Field Theory}},  {\em
  JHEP} {\bf 04} (2021) 279, [\href{http://arxiv.org/abs/2012.02779}{{\tt
  arXiv:2012.02779}}].

\bibitem{Arzt:1994gp}
C.~Arzt, M.~B. Einhorn, and J.~Wudka, {\it {Patterns of deviation from the
  standard model}},  {\em Nucl. Phys. B} {\bf 433} (1995) 41--66,
  [\href{http://arxiv.org/abs/hep-ph/9405214}{{\tt hep-ph/9405214}}].

\bibitem{Einhorn:2013kja}
M.~B. Einhorn and J.~Wudka, {\it {The Bases of Effective Field Theories}},
  {\em Nucl. Phys. B} {\bf 876} (2013) 556--574,
  [\href{http://arxiv.org/abs/1307.0478}{{\tt arXiv:1307.0478}}].

\bibitem{Gargalionis:2020xvt}
J.~Gargalionis and R.~R. Volkas, {\it {Exploding operators for Majorana
  neutrino masses and beyond}},  {\em JHEP} {\bf 01} (2021) 074,
  [\href{http://arxiv.org/abs/2009.13537}{{\tt arXiv:2009.13537}}].

\bibitem{DasBakshi:2021xbl}
S.~Das~Bakshi, J.~Chakrabortty, S.~Prakash, S.~U. Rahaman, and M.~Spannowsky,
  {\it {EFT diagrammatica: UV roots of the CP-conserving SMEFT}},  {\em JHEP}
  {\bf 06} (2021) 033, [\href{http://arxiv.org/abs/2103.11593}{{\tt
  arXiv:2103.11593}}].

\bibitem{Naskar:2022rpg}
W.~Naskar, S.~Prakash, and S.~U. Rahaman, {\it {EFT Diagrammatica. Part II.
  Tracing the UV origin of bosonic D6 CPV and D8 SMEFT operators}},  {\em JHEP}
  {\bf 08} (2022) 190, [\href{http://arxiv.org/abs/2205.00910}{{\tt
  arXiv:2205.00910}}].

\bibitem{Banerjee:2022thk}
U.~Banerjee, J.~Chakrabortty, C.~Englert, S.~U. Rahaman, and M.~Spannowsky,
  {\it {Integrating out heavy scalars with modified equations of motion:
  Matching computation of dimension-eight SMEFT coefficients}},  {\em Phys.
  Rev. D} {\bf 107} (2023), no.~5 055007,
  [\href{http://arxiv.org/abs/2210.14761}{{\tt arXiv:2210.14761}}].

\bibitem{Cepedello:2022pyx}
R.~Cepedello, F.~Esser, M.~Hirsch, and V.~Sanz, {\it {Mapping the SMEFT to
  discoverable models}},  {\em JHEP} {\bf 09} (2022) 229,
  [\href{http://arxiv.org/abs/2207.13714}{{\tt arXiv:2207.13714}}].

\bibitem{Banerjee:2023iiv}
U.~Banerjee, J.~Chakrabortty, S.~U. Rahaman, and K.~Ramkumar, {\it {One-loop
  Effective Action up to Dimension Eight: Integrating out Heavy Scalar(s)}},
  \href{http://arxiv.org/abs/2306.09103}{{\tt arXiv:2306.09103}}.

\bibitem{deBlas:2017xtg}
J.~de~Blas, J.~C. Criado, M.~Perez-Victoria, and J.~Santiago, {\it {Effective
  description of general extensions of the Standard Model: the complete
  tree-level dictionary}},  {\em JHEP} {\bf 03} (2018) 109,
  [\href{http://arxiv.org/abs/1711.10391}{{\tt arXiv:1711.10391}}].

\bibitem{Guedes:2023azv}
G.~Guedes, P.~Olgoso, and J.~Santiago, {\it {Towards the one loop IR/UV
  dictionary in the SMEFT: one loop generated operators from new scalars and
  fermions}},  \href{http://arxiv.org/abs/2303.16965}{{\tt arXiv:2303.16965}}.

\bibitem{Li:2023cwy}
X.-X. Li, Z.~Ren, and J.-H. Yu, {\it {A complete tree-level dictionary between
  simplified BSM models and SMEFT (d $\leq$ 7) operators}},
  \href{http://arxiv.org/abs/2307.10380}{{\tt arXiv:2307.10380}}.

\bibitem{Li:2022abx}
H.-L. Li, Y.-H. Ni, M.-L. Xiao, and J.-H. Yu, {\it {The bottom-up EFT: complete
  UV resonances of the SMEFT operators}},  {\em JHEP} {\bf 11} (2022) 170,
  [\href{http://arxiv.org/abs/2204.03660}{{\tt arXiv:2204.03660}}].

\bibitem{Jiang:2020rwz}
M.~Jiang, J.~Shu, M.-L. Xiao, and Y.-H. Zheng, {\it {Partial Wave Amplitude
  Basis and Selection Rules in Effective Field Theories}},  {\em Phys. Rev.
  Lett.} {\bf 126} (2021), no.~1 011601,
  [\href{http://arxiv.org/abs/2001.04481}{{\tt arXiv:2001.04481}}].

\bibitem{Shu:2021qlr}
J.~Shu, M.-L. Xiao, and Y.-H. Zheng, {\it {Constructing the general partial
  wave and renormalization in effective field theory}},  {\em Phys. Rev. D}
  {\bf 107} (2023), no.~9 095040, [\href{http://arxiv.org/abs/2111.08019}{{\tt
  arXiv:2111.08019}}].

\bibitem{Li:2020tsi}
H.-L. Li, Z.~Ren, M.-L. Xiao, J.-H. Yu, and Y.-H. Zheng, {\it {Low energy
  effective field theory operator basis at d \ensuremath{\leq} 9}},  {\em JHEP}
  {\bf 06} (2021) 138, [\href{http://arxiv.org/abs/2012.09188}{{\tt
  arXiv:2012.09188}}].

\bibitem{Li:2020zfq}
H.-L. Li, J.~Shu, M.-L. Xiao, and J.-H. Yu, {\it {Depicting the Landscape of
  Generic Effective Field Theories}},
  \href{http://arxiv.org/abs/2012.11615}{{\tt arXiv:2012.11615}}.

\bibitem{Li:2021tsq}
H.-L. Li, Z.~Ren, M.-L. Xiao, J.-H. Yu, and Y.-H. Zheng, {\it {Operator bases
  in effective field theories with sterile neutrinos: d \ensuremath{\leq} 9}},
  {\em JHEP} {\bf 11} (2021) 003, [\href{http://arxiv.org/abs/2105.09329}{{\tt
  arXiv:2105.09329}}].

\bibitem{Ma:2019gtx}
T.~Ma, J.~Shu, and M.-L. Xiao, {\it {Standard Model Effective Field Theory from
  On-shell Amplitudes}},  \href{http://arxiv.org/abs/1902.06752}{{\tt
  arXiv:1902.06752}}.

\bibitem{Shadmi:2018xan}
Y.~Shadmi and Y.~Weiss, {\it {Effective Field Theory Amplitudes the On-Shell
  Way: Scalar and Vector Couplings to Gluons}},  {\em JHEP} {\bf 02} (2019)
  165, [\href{http://arxiv.org/abs/1809.09644}{{\tt arXiv:1809.09644}}].

\bibitem{Arkani-Hamed:2017jhn}
N.~Arkani-Hamed, T.-C. Huang, and Y.-t. Huang, {\it {Scattering Amplitudes For
  All Masses and Spins}},  \href{http://arxiv.org/abs/1709.04891}{{\tt
  arXiv:1709.04891}}.

\bibitem{Criado:2018sdb}
J.~C. Criado and M.~P\'erez-Victoria, {\it {Field redefinitions in effective
  theories at higher orders}},  {\em JHEP} {\bf 03} (2019) 038,
  [\href{http://arxiv.org/abs/1811.09413}{{\tt arXiv:1811.09413}}].

\end{thebibliography}\endgroup

\end{document}